\documentclass[12pt]{article}
\pdfoutput=1
\usepackage{amsmath,amssymb,amsthm}
\usepackage{bm}
\usepackage{bbm}
\usepackage{pdflscape}
\usepackage[usenames,dvipsnames]{xcolor}
\usepackage{graphicx}
\usepackage[para]{threeparttable}
\usepackage{longtable}
\usepackage{booktabs}
\usepackage{float}
\usepackage{array}
\usepackage{arydshln}
\usepackage{multirow}
\usepackage{rotfloat}
\usepackage{caption}
\usepackage{subcaption}
\usepackage{bbold}
\usepackage[normalem]{ulem}
\usepackage{cite}
\usepackage{hyperref}
\usepackage{cleveref}
\usepackage{tikz}
\newcommand{\fsu}{\mathfrak{su}}
\newcommand{\fso}{\mathfrak{so}}
\newcommand{\fsp}{\mathfrak{sp}}
\newcommand{\fg}{\mathfrak{g}}
\newcommand{\ff}{\mathfrak{f}}
\newcommand{\fe}{\mathfrak{e}}

\numberwithin{equation}{section}
\numberwithin{figure}{section}

\usepackage{parskip}
\usepackage{setspace}
\graphicspath{{figures/}}

\DeclareMathOperator{\Tr}{Tr}
\DeclareMathOperator{\tr}{tr}
\DeclareMathOperator{\rk}{rk}

\theoremstyle{definition}
\newtheorem{definition}{Definition}

\def\node#1#2{\overset{#1}{\underset{#2}{\circ}}}
\def\bnode#1#2{\overset{\textcolor{blue}{#1}}{\underset{#2}{\textcolor{blue}{\bullet}}}}
\def\ver#1#2{\overset{{\llap{$\scriptstyle#1$}\displaystyle\circ{\rlap{$\scriptstyle#2$}}}}{\scriptstyle\vert}}
\def\iver#1#2{\overset{\scriptstyle\vert}{{\llap{$\scriptstyle#1$}\displaystyle\circ{\rlap{$\scriptstyle#2$}}}}}
\def\bver#1#2{\overset{{\llap{$\scriptstyle\textcolor{blue}{#1}$}\displaystyle\textcolor{blue}{\bullet}{\rlap{$\scriptstyle#2$}}}}{\scriptstyle\vert}}

\usetikzlibrary{arrows}
\tikzstyle{every picture}+=[remember picture]
\tikzstyle{na} = [baseline=-.5ex]
\tikzstyle{mine}= [arrows={angle 90}-{angle 90},thick]

\def\Llleftarrow{%
\lower2pt\hbox{\begingroup
\tikz
\draw[shorten >=0pt,shorten <=0pt] (0,3pt) -- ++(-1em,0) (0,1pt) -- ++(-1em-1pt,0) (0,-1pt) -- ++(-1em-1pt,0) (0,-3pt) -- ++(-1em,0) (-1em+1pt,5pt) to[out=-105,in=45] (-1em-2pt,0) to[out=-45,in=105] (-1em+1pt,-5pt);
\endgroup}
}

\begin{document}
 
~\vspace{2cm}
\begin{center}
\vspace*{-3cm} 
\begin{flushright}
{\tt ZMP-HH/24-10}

\end{flushright}

{\LARGE\bfseries Bounds and Dualities of\\[12pt] Type II Little String Theories}
\vspace{1.2cm}

Florent Baume$^{\,a,}$\footnote{\href{mailto:florent.baume@desy.de}{florent.baume@desy.de}}, 
Paul-Konstantin Oehlmann$^{\,b,}$\footnote{\href{mailto:p.oehlmann@northeastern.edu}{p.oehlmann@northeastern.edu}}, Fabian Ruehle$^{\,b,c,d,}$\footnote{\href{mailto:f.ruehle@northeastern.edu}{f.ruehle@northeastern.edu}}   \\[10mm]
\bigskip
{  
    {\it ${}^{\text{a}}$II. Institut fur Theoretische Physik, Universitat Hamburg, Hamburg 22607, Germany}\\[.5em]
	{\it ${}^{\text{b}}$Department of Physics, Northeastern University, Boston, MA 02115, USA}\\[.5em]
	{\it ${}^{\text{c}}$Department of Mathematics, Northeastern University, Boston, MA 02115, USA}\\[.5em]
	{\it ${}^{\text{d}}$NSF Institute for Artificial Intelligence and Fundamental Interactions, Boston, MA, USA}\\[.5em]
   
}
\end{center}
\setcounter{footnote}{0} 
\bigskip\bigskip

\begin{abstract}
		\noindent We explore the symmetry structure of Type II Little String Theories and their T-dualities. We construct these theories both from the bottom-up perspective starting with seed Superconformal Field Theories, and from the top-down using F-/M-theory. By exploiting anomaly inflow and unitarity of the LST worldsheet theory, we derive strong conditions on the possible 6D bulk theories and their flavor algebras. These constraints continue to apply if gravity is coupled to the theory. We also study the
        higher form symmetry structure of these theories and show how they get exchanged under T-duality. Finally, we comment on seemingly consistent bottom-up Little String Theories that cannot be constructed from the top-down approach.
\end{abstract}

\clearpage
\tableofcontents
\clearpage

\newpage

\section{Introduction} 
Six-dimensional supersymmetric Quantum Field Theories (SQFTs) can have two
different ultraviolet (UV) complete behaviors: one corresponds to theories that
have emergent superconformal symmetry and a local energy-momentum tensor, and are characterized by a tower of tensionless strings
\cite{Witten:1995zh,Strominger:1995ac,Seiberg:1996qx}. The second type of
theories admits strings in the spectrum that do not become tensionless in the
UV. The tension of these strings sets a characteristic scale $M_S$, which defines the UV cutoff above which a Hagedorn growth of states emerges
\cite{Aharony:1998ub}. This is very characteristic of the closed-string sector,
which contains the graviton, and although LSTs are not coupled to gravity,
the UV behavior of both theories are therefore very similar. Just like SCFTs, LSTs admit a
``tensor branch'' description below $M_s$, which may not be perturbative, but
allows for a description in terms of ordinary 6D quantum fields, such as vector, hyper-, and tensor multiplets.

SUGRA, LSTs and SCFTs do not exist in isolation from each other: Both LSTs and SCFTs can be obtained by a careful decoupling limit of a gravity theory that sends the Planck mass to infinity while preserving a non-trivial interacting
theory. In particular, one may view LSTs and SCFTs as a generic sub-sector of (almost any) gravity theory. The converse, however, is generally not true: Only a finite subset of theories lying in the
infinite landscape of SCFTs and LSTs can be coupled to gravity in a consistent
fashion. This is also clear from the expectation that the landscape of
supergravity theories is finite.\footnote{Although believed to belong to the
Swampland, new classes of models seemingly consistent with all known
field-theory constraints have been constructed in
\cite{Hamada:2023zol,Loges:2024vpz, Hamada:2024oap}.}

LSTs and SCFTs also very closely related: in fact, there are various ``tensor-decoupling limits'' in which the Little String scale is taken to infinity, that give rise to SCFTs. It has been proposed that a plethora of 6D SCFTs may be
obtained as a decoupling limit of a (not necessarily unique)
LST~\cite{Bhardwaj:2015oru}. Conversely, most SCFTs can be ``affinized''---in a
fashion described below---to at least one
LST.\footnote{In~\cite{Bhardwaj:2019hhd}, the author proposes the existence of
SCFTs that do not descend from LSTs.} Thus the landscape of a generic
UV-complete SQFT will consist mostly of LSTs and deformations thereof.

String theory---and by extension F-theory---provides an elegant and consistent
way of constructing and studying SUGRAs, LSTs, and SCFTs via geometry.  This
systematic approach may therefore give us hints about the nature of their
landscape, and the potential consistency conditions that are yet missing from a
low-energy point of view. Furthermore, our current understanding of symmetries
has recently undergone rapid acceleration, initiated in \cite{Gaiotto:2014kfa},
see e.g. \cite{Bah:2022wot, Schafer-Nameki:2023jdn, Bhardwaj:2023kri} and
references therein for an overview of recent progress. Guided by geometry, a
first step towards the study of LSTs and their non-local properties is 
to map out their symmetry and duality structures. In the context of six-dimensional QFTs, recent work shows that these theories generally admit a wide range of higher-form symmetries (HFS)
\cite{Bhardwaj:2020phs, Apruzzi:2021mlh, Hubner:2022kxr, Cvetic:2021sxm,
Apruzzi:2020zot, Dierigl:2020myk, Dierigl:2023jdp, Heckman:2022muc,
Heckman:2022suy, Apruzzi:2022Dlm}. In this regard, LSTs are special, since they
possess a unique continuous global 1-form symmetry mixing non-trivially with
other conventional 0-form symmetries to form a continuous 2-group
\cite{Cordova:2018cvg, Cordova:2020tij}.\footnote{In contrast, 6D SCFTs may
admit \textit{discrete} 2-group symmetries~\cite{Apruzzi:2021mlh}.} This
improved understanding of symmetries therefore yields a finer characterization
of the theories and provides novel invariants across dualities
\cite{DelZotto:2020sop}.

These invariants, together with their geometric engineering, give us additional leverage to chart the space of LSTs modulo T-duality, and has been used recently to study LSTs of Heterotic type \cite{DelZotto:2022ohj,DelZotto:2022xrh, DelZotto:2023ahf, DelZotto:2023nrb, Mansi:2023faa, Lawrie:2023uiu}. These theories are defined by a choice of either $G_{\text{Het}}=E_8 \times E_8$ or $G_{\text{Het}}=Spin(32)/\mathbbm{Z}_2$ Heterotic string theory, and by a stack of $M$ NS5 branes probing a transverse
ADE-type singularity $\mathbbm{C}^2/\Gamma$, whose worldvolume is described by the 6D theory.\footnote{For work on twisted T-duals in theories with eight supercharges, see \cite{Bhardwaj:2022ekc, Anderson:2023wkr, AhmedII}, and for generalization to frozen F-theory vacua see \cite{Oehlmann:2024cyn}.} We denote these theories by
 \begin{align}
     \mathcal{K}^{G_{\text{Het}}}(\fg, \mu)_M \, ,
 \end{align}
where $\mathfrak{g}$ is the algebra associated with the singularity, and $\mu$ is a homomorphism embedding $\Gamma\rightarrow G_{\text{Het}}$. These holonomies
act on the 10D gauge groups $G_{\text{Het}}$, which lives on non-compact flavor 9-branes in the Ho\v{r}ava--Witten or Type I dual pictures. Unsurprisingly, this flavor symmetry structure can be elegantly geometrized into the Picard
lattice of an elliptically polarized K3 fiber, from which other T-dual theories
can be systematically be derived~\cite{DelZotto:2022xrh,DelZotto:2023ahf}.
 
This work aims to extend this exploration to the second type of LSTs finding
their origin in Type II string theories. We will therefore refer to them as
\emph{Type II LSTs}, in contrast to the \emph{Heterotic LSTs} described above.
This type of theories is instead specified by two choices of singularities
$(\fg_F,\fg_B)$, that are also of ADE type (or possibly of the special Kodaira
Type II, III, IV). We denote these theories by
\begin{align}
    \mathcal{K}^{\text{II}}( \fg_F,\fg_B) \,.
\end{align}
In the Type IIB description, $\mathfrak{g}_F$ and $\mathfrak{g}_B$ may be
viewed as 7- and 5-brane singularities, respectively.

The absence of a 10D gauge group that could become generic 6D flavor branes severely restricts this class of models as compared to Heterotic LSTS. Moreover, it allows for non-trivial higher-form symmetries: first, there is a
diagonal 1-form symmetry sector acting diagonally on gauge group factors
\cite{Apruzzi:2020zot}, and secondly, there may be a non-trivial string defect
group \cite{DelZotto:2015isa}. In this work we show
explicitly that these higher-form symmetries $\mathcal{D}^{(1)}$ and $\mathcal{D}^{(2)}$ have a natural relation to the
defining singularities via the centers $Z$ of the associated algebras,
\begin{align}
    \mathcal{D}^{(1)}:~Z(\fg_F) \, , \qquad \mathcal{D}^{(2)}:~Z(\fg_B) \, .
\end{align} 
Being yet another set of generalized symmetries, it is natural to expect that
they are preserved under T-duality. Indeed, we explicitly show that T-duality
acts as an exchange of the defining singularities
\begin{align}
\label{eq:Tduality}
    \fg_F \overset{\text{T-Duality }}{ \longleftrightarrow} \fg_B
\end{align}
and hence, exchanges the higher form symmetry sectors.

Geometry gives both a beautiful and self-consistent way to construct either
kind of LSTs from the top down, which can be used to prove the relation
\eqref{eq:Tduality}.  From the bottom-up perspective, there are \emph{a
priori} substantially more theories that satisfy all known consistency conditions than those that can be realized via geometric engineering. This is true even
when considering the frozen phase of F-theory, see \cite{Witten:1997bs,
Morrison:2023hqx}, and when gravity is coupled to the theory. Novel consistency
conditions have recently been uncovered in the presence of extended BPS
strings, whose worldsheet theories receive current algebra contributions
from bulk symmetries via anomaly inflow. Unitarity considerations for these
extended objects then allow to constrain bulk symmetries in SUGRA theories
\cite{Kim:2019vuc,Lee:2019skh,Tarazi:2021duw}. Similar BPS strings are
also present in theories without gravity such as SCFTs and LSTs, and one might
therefore expect analogous consistency conditions to play a similarly important
role even when gravity is decoupled. Typically, many such BPS strings are
present in a generic 6D theory, which makes a general analysis rather
cumbersome, and the same is true for LSTs.\footnote{Rank-one SCFTs on the other
hand are simple enough to extract universal features of BPS strings
\cite{DelZotto:2018tcj}.}  However, as their name suggests, Little String
Theories have a universal characteristic string which receives contributions
only from the bulk flavor symmetries for which strong constraints can be
derived.

This work is structured as follows. In Section~\ref{sec:two}, we give a
general account of 6D LSTs from a field theory perspective: in
Section~\ref{ssec:review} we set notation and review basic concepts, and
Section~\ref{sec:HFSDuality} focuses in particular on the generalized symmetry
structure of Type II and Heterotic LSTs, and we propose that they are new T-duality invariants. In Section~\ref{sec:flavbounds}, we derive new
bounds on the flavor symmetry of these theories by demanding unitarity of the Little String worldsheet description. In Section~\ref{sec:BottomUp}, we give a
bottom-up construction of Type II LSTs and discuss their dualities and HFS
structures. In Section~\ref{sec:geometry} we give a top-down construction of
Type II LST and present a geometric framework in which T-duality is manifest via a double fibration structure. We give an outlook and conclusion in Section~\ref{sec:outlook}.
Appendix~\ref{app:tables} we summarize the possible Type II LSTs, Appendix~\ref{app:6D-anomalies} gives more details on six- and two-dimensional SCFTs, and Appendix~\ref{app:Enhanced} summarizes SUSY enhancements and possible obstructions in the 5D M-theory picture.

\textit{Note added: while writing up this work, a related
article~\cite{Zhang:2024oas} appeared on the arXiv, which has some overlap with our results of matching discrete higher-form symmetries across T-duality of LSTs as described in Section \ref{sec:HFSDuality}.}

\section{The Symmetry Structure of Little Strings}\label{sec:two}

Many Little String Theories (LSTs) can be engineered by fusing 6D SCFTs together
\cite{Bhardwaj:2015oru}. This vast landscape of theories can be ordered
by their symmetries. Moreover, different LSTs can be related by T-duality, either in pairs or in some cases in larger families of T-dual theories. The notion of T-duality for LSTs is the same
as for the 10D $E_8\times E_8$ and Spin(32)$/\mathbb{Z}_2$ Heterotic strings:
both theories are different in 10D, but after compactifying both theories on a circle and moving along their lower-dimensional moduli space, ones reaches a point where
both theories are identical.\footnote{Phrased differently, T-duality means that there is a KK theory that can be UV completed by two (or more) higher dimensional theories.} This, however, allows one to define certain 6D invariants
that must match across each T-dual theory:
\begin{enumerate}
		\item The 5D Coulomb branch dimension, $\text{dim}(\text{CB})=n_T+\rk(G)$, where $n_T$ is the number of tensor multiplets and $rk(G)$ is the rank of the gauge group.
    \item The universal 2-group structure constants: 
    $(\kappa_P,\kappa_R)$ \cite{DelZotto:2020sop}.
    \item The rank of the 0-form flavor symmetry $G_F$, $\rk(G_F) $ \cite{DelZotto:2020sop,Ahmed:2023lhj,Lawrie:2023uiu} . 
    \item The 5D 1-form symmetry sector $ \mathcal{D}^{(1)}_{6D}\times \mathcal{D}^{(2)}_{6D} \rightsquigarrow \mathcal{D}^{(1)}_{5D}$, which receives contribution from the 6D 1-form symmetry and the defect group as discussed in Section~\ref{sec:HFSDuality}. 
\end{enumerate}
These invariants---which will be reviewed in more detail below---needs to match individually across T-duality, and therefore severely constrain a candidate dual theory. Not all of the invariants listed above are independent; in fact, the most central one is the 2-group structure constant $\kappa_P$, which can only take two values~\cite{Bhardwaj:2015oru}:
\begin{align}
\begin{split}
    \kappa_P&=   2 \, \qquad \text{ for Heterotic LST}\, , \\
\kappa_P&=   0 \, \qquad \text{ for Type II LST}\, .
\end{split}
\end{align}
In terms of F-theory, this is equivalent to the statement that LSTs can be constructed from non-compact elliptically fibered Calabi--Yau threefolds.  These geometries can only have two types of birational bases,
\begin{align}
    B_{2}^\text{Het}= \mathbbm{P}^1 \times \mathbbm{C} \, , \quad \text{ and } \quad B_{2}^{\text{T-II}}= (\mathbbm{T}^2 \times \mathbbm{C})/\Lambda \,,
\end{align}
where $\Lambda \subset SU(2)$ is a discrete group associated with the algebra $\mathfrak{g}$. Physically, we may interpret the structure constants $\kappa_P$ in three different ways: 
\begin{enumerate}
    \item from their 10D LST origin, which is either Type II or Heterotic 5-branes;
    \item as the number of flavor 9-branes;
    \item on the LST 2D worldsheet, the quantity $8 \kappa_p$ counts the number of 3-7 string defects \cite{Lawrie:2016axq}.
\end{enumerate} 
All three perspectives are consistent with each other: $\kappa_p$ determines
possible non-trivial bulk flavor symmetries. The three perspectives will allow us to derive new consistency conditions from the LST worldsheet in
Section~\ref{sec:flavbounds}.

\subsection{Review}\label{ssec:review}

Before delving into the analysis of Type II LSTs and their symmetries, we shortly review the six-dimensional generalized quivers and the main properties of LSTs, which we use as an opportunity to set our notation and conventions. For a
more in-depth treatment of generalized quivers and their F-theory construction, we refer the reader to the reviews ~\cite{Heckman:2018jxk, DelZotto:2023ahf}. 

The tensor branch of a six-dimensional $\mathcal{N}=(1,0)$ theory can be
described in terms of weakly-coupled supermultiplets. In the presence of tensor
multiplets, there exist additional BPS strings which couple naturally to the
self-dual tensor fields, inducing a Dirac pairing $\eta^{IJ}$ for the strings.
In the F-theory description, they are obtained by wrapping branes on curves
$\mathcal{C}^I$ in the base of the elliptic fibration, and the pairing is given
by their intersection form\footnote{Note that the matrix $\eta$ is sometimes
defined as the intersection matrix itself rather than the Dirac pairing:
$\eta^{IJ} = + \mathcal{C}^I\cdot \mathcal{C}^J$. It is then negative
semi-definite, since the curve have to have negative self-intersection in order
for them to be shrinkable to zero size to reach the LST phase. In our
convention, the diagonal elements are always positive integers.}
\begin{equation}\label{def-eta}
		\eta^{IJ} = - \mathcal{C}^I\cdot \mathcal{C}^J \,.
\end{equation}
Since the matrix $\eta^{IJ}$ has an interpretation as the Dirac pairing of strings, it must be positive semi-definite. For LSTs, we furthermore demand the presence of a \emph{single} null eigenvector, which we refer to as the (integer-valued) LST charge $\ell^\text{LST}$,
\begin{equation}\label{def-LST-charge}
		\eta^{IJ}\ell^\text{LST}_J = 0\,,\qquad \ell^\text{LST}_I>0\,~\forall I\,, \qquad\text{gcd}(\ell^\text{LST}_1, \ell^\text{LST}_2,\dots) = 1\,.
\end{equation}
Here and throughout, we sum over repeated indices. The LST charge is then interpreted as the coefficients of the linear combination of self-dual two-forms that couple to the Little String:
\begin{equation}
		B_2^\text{LST} = \ell_I^\text{LST}\,B_2^I\,,
\end{equation}
where $B^I$ is the tensor associated with the curve $\mathcal{C}^I$. There then
remains $n_T$ dynamical tensors.

In addition to tensor multiplets, there can also be vector and hypermultiplets
charged under gauge symmetries. To ensure gauge anomaly cancellation, each
gauge algebra must be associated with a tensor multiplet, or equivalently its magnetic dual string. Indeed, in six dimensions, gauge anomalies generically
do not cancel and a Green--Schwarz--West--Sagnotti mechanism, or GS mechanism for short, mediated by the tensor fields is necessary to obtain a consistent theory \cite{Green:1984bx, Sagnotti:1992qw}. In F-theory, the gauge
sector arises naturally from non-trivial elliptic fibers over the curves
$\mathcal{C}^I$, and the dictionary between the matter content and the geometry
can be obtained straightforwardly \cite{Heckman:2013pva, Heckman:2015bfa}.
Gauge anomaly cancellation is then guaranteed if the theory descends from a well-defined elliptically fibered Calabi--Yau compactification.

It is common to summarize the spectrum of an $\mathcal{N}=(1,0)$ six-dimensional
theory on its tensor branch in a pictorial way called a (generalized) quiver.
The $I^\text{th}$ tensor multiplet with $\eta^{II}=n$ associated with an
algebra $\mathfrak{g}$ is denoted by
\begin{equation}
		\overset{\mathfrak{g}}{n}\,.
\end{equation}
If there is more than one tensor multiplet, they are arranged in a graph whose adjacency matrix is given by the Dirac pairing. If the algebra is trivial, the label $\mathfrak{g}$ is omitted. Furthermore, flavor symmetries of tensor multiplets are indicated with brackets, $[\mathfrak{f}]$. 

For instance, the following quiver represents a Heterotic LST:
\begin{equation}
		[\mathfrak{so}_{16}]\overset{\mathfrak{sp}_N}{1}\overset{\mathfrak{sp}_N}{1}[\mathfrak{so}_{16}]\,,\qquad 
		\eta = \begin{pmatrix}
			1&-1\\-1&1		
		\end{pmatrix}\,,
\end{equation}
The two strings of charge $1$ intersect once, and there are $2N+8$
hypermultiplets transforming in the fundamental representation of each
$\mathfrak{sp}_N$, necessary to ensure anomaly cancellation. Some of these
transform in the bifundamental representation $(\bm{2N},\bm{2N})$, and there
are sixteen remaining half-hypermultiplets on each side, rotated by an $\mathfrak{so}(16)$ flavor symmetry. As one can see, the quiver notation provides a convenient book-keeping device that (almost) uniquely encodes the spectrum upon demanding absence of anomalies.

In F-theory, two curves are depicted side by side if they have a normal crossing intersection. While this is enough when considering SCFTs, the requirement of a
unique null eigenvector for the Dirac pairing of an LST allows for different patterns of intersections: 
\begin{equation}\label{base-pattern-notation}
\begin{tabular}{lc}
				Normal intersection: &\qquad $\overset{\mathfrak{g}_1}{m}\,\overset{\mathfrak{g}_2}{n}$\,,\\[5pt]
				Tangential intersection: &\qquad $\overset{\mathfrak{g}_1}{m}||\overset{\mathfrak{g}_2}{n}$\,,\\[5pt]
				Triple intersection: &\qquad $\overset{\mathfrak{g}_1}{n_1} \overset{\displaystyle \overset{\mathfrak{g}_2}{n_2}}{\Delta} \overset{\mathfrak{g}_3}{n_3}$\,,\\[5pt]
				Loop configuration: &\qquad $//\overset{\mathfrak{g}_1}{n_1}\,\overset{\mathfrak{g}_2}{n_2}\,\dots\,  \overset{\mathfrak{g}_{k-1}}{n_{k-1}}\,\overset{\mathfrak{g}_k}{n_k}//$
\end{tabular}
\end{equation}
In the last case, the symbol $//$ denotes the intersection between the first
and last curve.

Note that the quiver notation summarizes the theory at a generic point in the tensor branch. At such a generic point, the strings, being BPS objects, have a tension $T_I$ that is given by the vacuum expectation value of the scalar field $t_I$ of the associated tensor multiplet, $T_I\sim\langle t_I\rangle$. For theories where $\eta$ does not have a null eigenvector, moving to a point where the vacuum expectation value of all tensor fields vanish, leads to a conformal fixed point where no scales can be present. However, the existence of the null eigenvector $\ell^\text{LST}$ means that not all curves can be shrunk and hence ensures that an intrinsic LST scale remains.

The two-form field $B_2^\text{LST}$ acts as a background field for the $U(1)_\text{LST}^{(1)}$ one-form symmetry, and	the Bianchi identity of the associated 3-form field strength $H_3^\text{LST}$ controls the 2-group structure
constant:
\begin{equation}\label{Bianchi-LST}
		dH_3^\text{LST} = \widehat{\kappa}_R c_2(R) - \frac{\kappa_P}{4} p_1(T) - \kappa_F^A c_2(F_A) \,,
\end{equation}
where $c_2(R)$, $c_2(F_A)$ are the one-instanton-normalized second Chern classes of the $R$- and  $\mathfrak{f}=\oplus_A\mathfrak{f}^A$ flavor symmetry
bundles respectively, and $p_1(T)$ is the first Pontryagin class of the spacetime tangent bundle. The two-group structure constants can then be obtained from the quiver data by expanding $B_2^\text{LST}$ in terms of the tensors $B_2^I$. At a generic point in the tensor branch, their
respective Bianchi identities are given by
\begin{equation}
		dH_3^I = h^I c_2(R) - \frac{a^I}{4} p_1(T) - B^{IA} c_2(F_A) \,,
\end{equation}
where $h^I=h^\vee_{\mathfrak{g}^I}$ is the dual Coxeter number of the $I^\text{th}$ algebra, $a^I = 2-\eta^{II}$, and $B^{IA}$ encodes the pairing between the $I^\text{th}$ string and the $A^\text{th}$ flavor algebra. Comparing with equation \eqref{Bianchi-LST}, we find that the
2-group structure constants are given by
\begin{equation}\label{def-kappas}
		\kappa_P = h^I\ell_I^\text{LST}\,,\qquad
		\kappa_R = a^I\ell_I^\text{LST}\,,\qquad
		\kappa_F^A = B^{IA}\ell_I^\text{LST}\,.
\end{equation}
In \cite{DelZotto:2020sop}, the first two structure constants were called \textit{universal structure constants}. They have to match across T-duality, which is not true for the third structure constant~$\kappa_{F}^A$.

\subsection{Higher Symmetry Dualities in LSTs}\label{sec:HFSDuality}

Beyond the 2-group symmetry structure, 6D QFTs may also admit general 
(discrete) higher form symmetries. In the following, we shortly review how such symmetries can arise and propose to use them as novel invariants across T-duality. Our main
interest lies in the following two types of generalized symmetries 
\begin{enumerate}
	\item \textbf{Discrete center 1-form
			symmetries $\boldsymbol{\mathcal{D}^{(1)}_{6D}}$} associated to Wilson lines not screened by dynamical
			gauge or matter fields.
	\item \textbf{Defect Group $\boldsymbol{\mathcal{D}^{(2)}_{6D}}$} associated to self-dual string defects that can not be screened by dynamical BPS strings \cite{DelZotto:2015isa}.
\end{enumerate}  
We propose that a suitable combination of the above higher symmetries serves as a novel invariant across T-dual LSTs. 

Since LSTs may be T-dual in 5D, we next review circle compactifications of higher form symmetries. Compactifying a 6D theory on a circle to 5D decomposes an $n$-form symmetry into $\mathcal{D}^{(n)}_{D-1}\times\mathcal{D}^{(n-1)}_{D-1}$
symmetries. From the perspective of a $(D-n-1)$-dimensional topological operator that generates the $D$-dimensional $n$-form symmetry, the two possibilities correspond to a case where the defect does or does not wrap the circle. 
Indeed, the 1-form symmetry $\mathcal{D}^{(1)}$ becomes a 1-form and a 0-form symmetry in 5D. 

As discussed in \cite{DelZotto:2015isa}, the 6D defect group 
$\mathcal{D}^{(2)}_{6D}$ is special as strings are self-dual. Upon compactification to five dimensions these objects remain either strings, or lines when they wrap the circle, and are related to one- and two-form symmetries. Due to the self-duality in six dimensions we can then make a choice of electric polarization in five dimensions which keeps only the 5D one-form symmetry  \cite{Bhardwaj:2020phs, Gukov:2020btk, Morrison:2020ool, Lawrie:2023tdz}. Thus we obtain two contributions to the 1-form symmetries in 5D,
\begin{align}
     \mathcal{D}^{(1)}_{6D} \times \mathcal{D}^{(2)}_{6D} ~\rightsquigarrow~ \mathcal{D}^{(1)}_{5D} \,.
\end{align}
We therefore expect T-dual LSTs to have a non-trivial 1-form and defect group structure that combines into a single 5D 1-form symmetry upon circle compactification.

\subsubsection*{6D Defect Group}
The defect group $\mathcal{D}^{(2)}_{6D}$ of LSTs can be obtained from the Dirac pairing matrix~$\eta^{IJ}$. The defect group receives contributions from string defects, modulo those that can be screened by dynamical ones, which we can represent as a lattice quotient. However, one of these combinations is exactly the little string itself appearing as a null vector in~$\eta^{IJ}$ and coupling to a non-dynamical tensor field.  We already attribute this to the $U(1)_\text{LST}^{(1)}$ symmetry, which has to be quotiented out so that the defect group is given by 
\begin{align}
\label{eq:2formSyms}
    \mathcal{D}^{(2)}_{6D}= \frac{\mathbbm{Z}^{n_T+1}}{ [\eta^{IJ}]\mathbbm{Z}^{n_T+1}} / \mathbbm{Z} = \prod_{i=1}^{n_T} \mathbbm{Z}_{m_i} \,  .
\end{align}
A convenient way to compute the defect group is via the Smith normal form of $\eta$,
\begin{align}
    M = S \cdot \eta \cdot T\,.
\end{align}
Here, $S$ and $T$ are invertible matrices and $M$ is diagonal. The integers
$m_i$ in equation~\eqref{eq:2formSyms} are given by the diagonal entries of $M$. Note that for LSTs there is always a null entry $m_0=0$ associated with the null eigenvector of $\eta$. The other diagonal element are then $m_i \geq 0$, $i=1\ldots n_T$. The defect groups of LSTs and SCFTs are therefore computed in the same way, up to the presence of $m_0$ in the former case.

It is easy to check that $\mathcal{D}^{(2)}_{6D}$ is invariant under blow-ups and blowdowns of smooth points \cite{DelZotto:2015isa}. It might therefore be simplest to compute the defect group for a maximal (smooth) blowdown of the quiver, a phase we will refer to as the \textit{endpoint configuration}. 
For LSTs, the respective endpoint configurations and possible defect groups are  
\begin{align}
\begin{split}
\begin{array}{lll}
	\text{Heterotic LSTs: }& \quad  \eta=(0) \, , \qquad & \mathcal{D}^{(2)}_{6D}= \emptyset \,,\\
	\text{~Type II~ LSTs: }& \quad\eta= \widehat{A}(G)\, , \qquad & \mathcal{D}^{(2)}_{6D}= Z(G) \,.
  \end{array}
  \end{split}
\end{align}  
Here, $\widehat{A}(G)$ is the affine extended Dynkin diagram of the gauge algebra $G$ and $Z(G)$ is its center. For all ADE groups, we have listed the respective center symmetries in  Table~\ref{tab:CenterSyms}. As already remarked in~\cite{DelZotto:2020sop}, Heterotic LSTs cannot have a non-trivial defect group. In contrast, Type~II LSTs generally have a non-trivial defect group which, however, is bounded by the ADE center symmetries. Note that this strongly constrains the higher symmetry structure of Heterotic LSTs: since the defect group is always trivial, two Heterotic LSTs can only be T-dual if they have the same global center 1-form symmetry. However, in all examples discussed in the literature~\cite{DelZotto:2020sop,DelZotto:2022ohj,DelZotto:2022xrh,DelZotto:2023ahf,Ahmed:2023lhj}, only gauged center 1-form symmetries were found.

\subsubsection*{1-form Symmetries}
Center 1-form symmetries are the simplest higher form symmetries and are hence well-explored and understood. A center 1-form symmetry is associated to line defects, such as Wilson lines of a gauge group $G$ with non-trivial center $Z(G)$. The 1-form symmetry $\mathcal{D}^{(1)}$ acts on Wilson lines that are in the root lattice of $G$ modulo the screening by dynamical W-bosons living in the co-root lattice. In $\mathcal{N}=(1,0)$ theories, there can be additional sources that can break the symmetry, such as hypermultiplets charged under $Z(G)$. Depending on their representations, Wilson lines may then end on those hypermultiplets and get screened, such that the 1-form symmetry is also broken. This can be generalized to non-simple gauge groups, $G=\prod_I G_I$ with center $Z(G)=\prod_I Z(G_I)$. This combined center symmetry cannot be fully realized, since various matter representations charged under~$G$ (typically bifundamentals of $G_I\times G_{I\pm 1}$) have to be present in the theory to ensure gauge anomaly cancellation. Nevertheless, there is generically a diagonal combination of center symmetries $\mathcal{Z} \in Z(G)$ that acts trivially on all matter representations, which can then be a center 1-form symmetry. 

The putative center 1-form symmetry identified above can be rendered inconsistent by BPS instanton strings that are charged under it. The consistency of the 1-form symmetry can be studied as follows \cite{Cordova:2019uob,Apruzzi:2020zot,Heckman:2022suy}: for a consistent 1-form symmetry $\mathcal{D}^{(1)}_{6D}=\mathcal{Z}\in Z(G)$, one should be able to treat it as a non-trivial background field. The background is parameterized by a twist vector $\vec{k}$ for $\mathcal{Z}$ whose entries are integral values in the centers 
\begin{align}
    k_I \in Z(G_I)\, .
\end{align}
Upon switching on the above background twist, the respective gauge instanton fractionalizes~\cite{Cordova:2019uob}. By performing large gauge transformations of the 2-form fields, we are then left with a fractional shift that leads to a phase in the path integral. This phase is trivial if
\begin{align}
\label{eq:obstruct}
   \sum_j \eta^{IJ}  k_J^2 \alpha_{G_J}   = 0 \text{ mod 1} \quad \forall I\,,
\end{align}
which signals consistency of the chosen 1-form symmetry background. Note that the $k_i$ have to be trivial for the flavor symmetry factors (which are mostly absent in Type II LSTs anyways), and hence do not appear in the sum~\eqref{eq:obstruct}.

The above consistency check is the same for center 1-form \textit{quotient symmetries} that appears in the total symmetry group $\widetilde{G}= ([G_F] \times G) / \mathcal{Z}$, which acts simultaneously on gauge and flavor group factors.  This is precisely what happens in Heterotic LSTs: the global gauge group structure is fixed by the flavor branes. The $\mathbbm{Z}_2$ quotient of the $Spin(32)/\mathbbm{Z}_2$ flavor group acts diagonally on all gauge group factors\footnote{
The global quotient structure can be engineered in F-theory, for both compact and non-compact cases via a non-trivial Mordell--Weil torsion group \cite{Aspinwall:1997ye,Aspinwall:1998xj,Dierigl:2020myk,Apruzzi:2020zot,Hubner:2022kxr}.} .

\begin{table}[t!]
\centering
\renewcommand{\arraystretch}{1.3}
\begin{tabular}{lcc}
$G$ & $Z(G)$ & $\alpha_{G}$ \\ 
\toprule
$SU(N)$ & $\mathbbm{Z}_N$ & $\frac{N-1}{2N}$ \\ 
\midrule 
$Sp(N)$ & $\mathbbm{Z}_2$ & $\frac{N}{4}$ \\ 
\midrule 
$\textit{Spin}(2N+1)$& $\mathbbm{Z}_2$ & $\frac12$ \\ 
$\textit{Spin}(4N+2)$& $\mathbbm{Z}_4$ & $\frac{2N+1}{4}$ \\ 
$\textit{Spin}(4N)$& $\mathbbm{Z}_2^s\times \mathbbm{Z}_2^c$ & $(\frac{N}{4},\frac{1}{2})$\\ 
\midrule
$G_2$ & $\mathbbm{Z}_1$ & $-$ \\ 
$F_4$ & $\mathbbm{Z}_1$ & $-$ \\ 
$E_6$ & $\mathbbm{Z}_3$ & $\frac23$ \\ 
$E_7$ & $\mathbbm{Z}_2$ & $\frac34$ \\ 
$E_8$ & $\mathbbm{Z}_1$ & $-$ \\ 
\bottomrule
\end{tabular} 
\renewcommand{\arraystretch}{1.0}
\caption{Summary of center symmetries and group theory coefficients for all simple groups, reproduced from~\cite{Oehlmann:2024cyn}. For $\textit{Spin}(4N)$, the two $\mathbbm{Z}_2$ factors act on the spinor and co-spinor, respectively; the vector representation is charged under both.}
    \label{tab:CenterSyms}
\end{table}

\subsubsection*{Examples}

We illustrate the above considerations with a few concrete examples. First, let
us consider a theory with a $G=SU(n)$ singularity, whose quiver has the
necklace shape of the corresponding affine Dynkin diagram with Dirac pairing
matrix
\begin{align}
    \eta=\left( \begin{array}{ccccc}
    2 & -1 & 0 & .. & -1 \\
    -1 & 2 & -1 & .. & 0 \\
    0 & -1 & 2 & ..  & 0 \\
    ..& .. & .. & .. & .. \\
     -1 & 0 & 0 & .. &2
    \end{array} \right)
\end{align}
and LST vector $\vec{\ell}^\text{\,LST} = (1,1,1,\ldots 1)$. The defect group can be computed using the Smith normal form, which is given by 
\begin{align}
    M = \text{diag}(0,1,1,\ldots, 1,n)\,.
\end{align} 
Hence, the defect group of the theory is
\begin{align}
\mathcal{D}^{(2)}=\mathbbm{Z}_n \, .
\end{align}
We may now decorate the theory with an $SU(m)$ gauge symmetry, which gives the quiver
\begin{align}
    // \underbrace{\overset{\fsu_m}{2} \, \,  \overset{\fsu_m}{2} \, \,  \overset{\fsu_m}{2} \ldots \overset{\fsu_m}{2} }_{ n\times } //\,.
\end{align}
The theory has a non-trivial diagonal $1$-form symmetry: all matter transforms in bifundamental representations $(\boldsymbol{m},\boldsymbol{\overline{m}})$, which leaves the diagonal $\mathbbm{Z}_m$ $1$-form symmetry unbroken upon choosing all twists $k_i=1$.
The 2-group structure constants and Coulomb branch dimension of this theory are
\begin{align}
	\kappa_P = 0 \,,\qquad  
	\kappa_R =m \cdot n\,,\qquad  
	\text{dim}(\text{CB}) = m \cdot n-1 \, ,
\end{align}
using the formulas given in equation \eqref{def-kappas}
Together with the higher form symmetries 
\begin{align}
   \mathcal{D}^{(2)}:~\mathbbm{Z}_n^{(2)} \, , \qquad \mathcal{D}^{(1)}:~\mathbbm{Z}_m^{(1)}    \,,
\end{align}
this gives the T-duality invariant data. An obvious T-dual of the above theory is given by exchanging fiber and base
singularities, which results just in a flip of the respective $1$-form and $2$-form symmetries, leaving CB and 2-group structure constant invariant. 

The situation at hand is in fact even more general: recall that the total space of the fibration can be represented by the orbifold of $X_3=(\mathbbm{T}^2 \times \mathbbm{T}^2 \times
\mathbbm{C}^1)/\Lambda$ with $\Lambda$ given by the two actions
\begin{align*}
    \mathbbm{Z}_n:\quad & g_n:(e^{2 \pi \text{i}/n }, \mathbbm{1},e^{-2 \pi \text{i}/n } )  \\ 
    \mathbbm{Z}_m:\quad & g_m:(\mathbbm{1}, e^{2 \pi \text{i}/m },e^{-2 \pi \text{i}/m } ).
\end{align*}
If $m$ and $n$ are co-prime, the Chinese Remainder Theorem tells us that we can write
\begin{align}
    \Lambda=\mathbbm{Z}_m \times \mathbbm{Z}_n = \mathbbm{Z}_{m \cdot n} \, .
\end{align}
leading to a $\mathbbm{Z}^{(1)}_{n \cdot m}$ 1-form symmetry in 5D. 
This hints at the possibility that there might be more T-dual Type II LSTs of type
$\mathcal{K}^\text{II}(\fsu_k, \fsu_l)$ with
\begin{align}
    k \cdot l = m \cdot n \, ,
\end{align}
as those have the same $\mathbbm{Z}_{n m}^{(1)}$ 1-form symmetry, Coulomb branch dimension and 2-group structure constants.
Indeed, it is has been proposed in \cite{Hohenegger:2016yuv,
Bastian:2017ing} that there exist flop transitions in the extended K\"ahler cone of the threefold which lead to multiple T-duals for each $k,l$ partition of $n,m$
. Since the 1-form symmetry group in 5D is insensitive to flop transitions, this perspective would be consistent with conditions imposed by the matching of higher form symmetries. Similarly, for $\text{gcd}(n,m)=k$ we can then write $
\Gamma=\mathbbm{Z}_{n \cdot m/k} \times \mathbbm{Z}_k $,
which gives the maximal 1-form symmetry in  5D. In fact, any other partition $n m= \tilde{n} \tilde{m}$ that leads to the above symmetry is also consistent, which implies the additional condition gcd$(\tilde{n},\tilde{m})=k$. In Section~\ref{sec:geometry}, we give an explicit geometric construction for the case of two inequivalent elliptic fibrations.

The computation of the defect group is generally straightforward by either moving to the endpoint configuration, where the base has an ADE singularity and taking its center, or by directly computing the Smith Normal Form on the full tensor branch geometry. Determining the center 1-form symmetry can be more involved, so we discuss two more examples in the following to illustrate the computation. Generically, we might need to identify a center symmetry transformation on the gauge group factors that acts faithfully on all representations.

Our next example is a slightly more complicated LST, which is the theory $\mathcal{K}^{\text{II}}(\fsu_n, \fso_{12})$ with tensor branch quiver
\begin{align} 
	\mathcal{K}^{\text{II}}(\fsu_n, \fso_{12}):\qquad
	\overset{\mathfrak{su}_{n}}{2} \, \overset{\displaystyle \overset{\mathfrak{su}_{n}}{2}}{\overset{\mathfrak{su}_{2n}}{2}} \, 
 \overset{\mathfrak{su}_{2n}}{2} \,
 \overset{\displaystyle \overset{\mathfrak{su}_{n}}{2}}{\overset{\mathfrak{su}_{2n}}{2}} \, \overset{\mathfrak{su}_{n}}{2}\,.
\end{align}
The total gauge group has a $\mathbbm{Z}_n^4 \times \mathbbm{Z}_{2n}^3 $ center, which is broken to a diagonal subgroup by the various bifundamental matter representations of the quiver. For example, the fundamental representation $\boldsymbol{n}$ of $\fsu_n$ has center charge 1 and transforms with phase  
\begin{align}
    \mathbbm{Z}_n: \phi( \mathbf{n} )= e^{\frac{2 \pi i}{n}} \, .
\end{align}
Bifundamental representations such as $(\boldsymbol{n},\boldsymbol{\overline{2n}})$ may leave a diagonal center symmetry $\mathcal{Z}=\mathbbm{Z}_n \subset \mathbbm{Z}_n \times \mathbbm{Z}_{2n}$ invariant, since
\begin{align}
 (\phi_1 \otimes \phi_2^2)(\boldsymbol{n},\boldsymbol{\overline{2n}}) =  e^{\frac{2\pi i}{n}}\cdot e^{\frac{-2\cdot  2\pi i}{ 2n}}=\mathbbm{1} \,.
\end{align}
The above $\mathbbm{Z}_n^{(1)}$ generator must be further extended to encompass all gauge symmetry factors in order to account for all bifundamental hypermultiplets. Furthermore, we need to check the BPS string obstruction in \eqref{eq:obstruct}. We find a consistent $\mathbbm{Z}_n$ background flux for the following twist choice for $k_i$ in \eqref{eq:obstruct}: $k_i=1$ for $G_i=SU(n)$ and $k_j=2$ for $G_j=SU(2n)$. In summary, the HFS structure of the above LST is
\begin{align}
  \mathcal{D}^{(2)}:  \mathbbm{Z}^{s\;(2)}_2 \times \mathbbm{Z}^{c\;(2)}_2 \, , \qquad   \mathcal{D}^{(1)}:  \mathbbm{Z}^{(1)}_n
\end{align}
We may proceed similarly for the $\mathcal{K}^{\text{II}}( \fso_{12},\fsu_n)$ theory, whose quiver has the necklace shape
\begin{align}
 \label{eq:SO12SUnExample}
 \mathcal{K}^{\text{II}}( \fso_{12},\fsu_n):\qquad //  \overset{\fso_{12}}{4}\, \overset{\fsp_2}{1} \, \overset{\fso_{12}}{4}\, \overset{\fsp_2}{1}  \, \ldots \overset{\fso_{12}}{4}\, \overset{\fsp_2}{1}//
\end{align}
with $n$ pairs of $\overset{\fso_{12}}{4}\, \overset{\fsp_2}{1}$. 
Each $\fso_{12}$ gauge factor has a $\mathbbm{Z}^s_2 \times \mathbbm{Z}^c_2$ center symmetry and each $\fsp_2$ factor has a $\mathbbm{Z}_2$ symmetry. The fundamentals of $\fso_{4N}$ are charged under both $\mathbbm{Z}^{s/c}_2$ factors, and $\fsp_M$ fundamentals are charged under the $\mathbbm{Z}_2$ center symmetry of $Sp(2)$. Hence, we obtain two invariant generators. One is the diagonal combination of the $\mathbbm{Z}_2^s\times\mathbbm{Z}_2$ generator, and the other of the $\mathbbm{Z}^{c}_2\times\mathbbm{Z}_2$ generator, $\phi_{\fso_{12}}^{s/c} \otimes \phi_{\fsp_{2}}$, which acts trivially on the bifundamental representations. With this, the full quiver has a
\begin{align}
\label{eq:ExampleCenter}
Z=\prod_{i=1}^{n}(\mathbbm{Z}_{2,i}^s \times \mathbbm{Z}_{2,i}^c \times \mathbbm{Z}_{2,i})
\end{align} 
center symmetry. The bifundamental matter will be invariant under a diagonal combination of the invariant generators of each block,
\begin{align}
\mathbbm{Z}_2^{s\;(1)} \times \mathbbm{Z}_2^{c\;(1)}: \qquad   \bigotimes_{i=1}^{n} \left( \phi_{\fso_{12},i}^{s/c}\otimes  \phi_{\fsp_{2},i} \right) \,,
\end{align}
Note that there are no spinor or co-spinor representations that would break the respective $\mathbbm{Z}^{s/c}_2$ center symmetries. Further we also have to check that the BPS strings are consistent with the center symmetries. 

Checking the BPS string obstruction works out just as in the previous case: there are two non-trivial center twists $(k^s,k^c)$ for an $SO(4N)$, each of which contributes a term~\cite{Heckman:2022suy}
\begin{align}
   \eta^{ij} \left(\frac{N_j}{4}(k_j^s+k_j^c)^2 + \frac12(k_j^s \cdot k_j^c) \right)
\end{align} 
to the condition~\eqref{eq:obstruct}. For each of the $n$ blocks $\overset{\fso_{12}}{4}\, \overset{\fsp_2}{1}$ in the quiver~\eqref{eq:SO12SUnExample}, we need to choose three $\mathbbm{Z}_2$-valued entries for $k_i$, which encode the $\mathbbm{Z}^s_{2,i} \times \mathbbm{Z}^c_{2,i}\times \mathbbm{Z}_{2,i}$ for the $SO(12)_i\times Sp(2)_i$ centers. There are two consistent choices: we can choose $k_i=(1,0,1)$ or $k_i=(0,1,1)$. Both choices correspond to a 1-form symmetry background generator of order two, and it is straightforward to check that the path integral phase~\eqref{eq:obstruct} is trivial for either of them. 

To summarize, the higher form symmetries of this example are
\begin{align}
  \mathcal{D}^{(2)}:  \mathbbm{Z}^{(2)}_n \, , \qquad   \mathcal{D}^{(1)}:  \left(\mathbbm{Z}^{(1)}_2\right)^2 \, .
\end{align}
The two Type II LSTs in the example are T-dual, which is consistent with the exchange of the higher form symmetry sectors we observe. 

\subsection{Flavor Bounds from the LST Worldsheet Theory}\label{sec:flavbounds}

We now turn to the 0-form flavor symmetries of LSTs. The presence of BPS
string in the spectrum is known to impose constraints on the flavor symmetries
demanding unitarity of the worldsheet theory
\cite{Kim:2019vuc,Lee:2019skh,Tarazi:2021duw}. 

From the perspective of the six-dimensional effective description at a generic
point of the tensor branch, the D3-branes wrapping two-cycles in the geometry
arise as strings with finite tension. Their worldsheet description is that of a
two-dimensional $\mathcal{N}=(0,4)$ gauge theory, which flows in the deep
infrared (IR) to a 2D SCFT. By studying the consequences of unitarity for such
SCFTs, one can obtain constraints on the allowed spectrum and the properties of
extended objects, since the two-dimensional central charges of the theory are inherited from protected quantities in six dimensions. We will follow the methods developed in
\cite{Kim:2019vuc}, which were used to obtain bounds for the rank of the
possible gauge groups appearing in supergravity theories containing BPS
strings.

The relevant quantities are encoded in the anomaly polynomial of the 2D worldsheet theory
\begin{equation}\label{I4-def}
		I_4 = \sum_{a} k^a c_2(F_a) - \frac{k_G}{24}p_1(T_2)\,,
\end{equation}
where $c_2(F_a)=\frac{1}{4}\Tr F_a^2$ are the one-instanton-normalized
second Chern classes of the worldsheet flavor symmetries, and $p_1(T_2)$ is the
first Pontryagin class of the worldsheet tangent bundle. Note that both the
gauge and flavor symmetries of the 6D bulk are seen as flavor symmetries from
the worldsheet point of view. The anomaly polynomial $I_4$ can be written in
term of the string charges and their intersection pairing, as well as other
quantities of the 6D theory~\cite{Berman:2004ew, Shimizu:2016lbw}. In the deep IR,
the emergent superconformal invariance fixes the left- and right-handed central
charges in terms of the level of the
$\mathfrak{su}(2)_l\oplus\mathfrak{su}(2)_r$ R-symmetry, and their difference
is furthermore fixed by the gravitational anomaly $k_G$.\footnote{We use
lowercase subscripts to indicate the worldsheet left and right R-symmetry,
reserving the letter $R$ for the bulk R-symmetry as the focus of this work is
on six-dimensional theories. In the literature, $c_2(I)$ is sometimes used for
the Chern class of the bulk R-symmetry bundle.} In the normalization defined in
equation~\eqref{I4-def}, we have
\begin{equation}\label{central-charges}
		c_l = 6 k_l\,,\qquad
		c_r = 6 k_r\,,\qquad c_r - c_l = k_G\,.
\end{equation}
On the worldsheet, the 6D bulk flavor and gauge symmetries $\mathfrak{g}_a$
manifest themselves through the presence of holomorphic currents associated
with a Kac--Moody algebra $\widehat{\mathfrak{g}}^a$ at level $k^a$ in the spectrum,
whose contribution to the left-handed central charge is given by
\begin{equation}
		c_F = \sum_{a}\frac{k^a\text{dim}(\mathfrak{g}^a)}{k^a+h^\vee_{\mathfrak{g}^a}}\,,
\end{equation}
where $h^\vee_{\mathfrak{g}^a}$ is the dual Coxeter number of the algebra $\mathfrak{g}^a$. We are therefore led to a lower bound on the central charge, which constrains the possibles flavor symmetries:
\begin{equation}\label{flavor-bound-general}
		c_F\leq c_l\,.
\end{equation}

While these constraints have been derived for supergravity theories in the context of the Swampland Program \cite{Kim:2019vuc, Lee:2019skh}, they are very generic, and we will now apply them to LSTs.\footnote{
Via generalized blow-up equations, the elliptic genus of LSTs has been computed recently in 
 \cite{Kim:2023glm}.} All the necessary quantities can be obtained from the anomaly polynomial of the six-dimensional effective theory --- which can be computed directly from the
tensor-branch description --- and are reviewed in Appendix~\ref{app:6D-strings} for convenience and to set our conventions. 

This gives us the spectrum on the D3-brane, and the anomaly polynomial of the CFT is obtained by removing the contributions of the center-of-mass universal hypermultiplet:
\begin{equation}
		I_4^\text{CFT} = I_4 - I_4^\text{CoM}\,.
\end{equation}
Furthermore, the left and right R-symmetries of the UV description might mix
with the other worldsheet flavor factor, changing that of the IR CFT. In our
case, the right-handed part of the R-symmetry $\mathfrak{su}(2)_r$ remains
unchanged. The left-handed central charges can then be inferred from the
gravitational anomaly, see equation~\eqref{central-charges}, which is easily
read off from the anomaly polynomial. We obtain
\begin{equation}
		c_l = 6 k_l = 6k_r + k_G = 3(3Q_Ia^I - Q_I\eta^{IJ}Q_J) + 2\,,
\end{equation}
where $Q_I$ denotes the charge of the string, and $a^I=2-\eta^{II}$. A short
summary and additional details on the derivation of the central charges can be
found in Appendix~\ref{app:6D-strings}. A bound on the possible symmetries then
depends on the values of the levels $k^a$. Those arise from the Green--Schwarz
coupling in the 6D bulk theory, and are given by
\begin{equation}
		k^a =  B^{a I} Q_I\,,
\end{equation}
where $B^{aI}$ is the pairing between the $a^\text{th}$ flavor and the $I^\text{th}$ string charge.

Let us now specialize to the little string itself. In that case, the charge of the string is by definition
\begin{equation}
		Q_I = \ell^\text{LST}_I\,,\qquad
		\eta^{IJ}\ell^\text{LST}_I = 0\,,
\end{equation}
see equation \eqref{def-LST-charge}. Moreover, the string charge
also fixes the two-group structure constant,
\begin{equation}
		\kappa_P = a^I\ell^\text{LST}_I\,,\qquad a^I = 2-\eta^{II}\,.
\end{equation}
On the worldsheet we can therefore simply rewrite the left-handed central
charge as
\begin{equation}
		c_l = 9\,\kappa_P + 2\,.
\end{equation}
The bound then follows from the value of the possible worldsheet flavor
symmetry. As already mentioned above, from the two-dimensional point of view,
both bulk gauge $\mathfrak{g}^I$ and flavor $\mathfrak{f}^A$ symmetries appear
as flavor symmetries. We must therefore distinguish between the two types of
possibilities, namely when $a=I$ where the flavor pairing is simply
$B^{IJ}=\eta^{IJ}$, and those of the bulk flavor symmetries
\begin{equation}
		k^I = \eta^{IJ}\ell_J^\text{LST} = 0\,,\qquad k^A = B^{AI}\ell_I^\text{LST}
\end{equation}
We then conclude that only bulk flavor symmetries can contribute to the bound
given in equation~\eqref{flavor-bound-general}. Moreover, in the geometric
picture, the flavor pairing matrix is given by the intersection of D7-branes
wrapping non-compact divisors $\overline{D}^A$ with quiver curves $w^I$,
$B^{AI} = \overline{D}^A\cdot w^I$, so that in the overwhelming majority of
cases $B^{AI}\in\{0,1\}$, and the level of the flavor symmetry is simply the
charge of the tensor the flavor is attached to in the quiver
description.\footnote{As pointed out in \cite{Baume:2021qho}, there are a few
cases where the flavor pairing entries can be larger than one, occurring for instance in the
presence of undecorated $(-1)$-curves in the quiver description, where the
$\mathfrak{e}_8$ flavor is decomposed into a non-maximal subalgebra.}
Putting everything together, we
therefore obtain the bound
\begin{equation}\label{flavor-bound-WZW}
		\sum_{A}\frac{k^A\text{dim}(\mathfrak{f}^A)}{k^A + h^\vee_A} \leq 9\,\kappa_P + 2\,,
\end{equation}
where we remind the reader that the sum is taken only over the flavor symmetry factors $\mathfrak{f}=\oplus_A\mathfrak{f}^A$ of the 6D theory. Equation \eqref{flavor-bound-WZW} can be used to derive a bound on the total
rank of the flavor symmetry. Indeed, a consequence of the strange formula of
Freudenthal and de Vries is that for any simple algebra
\begin{equation}
		\rk(\mathfrak{g})\leq\frac{\text{dim}(\mathfrak{g})}{(1+h^\vee_{\mathfrak{g}})}\,.
\end{equation}
This relation is saturated when $\mathfrak{g}$ is simply laced, and is easily found to be correct by inspection, see Table \ref{tab:algebra-values}. 

We conclude that given an LST with two-group structure constant
$\kappa_P$, the rank of its total flavor symmetry $\mathfrak{f}=\oplus_A\mathfrak{f}^A$ must satisfy
\begin{equation}\label{bound-flavor-general}
		\rk(\mathfrak{f}) \,\leq\, 9\, \kappa_P + 2\,.
\end{equation}
Note that this bound is weaker than the one given in equation
\eqref{flavor-bound-WZW}, as the levels $k^A$ related to the little string
charges $\ell_I^{\text{LST}}$ can be larger than one. At the end of this section, we give an example of a would-be LST that satisfies the bound given in equation~\eqref{bound-flavor-general}, but violates the stronger constraints of equation~\eqref{flavor-bound-WZW}.

Using the two possible values of $\kappa_P$ in the Heterotic and Type II cases, we obtain the bounds
\begin{equation}\label{bound-flavor-specific}
	\begin{aligned}
			\text{Heterotic LSTs:}&\qquad \rk(\mathfrak{f})\leq 20\,,\\
			\text{Type II  LSTs:}&\qquad \rk(\mathfrak{f})\leq 2\,.\\
	\end{aligned}
\end{equation}
For Type II LSTs, we see that there is very little room for a potential flavor
symmetry; the only possibilities are $\mathfrak{f}\in\{\mathfrak{u}(1),
\mathfrak{u}(1)^2, \mathfrak{u}(1) \oplus \mathfrak{su}(2), \mathfrak{so}(4),
\mathfrak{su}(3)\}$ flavor algebras. On the other hand, it is well known that a
wealth of other algebras are permitted for Heterotic LSTs.  For instance, a
rank-$16$ flavor arises from the 3-7 strings stretching between the LST curve
and the two (unbroken) M9-branes. When fixing $\mathfrak{e}_8^2$ as the flavor
symmetry, one has $k_G=16$ as expected, and there are very few possible
additional contributions, such as $\mathfrak{u}(1)^4, \mathfrak{su}(2)^4$ or a
combination thereof. Notably, a concrete case of an Heterotic LST with rank-18
flavor has been discussed in \cite{DelZotto:2022xrh}. We are however not aware
of any theory saturating the bounds above.

To obtain the bound in equation~\eqref{bound-flavor-general}, we have used the
anomaly polynomial for strings arising in six-dimensional $\mathcal{N}=(1,0)$
theories derived in \cite{Shimizu:2016lbw}. The numerical values can be
confirmed from the F-theory point of view by looking at the matter content
arising from D3-branes wrapping a curve $\mathcal{C}\subset B_2$. For an LST,
the curve has by definition self-intersection zero,
$\mathcal{C}\cdot\mathcal{C}=0$, and its genus $g$ is either $g=1$ for the
Heterotic cases, or $g=0$ for Type II cases. The multiplicity of the
$\mathcal{N}=(0,4)$ multiplets was derived in~\cite{Lawrie:2016axq} for curves
of arbitrary genera, which we have collected in Table~\ref{tab:WSContent} for
convenience.\footnote{We have used that by adjunction, $2(g-1)=\mathcal{C}^2 -
\mathcal{C} \cdot c_1(B)$ to write the multiplicities purely in terms of the
self-intersection of the curve and its genus.} The anomaly polynomial is then
obtained by summing the contributions of each supermultiplets. 
\begin{table}[t!]
    \centering
    \begin{tabular}{ccccc}
     \toprule 
       (0,4) Multiplet &  Multiplicity &$ (c_l, c_r)$& Het LST $(g=0)$ & Type II LST $(g=1)$ \\ \midrule
      Hyper  &  $\mathcal{C}^2+1-g$ & (4,6)& 1 & 0  \\ 
Twisted Hyper & 1 & (4,6)&1 & 1\\
Fermi & $g$ & (2,0) &0 &1   \\
Half-Fermi & $8\mathcal{C}^2+ 16(1-g)$& (1,0)& 16 & 0\\  
\bottomrule 
    \end{tabular}
	\caption{Summary of the $(0,4)$ 2D SCFT field content and their contributions to left- and right-handed central charges of a D3-brane wrapping a curve $\mathcal{C}$ of genus $g$. The multiplicities are evaluated for Heterotic ($g=0$) and Type II ($g=1$) LSTs using $\mathcal{C}^2=0$.}
    \label{tab:WSContent}
\end{table} 
It is then easy to see that, taking the center-of-mass modes into account, the
left- and right-handed central charges are those discussed above.
\section{LSTs from Minimal Affinization of SCFTs}\label{sec:BottomUp}

We have seen that the possible flavor symmetries of general LSTs are quite
constrained, in particular those of Type II LSTs. Conversely, Superconformal
Field Theories (SCFTs) can host a vast zoo of flavor symmetries, which can be used to
construct LSTs. Indeed, through an operation called \emph{fusion}~\cite{Heckman:2018pqx}, certain SCFTs can be glued together to obtain an LST.

From the field-theory point of view, this procedure gauges a
common flavor symmetry (or a subalgebra thereof) of two decoupled SCFTs to
obtain a new theory. The presence of a new vector multiplet in the spectrum may
lead to gauge anomalies that need to be cancelled through a
Green--Schwarz--West--Sagnotti mechanism \cite{Green:1984bx, Sagnotti:1992qw},
requiring the introduction of an additional tensor multiplet to mediate it.
The fusion procedure can, however, substantially alter the UV behavior of the new
theory. Depending on the (anti-)self duality of the new tensor, the SCFT can be converted to an LST, or even to a SUGRA theory \cite{Oehlmann:2024cyn}.

The reverse procedure, where a tensor multiplet and the associated gauge
symmetry are decoupled, is called \emph{fission}. Since the pairing matrix of an LST has by definition a single null eigenvector, see equation \eqref{def-eta},
fission leads a positive definite pairing matrix, i.e., that of an SCFT. This is known as the \emph{tensor-decoupling criterion}~\cite{Bhardwaj:2015oru}.

Geometrically, fusion can be understood as K\"ahler deformations rendering a curve in the base compact. Schematically, given two SCFTs $\mathcal{T}_1$ and
$\mathcal{T}_2$ with a common flavor symmetry $\mathfrak{f}$, they can be fused
together to obtain a new theory $\mathcal{K}$:
\begin{equation}
	\begin{aligned}
			\mathcal{T}_1\quad\oplus \quad\mathcal{T}_2 ~ ~\,\quad\qquad= &\qquad\qquad \,\,\mathcal{K}\,,\\
			\,\cdots \overset{\mathfrak{g}_1}{n_1}\,\overset{\mathfrak{g}_2}{n_2}\,[\mathfrak{f}] ~\oplus~ [\mathfrak{f}]\, \overset{\mathfrak{h}_1}{k_1}\, \overset{\mathfrak{h}_2}{k_2}\cdots\quad
		= &
		\quad\cdots \overset{\mathfrak{g}_1}{n_1}\, \overset{\mathfrak{g}_2}{n_2}\, \textcolor{blue}{\overset{\mathfrak{f}}{m}}\, \overset{\mathfrak{h}_1}{k_1}\, \overset{\mathfrak{h}_2}{k_2}\,,
	\end{aligned}
\end{equation}
where we have highlighted the new compact curve in blue, whose precise
self-intersection $(-m)$ is dictated by anomaly cancellation. We again stress
that the theory obtained after fusion must be free of any gauge anomaly, and
not all fusions lead to consistent theories. 

The SCFT $\mathcal{T}_2$ can be taken to be trivial, in which case we are
simply gauging the flavor symmetry:
\begin{equation}
		\,\cdots \overset{\mathfrak{g}_1}{n_1}\,\overset{\mathfrak{g}_2}{n_2}\,[\mathfrak{f}] 
		\qquad\longrightarrow\qquad
		\,\cdots \overset{\mathfrak{g}_1}{n_1}\, \overset{\mathfrak{g}_2}{n_2}\, \textcolor{blue}{\overset{\mathfrak{f}}{m}}\,. 
\end{equation}

To study LSTs, we can therefore start with SCFTs, and through fusion introduce
a null eigenvalue. However, this can drastically change the (higher) symmetry
structure of the new theory compared to the original SCFT.  First, we have
removed the flavor symmetry $\mathfrak{f}$, which may have an impact on the
one-form (gauge) symmetry.  Second, the introduction of a new compact curve, or
equivalently new dynamical BPS strings, may also modify the defect
group~$\mathcal{D}^{(2)}$. This in turn gives additional constraints on the set
of consistent fusions. For example, for Heterotic LSTs, the defect group must
be trivialized and the flavor group reduced to at most rank $20$.  For Type II
LSTs, the flavor rank must be almost trivialized to at most rank two, and
$\mathcal{D}^{(2)}$ reduced to the center of a simple ADE algebra. The
landscape of those theories is therefore much more sparse than their Heterotic
cousins, and the possible bases giving rise to the defect group have been
classified in \cite{Bhardwaj:2015oru}. Through the tensor-decoupling criterion,
one might then ask which SCFT is closest to being an LST in the sense that
their defect group agrees, and the only operation needed is to gauge the flavor
symmetry. We will define such an operation as follows:
\begin{definition}
		Given a 6D SCFT $\mathcal{T}$ with a flavor symmetry $\mathfrak{f}$, a
		Little String Theory $\mathcal{K}$ is reached through \emph{minimal
		affinization} if:
		\begin{itemize}
				\item At a generic point of the tensor branch, gauging the
						flavor symmetry $\mathfrak{f}$ introduces a single
						curve, and a null eigenvalue to the Dirac pairing,
						turning the resulting theory into an~LST.
				\item Only vector multiplets are added to the spectrum of the tensor branch theory.
				\item The defect group is preserved, $\mathcal{D}^{(2)}(\mathcal{T}) = \mathcal{D}^{(2)}(\mathcal{K})$.
		\end{itemize}
\end{definition}
In particular, this means that this procedure does not require additional
matter to cancel possible new gauge anomalies. Furthermore, the symmetry
$\mathfrak{f}$ could be part of a larger flavor
symmetry, as long as the resulting LST $\mathcal{K}$ satisfies the bound given
in equation \eqref{flavor-bound-WZW}.

To give an example, let us consider the minimal affinization of the so-called
$A_2^{\mathfrak{su}_k}$ SCFT, discussed in more detail below. It has a flavor
symmetry $\mathfrak{su}_k\oplus\mathfrak{su}_k$, which can be trivialized by
gauging the common diagonal subalgebra:
\begin{equation}\label{example-minimal-affinization}
		[\mathfrak{su}_k] \overset{\mathfrak{su}_k}{2}\,\overset{\mathfrak{su}_k}{2} [\mathfrak{su}_k]  \qquad\longrightarrow\qquad
		\overset{\mathfrak{su}_k}{2} \overset{\displaystyle\textcolor{blue}{\overset{\mathfrak{su}_k}{2}}}{\phantom{2}~} \overset{\mathfrak{su}_k}{2} \quad=\quad \textcolor{blue}{//}\overset{\mathfrak{su}_k}{2}\overset{\mathfrak{su}_k}{2}\textcolor{blue}{\overset{\mathfrak{su}_k}{2}//}\,,
\end{equation}
Fusion has introduced a new curve, depicted in blue, and we remind the reader
that the symbols $//$ indicate identification of the ends of the quiver, see equation~\eqref{base-pattern-notation}. Anomaly cancellation demands that the new curve has self-intersection $(-2)$, and it is easy to check that the new
pairing matrix now has a null eigenvalue. The resulting quiver is therefore
an LST, and the defect group of both theories is $\mathcal{D}^{(2)}=\mathbbm{Z}_2$.
Generalization of this example to more complicated quivers will be discussed in
Section \ref{sec:ADE-orbifolds}.

While there is a unique way to obtain an LST in the previous example, in general multiple SCFTs can lead to the same LST through fusion. Minimal affinization therefore defines a canonical way to introduce a null eigenvalue to the pairing matrix without modifying the matter spectrum of the theory beyond the changes coming from the trivialisation of the flavor symmetry.

Another by-product of minimal affinization is that it enables us to obtain the
two-group structure constants of an LST from the anomaly polynomial of the 6D
SCFT. The anomaly polynomial of an $\mathcal{N}=(1,0)$ theory is generically\footnote{We ignore possible Abelian flavor symmetries for simplicity.} of the
form \cite{Ohmori:2014kda}
\begin{equation}\label{eqn:I8general}
  \begin{aligned}
    I_8 &=  \frac{\alpha}{24} c_2(R)^2+ \frac{\beta}{24}  c_2(R) p_1(T) + \frac{\gamma}{24}  p_1(T)^2 + \frac{\delta}{24} p_2(T) \cr &\quad + \sum_a \Tr F_a^2 \left(k^a c_2(R) + \mu^a p_1(T) + \sum_b \rho^{ab} \Tr   F_b^2\right) + \sum_a \nu^a \Tr F_a^4  \,,
  \end{aligned}
\end{equation}
where the traces are one-instanton normalized, and we use the same notation as
around equation \eqref{Bianchi-LST}. Given a six-dimensional quiver describing
the tensor branch of the theory, there is a simple procedure to extract the
anomaly polynomial of the theory, see e.g.~\cite{Baume:2021qho} for a concise
review of the algorithm. Via 't Hooft anomaly matching, the anomaly
coefficients are preserved as we move in the tensor branch to reach the
singular point.

When performing a minimal affinization, we only introduce a vector multiplet
mediating the gauge interactions, and since the new curve cannot
participate in the Green--Schwarz--West--Sagnotti mechanism \cite{Green:1984bx, Sagnotti:1992qw}, the anomaly
polynomial of an LST $\mathcal{K}$ obtained from an SCFT $\mathcal{T}$ is
simply given by
\begin{equation}\label{I8-LST-minimal}
		I_8(\mathcal{K}) = I_8(\mathcal{T}) + I_8^\text{vec}(\mathfrak{f})\,,
\end{equation}
where $I_8^\text{vec}$ is the contribution of the vector multiplet, defined in
Appendix \ref{app:6D-anomalies}.

This gives us an alternative way to find the two-group structure constants defined in equation \eqref{def-kappas}. Indeed, for an LST,
they can also be read off directly from the anomaly polynomial~\cite{Cordova:2020tij},
\begin{equation}\label{I8-LST-minimal-2}
		I_8(\mathcal{K}) = \left(-\kappa_R \,c_2(R) - \frac{\kappa_P}{4}p_1(T) + \kappa_F^A\, c_2(F_A) \right) c_2(F) + \dots\,,
\end{equation}
where $c_2(F) = \frac{1}{4}\Tr F^2$ is the second Chern class of the
background field strength of the background symmetry of the $U(1)^{(1)}$ 1-form
symmetry. In the case of minimal affinization, this is simply that of the
(now gauged) symmetry $\mathfrak{f}$. 

Note that the above procedure is very much a feature of minimal affinization,
and not true in general: generic fusions might require the introduction of
additional matter, change the Green--Schwarz term, or $c_2(F)$ might not be
associated directly with that of the SCFT flavor symmetry. As a result, the
anomaly polynomial will not take the form given in equations~\eqref{I8-LST-minimal} and~\eqref{I8-LST-minimal-2} in those cases.

Using the expression for the contribution
of the vector multiplet, in the convention defined in equation
\eqref{eqn:I8general}, we find
\begin{equation}
		\kappa_G(\mathcal{K}) = -16\mu^{\mathfrak{f}}(\mathcal{T}) + \frac{1}{3}h^\vee_{\mathfrak{f}}\,,\qquad
		\kappa_R(\mathcal{K}) = -4k^{\mathfrak{f}}(\mathcal{T}) + h^\vee_{\mathfrak{f}}\,,\qquad
		\kappa_F^A(\mathcal{K}) = 16\rho^{A,\mathfrak{f}}(\mathcal{T})\,.
\end{equation}
Beyond their usefulness as a cross-check for the two-group structure constants
computed from equation~\eqref{def-kappas}, these expressions will also enable
us to explain some of the coincidences we observe for numerical factors for
certain T-dual dual pairs of LSTs. Indeed, for SCFTs with a base that is part
of an infinite series---so-called long quivers---closed-form expressions have
been found in \cite{Baume:2023onr}, relying only on the rank $N$ of the base
and group-theoretic data of $\mathfrak{g}$, possibly supplemented with
nilpotent orbit data if the theory is reached through a Higgs-branch
Renormalization Group (RG) flow. 

In the sequel, we exemplify minimal affinization by considering SCFTs with a
base associated to an ADE orbifold of $\mathbbm{C}^2$ in F-theory, and then
move on to other more general types of theories and their duals.

\subsection{LSTs from ADE Orbifolds}\label{sec:ADE-orbifolds}
Our goal is to study LSTs using SCFTs. In the F-theory picture, $\mathcal{N}=(1,0)$ SCFTs are obtained from compactification on an elliptically fibered Calabi--Yau with a non-compact base $B_2$. The possible bases have been classified, and are known to correspond to discrete orbifolds of
$\mathbbm{C}^2$ \cite{Heckman:2013pva, Heckman:2015bfa}:
\begin{equation}
		B_{2} = \mathbbm{C}^2 / \Lambda\,,\qquad \Lambda\subset U(2)\,,
\end{equation}

The simplest cases, which we will focus on first, are those where we further
restrict $\Lambda$ to be a discrete subgroup of $SU(2)$ rather than $U(2)$.
These are well known to follow an ADE classification:
\begin{equation}
	\Lambda\subset SU(2):\qquad A_{N-1}\,,\quad D_N\,,\quad E_6\,,\quad E_7\,,\quad E_8\,,
\end{equation}
The first two infinite series are given by the cyclic ($\mathbbm{Z}_N$) and
binary dihedral groups, while the exceptional series correspond to the
tetrahedral, octahedral and icosahedral finite groups, in ascending order of
rank. When the fibers are trivial, it is well known that 16 supercharges are
preserved and we find a collection of undecorated $(-2)$-curves intersecting in
the same pattern as the Dynkin diagram of the associated simple algebra: these
are the celebrated $\mathcal{N}=(2,0)$ theories
\cite{Witten:1995zh,Strominger:1995ac,Seiberg:1996qx}, see Appendix~\ref{app:Enhanced}.

Generically, to preserve eight supercharges in the six-dimensional effective
description, the fiber over the curves can be singular. However, for a given
choice of orbifold action $\Lambda$, the possible fibers are severely
constrained by demanding a well-defined elliptic fibration. For instance, when
$\Lambda=A_{N-1} $, only fibers associated with a simply-laced algebra
$\mathfrak{g}=\mathfrak{su}_k,\mathfrak{so}_{2k}, \mathfrak{e}_6,
\mathfrak{e}_7,\mathfrak{e}_8$ are consistent. Except for
$\mathfrak{g}=\mathfrak{su}_k$, the resulting geometry contains non-minimal
singularities, and a series of blow-ups is necessary, leading to a repeating
sequence of curves called rank-$N$ $(\mathfrak{g}, \mathfrak{g})$ conformal
matter, which we denote as $A_{N-1}^{\mathfrak{g}}$:
\begin{equation}\label{def-CM}
	\begin{aligned} 
			A_{N-1}^{\mathfrak{su}_k}:&\qquad [\mathfrak{su}_k]\overset{\mathfrak{su}_k}{2}\overset{\mathfrak{su}_k}{2} \dots \overset{\mathfrak{su}_k}{2}\overset{\mathfrak{su}_k}{2}[\mathfrak{su}_k]\,,\\
			A_{N-1}^{\mathfrak{so}_{2k}}:&\qquad [\mathfrak{so}_{2k}]\overset{\mathfrak{sp}_{k-4}}{1}\overset{\mathfrak{so}_{2k}}{4}\overset{\mathfrak{sp}_{k-4}}{1}\dots\overset{\mathfrak{sp}_{k-4}}{1}\overset{\mathfrak{so}_{2k}}{4}\overset{\mathfrak{sp}_{k-4}}{1}[\mathfrak{so}_{2k}]\,,\\
			A_{N-1}^{\mathfrak{e}_{6}}:&\qquad [\mathfrak{e}_{6}]1\overset{\mathfrak{su}_{3}}{3}1\overset{\mathfrak{e}_{6}}{6}1\overset{\mathfrak{su}_{3}}{3}1\overset{\mathfrak{e}_{6}}{6} \dots 1\overset{\mathfrak{su}_{3}}{3}1\overset{\mathfrak{e}_{6}}{6}1\overset{\mathfrak{su}_{3}}{3}1[\mathfrak{e}_{6}]\,,\\
			A_{N-1}^{\mathfrak{e}_{7}}:&\qquad [\mathfrak{e}_{7}]1\overset{\mathfrak{su}_{2}}{2}\overset{\mathfrak{so}_{7}}{3}\overset{\mathfrak{su}_{2}}{2}1\overset{\mathfrak{e}_{7}}{8} \dots \overset{\mathfrak{e}_{7}}{8}1\overset{\mathfrak{su}_{2}}{2}\overset{\mathfrak{so}_{7}}{3}\overset{\mathfrak{su}_{2}}{2}1[\mathfrak{e}_{7}]\,,\\
			A_{N-1}^{\mathfrak{e}_{8}}:&\qquad [\mathfrak{e}_{8}]1\,2\overset{\mathfrak{su}_{2}}{2}\overset{\mathfrak{g}_2}{3}1\overset{\mathfrak{f}_{4}}{5}1\overset{\mathfrak{g}_2}{3}\overset{\mathfrak{su}_{2}}{2}2\,1\overset{\mathfrak{e}_{8}}{(12)} \dots \overset{\mathfrak{e}_{8}}{(12)}1\,2\overset{\mathfrak{su}_{2}}{2}\overset{\mathfrak{g}_2}{3}1\overset{\mathfrak{f}_{4}}{5}1\overset{\mathfrak{g}_2}{3}\overset{\mathfrak{su}_{2}}{2}2\,1[\mathfrak{e}_{8}]\,.
	\end{aligned}
\end{equation}
In each case, the curve with the lowest self-intersection---that is
$-2\,,-4\,,-6\,,-8\,,-12$ for
$\mathfrak{g}=\mathfrak{su}_k\,,\mathfrak{so}_{2k}\,, \mathfrak{e}_6\,,
\mathfrak{e}_7\,, \mathfrak{e}_8$, respectively---appears $(N-1)$ times and is
often referred to as a node of the quiver. With the links of minimal
conformal matter, the theory $A_0^{\mathfrak{g}}$, this lends the name to the
notion of generalized quivers, where bifundamental hypermultiplets are
replaced by a generalization of matter with $\mathfrak{g}\oplus\mathfrak{g}$
flavor, glued together via fusion \cite{DelZotto:2014hpa}. 

Up to a few outliers, the overwhelming majority of 6D SCFTs can be obtained
from deformations of a few so-called ``parent'' theories with a base associated
with the discrete ADE group $\Lambda$ and a flavor symmetry dictated by an
algebra $\mathfrak{g}$. We will follow the notation used extensively
in~\cite{Baume:2023onr} and denote the parent theories by
$\Lambda^{\mathfrak{g}}$. The rank $N$ conformal matter theories
$A_{N-1}^{\mathfrak{g}}$ described above are an example of parent theories. If
an SCFT is obtained by moving onto the Higgs branch of the parent theory via a
deformation encoded in e.g.\ a nilpotent orbit $\mathcal{O}$, we label it as
$\Lambda^{\mathfrak{g}}(\mathcal{O})$, see below for an example. The LST
reached through minimal affinization is then denoted as
$\widehat{\Lambda}^{\mathfrak{g}}(\mathcal{O})$, in the obvious notation. 

For conformal matter, minimal affinization is a generalization of the example
given in equation~\eqref{example-minimal-affinization}. In every case, we gauge
the diagonal subalgebra of the $\mathfrak{g}\oplus\mathfrak{g}$ flavor, and one
needs to introduce a new node with the appropriate self-intersection, as given
below equation~\eqref{def-CM}. 

This can also be seen on the partial tensor branch, also called the endpoint of
the theory, which is the point where all $(-1)$-curves have been successively
blown down. There, we obtain a collection of $(-2)$-curves intersecting like
the Dynkin diagram of the affine simple algebra $\widehat{A}_{N-1}$:
\begin{equation}\label{PTB-CM}
	A_{N-1}^{\mathfrak{g}}:\qquad 
	[\mathfrak{g}]\,\underbrace{\overset{\mathfrak{g}}{2}\,\cdots\, \overset{\mathfrak{g}}{2}}_{N-1}\,[\mathfrak{g}]  \qquad\longrightarrow\qquad
	\widehat{A}_{N-1}^{\mathfrak{g}}:\qquad 
	\textcolor{blue}{//}\underbrace{\overset{\mathfrak{g}}{2}\,\cdots\, \overset{\mathfrak{g}}{2}}_{N-1}\textcolor{blue}{\overset{\mathfrak{g}}{2}//}\,,
\end{equation}
Due to their loop configuration, these theories are sometimes referred to as
necklace LSTs, and we will denoted them as $\widehat{A}_{N-1}^\mathfrak{g}$, in
the notation summarized above. At a generic point in the tensor branch,
or in geometric terms when enough blow-ups have been performed to obtain a
configuration with only non-minimal singularities, we have for each algebra
\begin{equation}\label{def-necklace}
	\begin{aligned} 
			\widehat{A}_{N-1}^{\mathfrak{su}_k}:&\qquad \textcolor{blue}{//}\overset{\mathfrak{su}_k}{2}\overset{\mathfrak{su}_k}{2} \dots \overset{\mathfrak{su}_k}{2}\overset{\mathfrak{su}_k}{2}\textcolor{blue}{\overset{\mathfrak{su}_k}{2}//}\,,\\
			\widehat{A}_{N-1}^{\mathfrak{so}_{2k}}:&\qquad \textcolor{blue}{//}\overset{\mathfrak{sp}_{k-4}}{1}\overset{\mathfrak{so}_{2k}}{4}\overset{\mathfrak{sp}_{k-4}}{1}\dots\overset{\mathfrak{sp}_{k-4}}{1}\overset{\mathfrak{so}_{2k}}{4}\overset{\mathfrak{sp}_{k-4}}{1}\textcolor{blue}{\overset{\mathfrak{so}_{2k}}{4}//}\,,\\
			\widehat{A}_{N-1}^{\mathfrak{e}_{6}}:&\qquad \textcolor{blue}{//}1\overset{\mathfrak{su}_{3}}{3}1\overset{\mathfrak{e}_{6}}{6}1\overset{\mathfrak{su}_{3}}{3}1\overset{\mathfrak{e}_{6}}{6} \dots 1\overset{\mathfrak{su}_{3}}{3}1\overset{\mathfrak{e}_{6}}{6}1\overset{\mathfrak{su}_{3}}{3}1\textcolor{blue}{\overset{\mathfrak{e}_6}{6}//}\,,\\
			\widehat{A}_{N-1}^{\mathfrak{e}_{7}}:&\qquad \textcolor{blue}{//}1\overset{\mathfrak{su}_{2}}{2}\overset{\mathfrak{so}_{7}}{3}\overset{\mathfrak{su}_{2}}{2}1\overset{\mathfrak{e}_{7}}{8} \dots \overset{\mathfrak{e}_{7}}{8}1\overset{\mathfrak{su}_{2}}{2}\overset{\mathfrak{so}_{7}}{3}\overset{\mathfrak{su}_{2}}{2}1\textcolor{blue}{\overset{\mathfrak{e}_7}{8}//}\,,\\
			\widehat{A}_{N-1}^{\mathfrak{e}_{8}}:&\qquad \textcolor{blue}{//}1\,2\overset{\mathfrak{su}_{2}}{2}\overset{\mathfrak{g}_2}{3}1\overset{\mathfrak{f}_{4}}{5}1\overset{\mathfrak{g}_2}{3}\overset{\mathfrak{su}_{2}}{2}2\,1\overset{\mathfrak{e}_{8}}{(12)} \dots \overset{\mathfrak{e}_{8}}{(12)}1\,2\overset{\mathfrak{su}_{2}}{2}\overset{\mathfrak{g}_2}{3}1\overset{\mathfrak{f}_{4}}{5}1\overset{\mathfrak{g}_2}{3}\overset{\mathfrak{su}_{2}}{2}2\,1\textcolor{blue}{\overset{\mathfrak{e}_8}{(12)}//}\,.
	\end{aligned}
\end{equation}
A similar behavior occurs for the other types of SCFTs with ADE bases. For
those of exceptional types $\Lambda=E_r$, each $(-2)$-curve can be decorated with an
$\mathfrak{su}(d^i k)$ algebra, where $d^i$ is the Kac label corresponding node
of the base algebra, see Table \ref{tab:affinized-bases-LST}. The resulting
theory has a flavor symmetry $\mathfrak{su}_k$ which can be gauged in order
to obtain a well-defined LST, where the $(-2)$-curves now intersect like the
Dynkin diagram of the corresponding affine algebra. For instance, for an $E_6$
base, we have:
\begin{equation}
\label{eq:E6SUkQuiver}
		E_{6}^{\,\mathfrak{su}_k} :\qquad \overset{\mathfrak{su}_{k}}{2}\overset{\mathfrak{su}_{2k}}{2} \overset{\displaystyle[\mathfrak{su}_{k}]}{\overset{\displaystyle \overset{\mathfrak{su}_{2k}}{2}}{\overset{\mathfrak{su}_{3k}}{2}}}\,\overset{\mathfrak{su}_{2k}}{2} \overset{\mathfrak{su}_{k}}{2} \qquad\longrightarrow\qquad
		\widehat{E}_{6}^{\,\mathfrak{su}_k} :\qquad \overset{\mathfrak{su}_{k}}{2}\overset{\mathfrak{su}_{2k}}{2} \overset{\textcolor{blue}{\displaystyle\overset{\mathfrak{su}_{k}}{2}}}{\overset{\displaystyle \overset{\mathfrak{su}_{2k}}{2}}{\overset{\mathfrak{su}_{3k}}{2}}}\,\overset{\mathfrak{su}_{2k}}{2} \overset{\mathfrak{su}_{k}}{2} \,.
\end{equation}
The same can be done for the other two cases, giving rise to the following LST
quivers:
\begin{equation}
			\widehat{E}_{7}^{\,\mathfrak{su}_k} : \quad \textcolor{blue}{\overset{\mathfrak{su}_{k}}{2}}\overset{\mathfrak{su}_{2k}}{2}\overset{\mathfrak{su}_{3k}}{2} \overset{\displaystyle \overset{\mathfrak{su}_{2k}}{2}}{\overset{\mathfrak{su}_{4k}}{2}}\,\overset{\mathfrak{su}_{3k}}{2}\overset{\mathfrak{su}_{2k}}{2} \overset{\mathfrak{su}_{k}}{2}\qquad\quad
			\widehat{E}_{8}^{\,\mathfrak{su}_k} :\quad \overset{\mathfrak{su}_{2k}}{2}\overset{\mathfrak{su}_{4k}}{2} \overset{\displaystyle \overset{\mathfrak{su}_{3k}}{2}}{\overset{\mathfrak{su}_{6k}}{2}}\,\overset{\mathfrak{su}_{5k}}{2}\overset{\mathfrak{su}_{4k}}{2} \overset{\mathfrak{su}_{3k}}{2}\overset{\mathfrak{su}_{2k}}{2}\textcolor{blue}{\overset{\mathfrak{su}_{k}}{2}}
\end{equation}
Finally, the case of D-type bases is slightly more subtle. With fibers of type
$\mathfrak{su}_L$, gauge-anomaly-cancellation conditions demand that on the
arbitrarily-long spine, $L$ must be even:
\begin{equation}\label{D_su2k}
	D_N^{\mathfrak{su}_{2k}}:\qquad
	\overset{\mathfrak{su}_{k}}{2} \overset{\displaystyle \overset{\mathfrak{su}_{k}}{2}}{\overset{\mathfrak{su}_{2k}}{2}}\underbrace{\,\overset{\mathfrak{su}_{2k}}{2} \cdots \overset{\mathfrak{su}_{2k}}{2} \,}_{N-3}[\mathfrak{su}_{2k}] \,.
\end{equation}
Simply gauging the $\mathfrak{su}_{2k}$ flavor of the
$D_{N}^{\mathfrak{su}_{2k}}$ theory does not give rise to an LST but rather
another SCFT, $D_{N+1}^{\mathfrak{su}_{2k}}$. We can, however, break the
$\mathfrak{su}_{2k}$ flavor algebra to $\mathfrak{su}_k$ by moving onto the
Higgs branch of the SCFT. These types of deformations are classified by
nilpotent orbits~\cite{Heckman:2016ssk}, and for $\mathfrak{su}_L$ algebras,
they can be labelled by partitions of $L$. We denote a partition $L =
\sum_{i=1}^Li m_i$ as $[1^{m_1},2^{m_2},\dots, L^{m_L}]$, and we omit
entries where $m_i=0$ for ease of reading. 

Given a nilpotent orbit, the quiver configuration can be read off directly, and
some of the data of the resulting SCFT follows straightforwardly from the
associated group theory ~\cite{Mekareeya:2016yal, Heckman:2016ssk,
Baume:2023onr}. For the nilpotent orbit $[2^k]$ of $\mathfrak{su}_{2k}$, the
quiver is given by
\begin{equation}
	D_N^{\mathfrak{su}_{2k}}([2^k]):\qquad
	\overset{\mathfrak{su}_{k}}{2} \overset{\displaystyle \overset{\mathfrak{su}_{k}}{2}}{\overset{\mathfrak{su}_{2k}}{2}}\underbrace{\,\overset{\mathfrak{su}_{2k}}{2} \cdots \overset{\mathfrak{su}_{2k}}{2} \,}_{N-5}\overset{\displaystyle[\mathfrak{su}_{k}]}{\overset{\mathfrak{su}_{2k}}{2}}\overset{\mathfrak{su}_{k}}{2} \,.
\end{equation}
The LST is then obtained via minimal affinization by gauging the flavor
symmetry, which is achieved without changing the defect group:
\begin{equation}
	\widehat{D}_N^{\mathfrak{su}_{2k}}([2^k]):\qquad
	\overset{\mathfrak{su}_{k}}{2} \overset{\displaystyle \overset{\mathfrak{su}_{k}}{2}}{\overset{\mathfrak{su}_{2k}}{2}}\underbrace{\,\overset{\mathfrak{su}_{2k}}{2} \cdots \overset{\mathfrak{su}_{2k}}{2} \,}_{N-5}\overset{\displaystyle\textcolor{blue}{\overset{\mathfrak{su}_{k}}{2}}}{\overset{\mathfrak{su}_{2k}}{2}}\overset{\mathfrak{su}_{k}}{2} \,.
\end{equation}
Note that as for the exceptional bases---and trivially for
$A_{N-1}^{\mathfrak{su}_k}$---all fibers are now of the form $\mathfrak{su}(d^i
\cdot k)$ with $d^i$ the Kac labels of $\mathfrak{so}_{2N}$, see Table \ref{tab:affinized-bases-LST}.

To summarize, in all ADE cases, we start with an SCFT $\Lambda^{\mathfrak{g}}$
where $\Lambda$ is a discrete ADE group associated to a simple algebra
$\mathfrak{g}_B$ that describes the singularity structure of the base of the
F-theory geometry (and thus, after resolution, the intersection properties of
the base divisors), and $\mathfrak{g}$ describe both the generic fibers at the
origin of the tensor branch and the flavor of the parent theory. Up to additional possible
deformations, the SCFT has flavor symmetry $\mathfrak{g}_F$, and after the
minimal affinization procedure we obtain an LST labelled uniquely by the two
algebras $\mathfrak{g}_F$ and~$\mathfrak{g}_B$. Pictorially, we have
\begin{equation}
		\text{SCFT:\quad}\Lambda^{\mathfrak{g}}(\mathcal{O})\qquad
		\xrightarrow{\quad\text{min. affinization}\quad} \qquad 
		\text{LST:}\quad \mathcal{K}^{\text{II}}(\mathfrak{g}_F, \mathfrak{g}_B) = \widehat{\Lambda}^{\mathfrak{g}}(\mathcal{O})\,.
\end{equation}
The three LSTs with ADE base are given by
\begin{equation}
	\mathcal{K}^{\text{II}}(\mathfrak{g}_F, \mathfrak{su}_N) = \widehat{A}_N^{\mathfrak{g}_F}\,,\qquad 
	\mathcal{K}^{\text{II}}(\mathfrak{su}_k, \mathfrak{so}_{2N}) = \widehat{D}_N^{\mathfrak{su}_{2k}}([2^k])\,\qquad 
    \mathcal{K}^{\text{II}}(\mathfrak{su}_k, \mathfrak{e}_N) = \widehat{E}_N^{\mathfrak{su}_k}\,.
\end{equation}
For all these theories, one can show by direct computation using equation
\eqref{def-kappas} that the two-group invariants are arrange neatly in terms of
quantities related to $\mathfrak{g}_F$ and $\mathfrak{g}_B$:
\begin{equation}\label{kappas-CM}
		\mathcal{K}^{\text{II}}(\mathfrak{g}_F, \mathfrak{g}_B): \qquad 
		\kappa_R = \Gamma_{\mathfrak{g}_F}\Gamma_{\mathfrak{g}_B}\,,\qquad
		\kappa_P = 0\,,\qquad
		\text{dim}(\text{CB}) = h^\vee_{\mathfrak{g}_F}h^\vee_{\mathfrak{g}_B} - 1\,,
\end{equation}
where $\Gamma_\mathfrak{g}$ is the order of the discrete ADE group associated
with the algebra $\mathfrak{g}$. Their values are given in
Table~\ref{tab:algebra-values}. The factorization of these quantities in terms
of group-theoretic data of the fiber and base makes the T-duality between the
two classes of LSTs manifest:
\begin{equation}
	\mathcal{K}^{\text{II}}(\mathfrak{su}_N, \mathfrak{g}) 
	\qquad \overset{\quad\text{T-duality}\quad}{\longleftrightarrow} \qquad
	\mathcal{K}^{\text{II}}(\mathfrak{g}, \mathfrak{su}_N)\,.
\end{equation}
It is quite instructive to see how these structure constants arise from the
point of view of minimal affinization and the anomaly polynomial of the
associated SCFTs. One of the corollaries of fusion is that the anomaly
polynomial of the vast majority of SCFTs can be obtained from that of minimal
conformal matter, $A_{0}^\mathfrak{g}$ or a deformation thereof
\cite{Ohmori:2014kda, Mekareeya:2017sqh, Baume:2023onr}. This can be used to write the anomaly polynomial of conformal matter in a compact form,
depending only on the rank $N$ of the base and group-theoretic quantities
related to the algebra~$\mathfrak{g}$ \cite{Ohmori:2014kda}:
\begin{equation}\label{I8-CM}
		\begin{aligned}
			I_8(A^{\mathfrak{g}}_{N-1}) =&\,  
			\frac{N^3}{24}(c_2(R)\Gamma_{\mathfrak{g}})^2
			- \frac{1}{2}(c_2(R)\Gamma) \left(J(F_L) + J(F_R)\right) - \frac{1}{2N}\left[J(F_L) - J(F_R)\right]^2\\
		&
		+ I_\text{sing} - I_8^\text{tensor} - \frac{1}{2}\left[I_8^\text{vec}(F_L)- I_8^\text{vec}(F_R)\right]\,,
		\end{aligned}
\end{equation}
where $F_L, F_R$ refer to the background fields for the two flavor symmetries
$\mathfrak{g}$. The contributions $I_8^\text{tensor}$ and $I_8^\text{vec}$ of the tensor and vector multiplets, together with $I_\text{sing}$, are given in Appendix~\ref{app:6D-anomalies}. Furthermore, the four-form $J(F)$ is
\begin{gather}
		J(F) = \frac{\chi}{48}\left(4 c_2(R) + p_1(T)\right) + c_2(F)\,,\qquad
		\chi = \rk(\mathfrak{g}) + 1 - \frac{1}{\Gamma_{\mathfrak{g}}}\,.
\end{gather}
Since the Bianchi identity of the associated two-form is $dH_3^\text{CM}\sim J(F)$, it can be understood as the charge of the BPS string associated with conformal matter at the conformal fixed point. This means that the first line of equation \eqref{I8-CM} is generated entirely by the Green--Schwarz term. Furthermore, this equation makes it clear that minimal affinization for conformal matter sets $F_L=F_R$, and adds the contribution of the vector multiplet introduced through the gauging of the flavor, as explained around equation \eqref{I8-LST-minimal}. Doing so automatically removes all quartic
contributions $(\Tr F^2)^2$ and $\Tr F^4$, which would otherwise lead to
gauge anomalies, and the two-group constants are readily found to be given by
equation \eqref{kappas-CM}. 

As a consequence of 't Hooft anomaly matching, the anomaly polynomial does not
change as we move in the tensor branch by successively blowing down all
$(-1)$-curves. We can therefore work directly on the partial tensor branch:
even though the quiver consists of multiple tensor, hyper- and vector
multiplets, we can treat each of the fused $N$ minimal conformal
matter links $A_0^{\mathfrak{g}}$ as a single entity. By the argument above,
there are no quartic gauge anomalies, and by carefully tracking the changes in
the different quantities, as explained in Appendix \ref{app:6D-anomalies}, one
finds that the Green--Schwarz contribution of the SCFT \emph{on the partial
tensor branch} takes the form 
\begin{equation}\label{I8-GS-ADE}
		I_8^\text{GS} = \frac{1}{2}A_{ij} I^iI^j\,,\qquad I^i = -A^{ij}c_2(F_i) + B^{ia}J(F_a) + d^i\, \Gamma_{\mathfrak{g}} c_2(R)\,.
\end{equation}
Here, $A^{ij}$ is the pairing matrix at the conformal fixed point and $A_{ij}$
is its inverse. This pairing matrix is simply given by the Cartan matrix of
$A_{N-1}$, and should be distinguished from the pairing matrix at a generic point, which we have denoted by $\eta$. Moreover, we have $d^i=1$ for all $i$, and the flavor pairing matrix can be written as
\begin{equation}
		B^{ia} = (A^{ij}d_j)\, \delta^{ia}\,,\qquad A\cdot d = (1,0,\;\dots\;,0,1)\,.
\end{equation}
While writing the pairing matrix in this form might seem \emph{ad hoc}, if we
interpret $d^i$ as the Kac labels of the base $A_{N-1}$, the factorization of the structure constants constants shown in equation
\eqref{kappas-CM} becomes apparent. Indeed, using that for a simply-laced algebra
$\mathfrak{g}$, we have the relation
\begin{equation}
		\Gamma_{\mathfrak{g}} = 1 + \sum_{i=1}^{\rk(\mathfrak{g})}(d^i)^2\,,
\end{equation}
and the Green--Schwarz term of the SCFT can be found to be
\begin{equation}
		I^\text{GS}_8 = -\frac{1}{2}\Gamma_{\mathfrak{su}_k}\Gamma_{\mathfrak{g}} c_2(R)\big(J(F_L)+J(F_R)\big) + \dots
\end{equation}

This generalizes to all three families of SCFTs with an ADE base discussed
above. The Green--Schwarz term in equation \eqref{I8-GS-ADE} is given in terms
of the Kac labels $d^i$ of the algebra of base, in the convention of Table
\ref{tab:affinized-bases-LST}.\footnote{Note that here, the coefficients $d^i$ are Kac labels of the simple algebra $\mathfrak{g}$, and not its affine version. For simply-laced algebras, they also correspond to the highest root of $\mathfrak{g}$ in the Serre--Chevalley basis. Moreover, they are sometimes also referred to as Dynkin multiplicities in the literature.} The
flavor pairing matrix is again given in terms of $(A\cdot d)$, and a
similar reasoning explains the numerical coincidences observed in the structure
constants.\footnote{In the case of $D_N^{\mathfrak{su}_{2k}}([2^k])$,
$\Gamma_{\mathfrak{su}_{2k}}$ must be substituted for that of the flavor
symmetry $\mathfrak{su}_k$.}. The Kac labels will similarly play an important role in the geometric engineering of the Type II LSTs in Section~\ref{sec:geometry}.

We find that not only the structure constants, but the complete anomaly polynomial, can be described purely in terms of group-theoretic data of the base and fiber algebras,
\begin{equation}\label{I8-ADE-LST}
	\begin{aligned}
	I_8\left(\mathcal{K}(\mathfrak{g}_F, \mathfrak{g}_B)\right) =& \frac{\rk(\mathfrak{g}_B)+1}{24}(\Gamma_{\mathfrak{g}_B}\Gamma_{\mathfrak{g}_F}\,c_2(R))^2 - \Gamma_{\mathfrak{g}_B}\Gamma_{\mathfrak{g}_F}\,c_2(R)c_2(F)+ (\rk(\mathfrak{g}_B)+1)I_8^\text{sing}\\
	&- \frac{1}{48}\left[\Gamma_{\mathfrak{g}_B}\Gamma_{\mathfrak{g}_F}^2-(\rk(\mathfrak{g}_B)+1)\right]c_2(R)(4c_2(R) + p_1(T)) - I_8^\text{tensor}\,.
	\end{aligned}
\end{equation}
This generalizes the minimal affinization of the anomaly polynomial of necklace
quivers given in equation \eqref{I8-CM} to the dual LSTs, and can be checked
explicitly for $\widehat{E}_N^{\mathfrak{su}_k}$ and
$\widehat{D}_N^{\mathfrak{su}_{2k}}([2^k])$ by using the algorithm to determine
the anomaly polynomial on the tensor branch. 

A similar analysis can be performed for the dimension of the Coulomb branch. It
is easy to check that for minimal conformal matter, the dimension can be written similarly compactly in terms of group-theory data of the fiber algebra:
\begin{equation}
		\text{dim}(\text{CB}_{A_0^{\mathfrak{g}}}) = h^\vee_{\mathfrak{g}}-\rk(\mathfrak{g})-1\,.
\end{equation}
On the partial tensor branch, the $i^\text{th}$ curve of the base is decorated with an algebra $\mathfrak{g}^i$ of rank $\rk(\mathfrak{g}^i) =
d^i\,\rk(\mathfrak{g}_F)$. Taking into account the vector multiplets
associated with each node, we then recover the expected result given in
equation \eqref{kappas-CM}.

We can therefore explain the numerical coincidences behind T-duality at the
level of the anomaly polynomial from group theory, and in particular by looking on the partial tensor
branch directly. In equation \eqref{I8-ADE-LST}, the duality is realized easily
by simply exchanging the two algebras
$\mathfrak{g}_F\leftrightarrow\mathfrak{g}_B$. 

Note that while we mainly used the language of the F-theory construction
throughout this section, every step has an interpretation in the
field-theory description. Fusion is a well-defined six-dimensional operation
for the gauge theories even at the non-perturbative level, and blowing down
curves corresponds to moving to a specific point of the tensor branch. The
expressions we have used are then guaranteed to be correct by 't Hooft anomaly
matching. This remains true for other types of LSTs as well: we can
always start by computing the anomaly polynomial at a generic, weakly-coupled
point of the tensor branch, and track how quantities change at the conformal
fixed point. Minimal affinization then simply adds a contribution from a single
vector multiplet to the anomaly polynomial from which the two-group structure
constants can be extracted.

Before moving on to LSTs beyond necklace quivers and their T-dual theories, let
us comment on the possibility of minimal affinization for theories of type
$\widehat{\Lambda}^{\mathfrak{su}_k}$, but where the rank of every
$\mathfrak{su}$-type algebra over any curve is not fixed by the value of the
Kac labels of the base algebra. In other words, can we pick any choice of
fibers for ADE bases? The possibilities are of course severely constrained by
anomaly-cancellation conditions: allowing for different gauge algebras
$\mathfrak{su}_{k_I}$, quartic traces $\Tr F_I^4$ must be cancelled. This
imposes that there must be $2k_I$ hypermultiplets in the fundamental
representation of $\mathfrak{su}_{k_I}$. Taking into account bifundamental
hypermultiplets $(\bm{k}_I,\bar{\bm{k}}_J)$ arising at the intersection of two
curves, there generically remains $f^I$ hypermultiplets, which are rotated by
an $\mathfrak{su}_{f^I}$ flavor symmetry. It is easy to check that the
constraints on the possible values for $f^I$ is given in terms of the Cartan
matrix of $\widehat{A}^{IJ}$ associated with the base $\widehat{\Lambda}$: 
\begin{equation}
		f^I = \widehat{A}^{IJ}k_J\,.
\end{equation}
While there are non-trivial integer solutions to this equation for SCFTs, the
only consistent choice of $k_I$ for LSTs are null vectors of
$\widehat{A}^{IJ}$, so that $f^I=0$. This forces the $k^I$ to be a multiple of
the Kac labels of the affine algebra of the base, and the consistent choices
reduce to the cases discussed above. For $\widehat{A}_{N-1}$ bases, the
generalization to fiber algebras of type DE is straightforward, and we conclude
that there cannot be any necklace LST with non-trivial flavor symmetry, and
similarly for their T-duals. Here, we have considered only non-Abelian algebras
associated with a particular curve $I$ as we have relied on cancellation of
quartic traces, and there is therefore still a possibility for Abelian
symmetries, or flavor factors ``delocalized'' along the quiver rotating
composite gauge-invariant operators. The possibilities are, however, severely
constrained by the worldsheet bound.

\subsection{Beyond Necklace Quivers}

We have seen that for ADE orbifolds, minimal affinization simply corresponds to
considering the affine extension of the Dynkin diagram of the discrete group
$\Lambda\subset SU(2)$, which is obtained by gauging the flavor symmetry. All
the relevant quantities can then be directly computed on the partial tensor
branch.

SCFTs whose bases are associated with discrete groups $\Lambda\subset U(2)$ that
are not of ADE type are usually referred to as generalized orbifolds. In the
M-theory picture, these SCFTs are---up to a few outliers---associated with
fractional M5-branes probing frozen singularities \cite{deBoer:2001wca,
Witten:1997bs, Tachikawa:2015wka, Ohmori:2015pua, Ohmori:2015pia,
Mekareeya:2017sqh}. However, these non-ADE bases cannot be affinized minimally. 

Let us give an example. The affinization of the base
\begin{equation}\label{non-ADE-affinization}
		\Lambda: 3\underbrace{2\,2\dots 2\,2}_{N-2}3 \qquad \longrightarrow \qquad \widehat{\Lambda}:\qquad \textcolor{blue}{//}3\underbrace{2\,2\dots 2\,2}_{N-2}3\textcolor{blue}{1//}
\end{equation}
does not preserve the defect group. Indeed, when blowing down the new
$(-1)$-curve, we get $\widehat{\Lambda}\simeq \widehat{A}_{N-2}$, and the
resulting LST has a defect group
$\mathcal{D}^{(2)}(\widehat{\Lambda})=\mathbbm{Z}_{N-1}$. However, the SCFT has
$\mathcal{D}^{(2)}(\Lambda)=\mathbbm{Z}_{4N}$, and this violates one of the defining
conditions of minimal affinization.

It is a straightforward exercise to show that only ADE bases can be affinized
minimally. One of course recovers the original geometric classification
\cite{Bhardwaj:2015oru}: on the partial tensor branch, the pairing matrix of
all LSTs corresponds the Dynkin diagram of an affine ADE algebra, see Table~\ref{tab:affinized-bases-LST}. There are two exceptions, associated with the
affinization of an (endpoint) $(-1)$-curve:
\begin{equation}
\label{eq:EndPoint-1}
		1\textcolor{blue}{1} \to \textcolor{blue}{0}\,,\qquad 
		1\textcolor{blue}{||4} \to \textcolor{blue}{0\cdot}\,,
\end{equation}
The former corresponds to the endpoint of Heterotic LSTs $\text{I}_0$, while
the latter gives rise to a Type II LST with a base associated with a Kodaira
singularity of Type II. We differentiate these two bases by including an
additional dot in the latter case, and write $0\cdot\,$, as
in equation~\eqref{eq:EndPoint-1}. In two further cases corresponding to Type III and IV
singularities in the Kodaira classification, the bases of the LST have the same
pairing matrix as those of $A_2$ and $A_3$, respectively, but the geometry of
the base is different. They correspond to a tangential intersection of two
curves or a triple intersection point,
\begin{equation}\label{base-III-and-IV}
		\text{III}:\quad 2\textcolor{blue}{||2}\,,\qquad\qquad
		\text{IV}:\quad 2\overset{\displaystyle\textcolor{blue}{2}}{\Delta}2\,.
\end{equation}

\begin{table}[p]
    \centering
    \begin{threeparttable}   
        \begin{tabular}{ccccc}
				Kodaira& $G$ & Endpoint & $\mathcal{D}^{(2)}$ & Affine Dynkin diagram\\
            \toprule
			$\mathbbm{P}^1$ & $\varnothing$ & $1\textcolor{blue}{1}\,\to\, \textcolor{blue}{0}$ & $\varnothing$ & $\bnode{}{1}$\\
				\midrule
				$\text{I}_0$ & $\varnothing$ & $\textcolor{blue}{0}$ & $\varnothing$ & $\bnode{}{1}$ \\
				$\text{I}_3$ & $A_2$ & $\textcolor{blue}{//}22\textcolor{blue}{2//}$ = $2\overset{\displaystyle\textcolor{blue}{2}}{\phantom{1}}2$ & $\mathbbm{Z}_3$ & 
				$\begin{array}{c}
\raisebox{-12pt}{\rotatebox{60}{$-$}}\hspace{-4pt}\bnode{}{1}\hspace{-4pt}\raisebox{-4pt}{\rotatebox{-60}{$-$}} \\[-7pt]
\node{}{1}-\node{}{1} 
\end{array}$\\
				$\text{I}_N$ & $A_{N-1}$ & $\textcolor{blue}{//}\underbrace{22 \cdots  22}_{N-1}\textcolor{blue}{2//}$ & $\mathbbm{Z}_n$ & $\begin{array}{c}
\raisebox{-12pt}{\rotatebox{30}{$-\!\!-\!\!-$}}\bnode{}{1}\raisebox{0pt}{\rotatebox{-30}{$-\!\!-\!\!-$}} \\[-7pt]
\node{}{1}-\node{}{1}-\cdots-\node{}{1} 
\end{array}$\\
				\midrule
				$\text{I}^\ast_{0}$ & $D_4$ & $2\underset{\displaystyle \textcolor{blue}{2_{\vphantom{|}}}}{\overset{\displaystyle2}{2}}2$ & $\mathbbm{Z}_{2}\times\mathbbm{Z}_2$ & $\node{}{1}-\underset{\iver{}{1}}{\node{\bver{}{1}}{2}}-\node{}{1}$ \\
				$\text{I}^\ast_{N-4}$ & $D_N$ & $2\overset{\displaystyle2}{2}\underbrace{22\dots 22}_{N-5}\overset{\displaystyle\textcolor{blue}{2}}{2}2$ & $\mathbbm{Z}_{2}\times\mathbbm{Z}_2$ or $\mathbbm{Z}_4$  & $\node{}{1}-\node{\ver{}{1}}{2}-\node{}{2}-\cdots-\node{\bver{}{1}}{2}-\node{}{1}$ \\
				\midrule
				\text{II} & $H_1 \simeq A_0$ & $1\textcolor{blue}{||4}\,\to\,\textcolor{blue}{0\cdot}$ & $\varnothing$ & $\bnode{}{1}$ \\[8pt]
				$\text{III}$ & $H_2 \simeq A_1$ & $2\textcolor{blue}{||2}$ & $\mathbbm{Z}_2$  & 
				$\node{}{1}=\bnode{}{1} $\\
				$\text{IV}$ & $H_3 \simeq A_{2}$ & $2\overset{\displaystyle\textcolor{blue}{2}}{\Delta}2$ & $\mathbbm{Z}_3$ & 
				$\begin{array}{c}
\raisebox{-12pt}{\rotatebox{60}{$-$}}\hspace{-4pt}\bnode{}{1}\hspace{-4pt}\raisebox{-4pt}{\rotatebox{-60}{$-$}} \\[-7pt]
\node{}{1}-\node{}{1} 
\end{array}$\\
				\midrule
			    $\text{IV}^\ast$ & $E_6$ & $22 \overset{\displaystyle\textcolor{blue}{2}}{\overset{\displaystyle 2}{2}}22$ & $\mathbbm{Z}_3$ & $\node{}{1}-\node{}{2}-\node{\overset{\bver{}{1}}{\ver{}{2}}}{3}-\node{}{2}-\node{}{1}$\\[12pt]
				$\text{III}^\ast$ & $E_7$ & $\textcolor{blue}{2}22 \overset{\displaystyle 2}{2}222$ & $\mathbbm{Z}_2$ & $\bnode{}{1}-\node{}{2}-\node{}{3}-\node{\ver{}{2}}{4}-\node{}{3}-\node{}{2}-\node{}{1}$\\[12pt]
				$\text{II}^\ast$ & $E_8$ & $22 \overset{\displaystyle 2}{2}2222\textcolor{blue}{2}$ & $\varnothing$ & $\node{}{2}-\node{}{4}-\node{\ver{}{3}}{6}-\node{}{5}-\node{}{4}-\node{}{3}-\node{}{2}-\bnode{}{1}$\\
            \bottomrule
        \end{tabular}
    \end{threeparttable}
	\caption{
			Possible (endpoint) bases associated with LSTs obeying the
			tensor-decoupling condition and preserving the defect groups of the
			associated SCFTs. We show those of the Heterotic LSTS in the first
			line for completeness. The algebras for the Kodaira singularities
			of type II, III, IV  are sometimes denoted $H_i$ to differentiate
			them from those of the $\text{I}_n$ series. Each node of the
			Dynkin diagram is denoted by its Kac label $d^i$. The curve defined
			by minimal affinization and the corresponding affine node in the Dynkin
			diagram are colored blue.
	}
    \label{tab:affinized-bases-LST}
\end{table}

Given a Type II LST base, there are very few choices of compatible algebras.
From the minimal affinization point of view, this is partly explained from the
discussion above. $A_N$-type bases only give rise to conformal matter, and we
have seen that their nilpotent deformations cannot lead to a gauge-anomaly free
LST, and similarly for the three exceptional classes
$\widehat{E}_N^{\mathfrak{su}_k}$. Leaving out the special bases given in
equations \eqref{eq:EndPoint-1} and \eqref{base-III-and-IV} for a moment, we
are then left with $D_N$ bases. The trivalent pattern severely constrains the
possible SCFT \cite{Morrison:2016djb, Merkx:2017jey}, which in addition to the
$\mathfrak{su}_{2k}$ treated above, can only have
$\mathfrak{g}\in\{\mathfrak{su}_3, \mathfrak{so}_8, \mathfrak{e}_6\}$. As with
$\mathcal{K}(\mathfrak{su}_k, \mathfrak{so}_{2L})$, the corresponding flavor
must first be broken via a nilpotent deformation before minimal affinization
can be performed. This once again fixes the choice of nilpotent orbit, and
there is a unique possibility for each SCFT, leading to the following three
infinite series of LSTs:
\begin{equation}
	\begin{gathered}
		\widehat{D}_{N}^{\mathfrak{su}_3}([3^1]):\qquad 2\, \overset{\displaystyle 2}{\overset{\mathfrak{su}_2}{2}}\, \underbrace{\overset{\mathfrak{su}_3}{2}\, \overset{\mathfrak{su}_3}{2} \cdots \overset{\mathfrak{su}_3}{2} }_{N-5}\, \overset{\displaystyle\textcolor{blue}{2}}{\overset{\mathfrak{su}_2}{2}}\, 2\,,\qquad
		\widehat{D}_{N}^{\mathfrak{so}_8}([4^2]):\qquad \overset{\mathfrak{su}_2}{2} \overset{\displaystyle\overset{\mathfrak{su}_2}{2}}{\overset{\mathfrak{so}_7}{3}} \, \underbrace{1 \, \overset{\mathfrak{so}_8}{4}\, 1 \, \cdots 1 \, \overset{\fso_8}{4}\, 1}_{N-5} \overset{ \displaystyle\textcolor{blue}{\overset{\mathfrak{su}_2}{2}}}{\overset{\mathfrak{so}_7}{3}} \, \overset{\mathfrak{su}_2}{2}\,,\\
		\widehat{D}_{N}^{\mathfrak{e}_6}(D_4):\qquad \overset{\mathfrak{su}_{3}}{3}1\overset{\displaystyle \overset{\mathfrak{su}_3}{3}}{\overset{\displaystyle 1}{\overset{\mathfrak{e}_6}{6}}}\underbrace{1\overset{\mathfrak{su}_3}{3}1\overset{\mathfrak{e}_6}{6}\cdots \overset{\mathfrak{e}_6}{6}1\overset{\mathfrak{su}_3}{3}1}_{N-5} \overset{\displaystyle\textcolor{blue}{\overset{\mathfrak{su}_{3}}{3}}}{\overset{\displaystyle 1}{\overset{\mathfrak{e}_6}{6}}}1\overset{\mathfrak{su}_{3}}{3}\,.
	\end{gathered}
\end{equation}
Note that in the case of $D_{N}^{\mathfrak{e}_6}$, the breaking is done with
the so-called $D_4$ nilpotent orbit in the Bala--Carter notation~\cite{bala1976classesI, bala1976classesII}. More information about these three
families of SCFTs and their nilpotent breaking can be found in~\cite{Baume:2023onr}.

For the case of $\widehat{D}_{N}^{\mathfrak{su}_3}([3^1])$, we see that
the minimal affinization does not involve any flavor. Indeed, the nilpotent orbit
associated with the partition $[3^1]$ breaks the original $\mathfrak{su}_3$
flavor completely. The new curve, however, has a fiber with a singularity of Type
II, which we will explore in more detail in Section \ref{sec:geometry}. The
other two series of LSTs also have special singularities of Type III and IV
rather than their ADE counterparts. Their 2-group invariants are
straightforwardly found to be
\begin{equation}
		\begin{aligned}
				\mathcal{K}(\text{II}, \mathfrak{so}_{2N}) = \widehat{D}_{N}^{\mathfrak{su}_3}([3^1]):& \qquad \kappa_R = \phantom{1}6(N-3)\,,\qquad \text{dim}(\text{CB}) = \phantom{1}3(N-2)-2\,,\\
				\mathcal{K}(\text{III}, \mathfrak{so}_{2N}) = \widehat{D}_{N}^{\mathfrak{so}_8}([4^2]):& \qquad \kappa_R = 16(N-3)\,,\qquad \text{dim}(\text{CB}) = \phantom{1}6(N-2)-2\,,\\
				\mathcal{K}(\text{IV}, \mathfrak{so}_{2N}) \,=\, \widehat{D}_{N}^{\mathfrak{e}_6}(D_4)~:& \qquad \kappa_R = 48(N-3)\,,\qquad \text{dim}(\text{CB}) = 12(N-2)-2\,.\\
		\end{aligned}
\end{equation}

We see that minimal affinization gives us precious information about the dual
theory. Since the SCFTs have a flavor algebra $\mathfrak{su}_3$,
$\mathfrak{so}_8$, or $\mathfrak{e}_6$, as can be seen from the spine, one
might naively expect that we should look for LSTs with such bases. However, the
correct bases are of Kodaira type II, III, IV. This will become clear when we consider the geometric engineering of these theory in Section \ref{sec:geometry}. Moreover, knowing that the affine curve
should have an algebra $\mathfrak{g}=\mathfrak{so}_{2N}$ is helpful to find the
correct T-dual theories,
\begin{equation}
	\begin{aligned}
			\mathcal{K}(\mathfrak{so}_{2N}, \text{II}): &\qquad \,~\quad\textcolor{blue}{\overset{\mathfrak{so}_{2N}}{0\cdot}} \qquad&\longrightarrow&\qquad \qquad
		\quad\overset{\mathfrak{su}_{2(N-4)}}{1}||\textcolor{blue}{\overset{\mathfrak{so}_{2N}}{4}}\\
		\mathcal{K}(\mathfrak{so}_{2N}, \text{III}): & ~\qquad \overset{\mathfrak{so}_{2N}}{2} \textcolor{blue}{|| \overset{\mathfrak{so}_{2N}}{2}} \qquad&\longrightarrow&\qquad\qquad
		~ ~\overset{\mathfrak{so}_{2N}}{4} \overset{ \displaystyle \overset{\mathfrak{su}_{2(N-4)}}{2}}{ \overset{\mathfrak{sp}_{2(N-4)}}{1}} \textcolor{blue}{\overset{\mathfrak{so}_{2N}}{4}} \\
		\mathcal{K}(\mathfrak{so}_{2N}, \text{IV}): &\qquad 
		\overset{\mathfrak{so}_{2N}}{2} \overset{\displaystyle \overset{\textcolor{blue}{\mathfrak{so}_{2N}}}{2}}{\Delta}  \overset{\mathfrak{so}_{2N}}{2} \qquad&\longrightarrow&\qquad
		\overset{\mathfrak{so}_{2N}}{4} \overset{\fsp_{2(N-4)}}{1} \overset{\displaystyle \textcolor{blue}{\overset{\mathfrak{so}_{2N}}{4}}}{\overset{\displaystyle\overset{\mathfrak{sp}_{2(N-4)}}{1}}{\overset{\mathfrak{so}_{6N-16}}{4}}} \, \, \overset{\mathfrak{sp}_{2(N-4)}}{1}\,\overset{\fso_{2N+8}}{4} 
	\end{aligned}
\end{equation}
We have chosen $N>4$ as there are certain special enhancements for
$\mathfrak{so}_8\simeq D_4$ bases which do not fit the patterns above. We will
treat those cases in more detail in the next section. 

The LST $\mathcal{K}^\text{II}(III, \mathfrak{so}_{2N})$ and its dual exhibit
an additional peculiar feature. The LST is reached from the
$D^{\mathfrak{so}_8}_N$ SCFT via the nilpotent orbit associated with the
very-even partition $[4^2]$. However, there are \emph{two} nilpotent orbits
related to this partition, usually denoted $[4^2]_I$ and $[4^2]_{II}$. While
the resulting quiver is the same, the Nambu--Goldstone modes that decouple
along the RG flow are different, and this leads to two inequivalent SCFTs. The
difference stems from the fact that the SCFT
$\overset{\mathfrak{sp}_{m-4}}{1}[\mathfrak{so}_{4m}]$ has a Higgs-branch
operator in its spectrum transforming in one of the two spin representations of
the $\mathfrak{so}_{4m}$ flavor symmetry \cite{Mekareeya:2017jgc, Distler:2022yse}. This choice of spin representation
selects one of the two theories. When that curve configuration is part of a
larger quiver there can also be physically different theories, which is revealed after a careful analysis of certain gauge-invariant operators. This was shown to be the case for nilpotent deformations of D-type conformal matter associated with very-even partitions \cite{Distler:2022yse}.
Similarly, since the SCFT associated with the LST
$\mathcal{K}^{\text{II}}(\mathfrak{so}_{2N}, \text{III})$ also has a single
$(-1)$-curve decorated with an $\mathfrak{sp}_m$ algebra, we again have two
different theories. Performing minimal affinization, it is natural to
expect two different LSTs as well, giving a nice additional cross-check of
T-duality. 

Let us comment on the possible flavor symmetries arising in Type
II LSTs, which we know to be constrained from the bound given equation
\eqref{bound-flavor-specific}. We find that there is single pair of T-dual LSTs
with a flavor symmetry. They have an $\mathfrak{su}_2$ symmetry, and are given
by the two quivers
\begin{equation}
		\mathcal{K}(\mathfrak{e}_6,\text{II}):\qquad [\mathfrak{su}_2]\,2\,\overset{\displaystyle \textcolor{blue}{\overset{\fe_6}{6}}}{1}\, \overset{\fsu_3}{3}\,,\qquad\qquad
		\mathcal{K}(\text{II}, E_6):\qquad \begin{array}{c}\textcolor{blue}{2} \\ \overset{\fsu_2}{2}  \\ 2 \overset{\fsu_2}{2}  \underset{\displaystyle[\mathfrak{su}_2]}{\overset{\fg_2}{2}} \overset{\fsu_2}{2} 2  \end{array}\,.
\end{equation}
The first $\mathcal{K}(E_6, \text{II})$ has a defect group
$\mathcal{D}^{(2)}=\mathbbm{Z}_3^{(2)}$, converted into a one-form symmetry
under the duality. Moreover, the pair has the following duality invariants:
\begin{equation}
		\kappa_G = 0\,,\qquad 
		\kappa_R = 27\,,\qquad
		\rk(\fg)= \rk(\mathfrak{su}_2) = 1\,,\qquad 
		\text{dim}(\text{CB}) = 11 \,.
\end{equation}
In addition, we find that in both cases, $\kappa_F^{\mathfrak{su}_2}=3$.

Minimal affinization therefore gives us the correct information to find the
T-dual theory. Indeed, we can systemtically scan over all possible SCFT bases and
affinize them to LSTs. Demanding minimal affinization then greatly simplifies the search for dual pairs. All Type II LSTs and their two-group invariants are collected in
Appendix \ref{app:tables}, which we will now construct geometrically.

\section{The Geometry of Type II LSTs}\label{sec:geometry}

In this section, we first review the geometric construction of Type II LSTs, and illustrate the procedure using F-theory on elliptic Calabi--Yau threefolds with multiple fibrations. For a givem choice of algebras associated with fiber and base singularities, this will allow us to construct the corresponding LST and to identify T-dual pairs.

\subsection{LSTs and Double Fibrations}
We first take a more general perspective on the F-theory geometry $X_3$ and discuss which type of LSTs can be obtained on general grounds. Our starting point will be compact threefolds $X_3$ and their possible elliptic fibration structures. By taking a decompactification limit of $X_3$ which keeps a curve of self-intersection $0$ in the base compact, we can decouple gravity and obtain an LST. Note that there are infinite classes of LSTs that cannot be coupled consistently to gravity, and therefore cannot be obtained via a decompactification limit of a supergravity theory. 

Indeed, we want to argue that only Heterotic and Type II LSTs can be obtained by decoupling gravity. We start by imposing two conditions on $X_3$ that are necessary to give rise to an LST via F-theory:
\begin{enumerate}
		\item \textbf{F-theory condition:} $X_3$ needs to have a torus fibration with base $B_2$.
    \item \textbf{Little String condition:} $B_2$ has to have a curve $D_b$ of self-intersection $D_b^2=0$. 
\end{enumerate} 
Both conditions have appeared in works by Koll\'ar, Oguiso, and Wilson ~\cite{Kollar:2012pv,Oguiso1993ONAF,Wilson1994} in their general study of the possible fibration structures of Calabi--Yau threefolds. For each linear independent nef divisor $D_F$ that satisfies $D_F^3
=0$ but $D_F^2\neq 0$, there exists an inequivalent torus fibration.\footnote{A divisor $D_F$ is nef if $D_F \cdot \mathcal{C} \geq 0$ for every algebraic curve $\mathcal{C} \in X_3$.} Intuitively, one may view
$D_F$ as a vertical divisor $D_B \in B_2$. Such divisors identify a fibration structure and have vanishing triple self-intersection in the full threefold. 

In his proof of Koll\'ar's conjecture for threefolds \cite{Oguiso1993ONAF}, Oguiso classified the general fibration 
structures $X_3$ can have, employing only the two following intersection
properties of an ample nef divisor $D_B$:
\begin{align}
\label{eq:OguisoCri}
    D_B^{\nu+1}=0 \, , \quad D_B \cdot c_2(X_3) = \delta \, ,
\end{align}
with $c_2(X_2)$ the second Chern class of the tangent bundle of $X_3$. 
Depending on the values of $\nu$ and $\delta$, $X_3$ admits a fibration structure with $\pi(X_3)=B_2$ and with general fibers $\pi^{-1}(B_2)=F$, as summarized in Table~\ref{tab:OguisoClassification}.
\begin{table}[t!]
    \centering
    \begin{tabular}{cccc}
    $\quad$ Type $\quad$  & $\quad\nu\quad$ & $\quad \delta \quad$  & $\quad$ Structure of $B$ and $F$ $\quad$  \\ 
    \toprule
    I$_+$ & 1 & $\delta >0$ & $B_2=\mathbbm{P}^1 \text{ and  } F=K3$ \\[4pt]
    I$_0$ & 1 & $\delta =0$ & $B_2=\mathbbm{P}^1 \text{ and  } F=\mathbbm{T}^4$  \\[4pt]
    II$_+$ & 2 & $\delta >0$ & $B_2=\Sigma_2 \text{ and  } F=\mathbbm{T}^2$ \\ \bottomrule 
    \end{tabular}
    \caption{Fibration structures of a threefold $X_3$ with a divisor $D_B$ satisfying~\eqref{eq:OguisoCri}. There are also fibrations of Oguiso Type II$_0$ and III, which we omitted since they are not relevant in this work.
    }
    \label{tab:OguisoClassification}
\end{table}

Oguiso Type II corresponds to the F-theory condition, i.e., $X_3$ has a torus fibration with $B=\Sigma_2$ a rational surface. The LST condition requires a curve $\mathcal{C} \in B_2 $ with $\mathcal{C}^2= 0$, which, when pulled back to a surface $S\in X_3$, satisfies $S^2\cdot D=0$ for any divisor $D\in X_3$. This tells us that we have a divisor of Oguiso Type I. Thus, when combining these two conditions, we require a divisor $D_F$ with $D_F^3=0$ and a divisor $S$ with $S^2=0$, compatible with the double fibration structure. The two subtypes of Oguiso Type I divisors correspond to different fibration structures and hence two different LSTs: Type I$_+$ has a K3 fibration and together with a compatible Oguiso Type II structure, it must be an elliptic K3, while for the Type I$_0$ case the fiber is $\mathbbm{T}^2 \times \mathbbm{T}^2$. We conclude that there can be only two types of LSTs that can be obtained by decoupling gravity in F-theory compactification, where the possible bases $B_2$ of the elliptic fibration are
\begin{align}
    \label{eq:BaseTopology}
    \begin{array}{|c|c|l|}
       \hline
       \text{Oguiso Type}  & \text{LST Type} & \text{Base Topology}     \\ \hline
       \text{I}_+ & \text{Heterotic} & B_2 \sim \mathbbm{P}^1 \times \mathbbm{C}
       \\ \hline
       \text{I}_0 & \text{Type II} & B_2 \sim (\mathbbm{T}^2 \times \mathbbm{C})/\Lambda  \\ \hline
    \end{array}
\end{align}
Note that we have added the possibility of a quotient $\Lambda$, as the fibration may have Kodaira fibers. Typical examples of Oguiso Type $\text{I}_0$ bases are rational elliptic surfaces, such as $dP_9$. These may have reducible Kodaira fibers at the origin of $\mathbbm{C}$.

\subsection{Type II LSTs From Gravity Decoupling}\label{sec:LST-double-fibration}

To describe an elliptically fibered threefold with a base that is itself and elliptic fibration, we can start with a compact geometry and subsequently decompactify it. A good starting point for Type II LSTs with a double elliptic fibration structure is the Schoen manifold,\footnote{In fact, in \cite{Anderson:2017aux} it was shown that there exists an infinite family of elliptic fibration structures for the Schoen manifold.} which features prominently in string constructions. The simplest description for the Schoen manifold is given by the split bi-cubic 
\begin{align}
\left[
  \begin{array}{ c|cc}
\mathbbm{P}^2 & 3 & 0 \\ 
\mathbbm{P}^2 & 0 & 3 \\ 
\mathbbm{P}^1 & 1 & 1 \\ 
\end{array}
\right]
\end{align}
We can then define two hypersurfaces $P_F$ and $P_B$ which are cubics in either of the two ambient $\mathbbm{P}^2$, and linear in their common $\mathbbm{P}^1$. Via a change of coordinates, this can be mapped into a Tate model given by the GSLM matrix
\begin{align}
    \begin{array}{|ccc|ccc|cc||cc|}\hline
   X & Y & Z &  \hat{X} & \hat{Y} & \hat{Z} & t_0 & t_1 & P_B & P_F \\ \hline   
   2 & 3 & 1 & 0 & 0 & 0 & 0 & 0 & 0 & 6 \\ \hline
   0 & 0 & 0 & 2 & 3 & 1 & 0 & 0 & 6 & 0 \\ \hline
   0 & 0 & -1& 0 & 0 & -1& 1 & 1 & 0 & 0 \\ \hline
    \end{array}
\end{align}
The two hypersurface equations are simply the following two polynomials in Tate form:  
\begin{align}
\begin{split}
    P_{F}&= Y^2 + X^3+ XYZ a_1 + X^2 Z^2 a_2 +Y Z^3 a_3 + X Z^4 a_4 + Z^6 a_6\,,  \\
P_{B}&= \hat{Y}^2 + \hat{X}^3+ \hat{X}\hat{Y}\hat{Z} b_1 + \hat{X}^2 \hat{Z}^2 b_2 +\hat{Y} \hat{Z}^3 b_3 + \hat{X} \hat{Z}^4 b_4 + \hat{Z}^6 b_6\,, 
\end{split}
\end{align}
with $a_i$ and $b_i$ generic polynomials in the $\mathbbm{P}^1$ coordinates $[t_0:t_1]$ of degree $i$.
The split bi-cubic and the double Tate model differ slightly in their singularity structures. However, this difference will disappear upon decompactification. The description makes the double fibration over $\mathbbm{P}^1$ manifest: we can treat either $\{p_F=0\}$ or $\{p_B=0\}$ as the F-theory torus. In either case, the discriminant of the base is of degree deg$(\Delta_b)=12$ and hence a rational elliptic surface. 

For concreteness, we take $\{p_B=0\}$ to be the base of the fibration. There are two divisors in the base $B_2$ that are of main interest: first, $\{\hat{Z}=0\}$ is the section of the rational elliptic surface and hence a copy of  the base $\mathbbm{P}^1_{t_0,t_1}$. Secondly, the two linear equivalent divisors $\{t_i=0\}$ are points in the base $\mathbbm{P}^1$ and thus a copy of the torus fiber, which in the full threefold is $\mathbbm{T}^2 \times \mathbbm{T}^2$. 
Before engineering various singularities in the fiber and the base, we need to decompactify the theory to a Type II LST.
We can perform this decompactification at the level of the ambient space by taking the volume of the $\mathbbm{P}^1_t$ factor to infinity, replacing 
\begin{align}
    \mathbbm{P}^1_{[t_0,t_1]} \rightarrow \mathbbm{C}_{t_0}\, .
\end{align}
We can therefore simply focus on the patch $t_1=1$, where we rename $t_0=t$. We can then factorize $a_i$ and $b_i$ in $t$ to obtain singularities in either F-theory fiber or base.

\subsection{Fiber and Base Singularity Structure}

A Type II LST $\mathcal{K}^{\text{II}}(\fg_F,\fg_B)$ is specified by a fiber and base with a singularity in the Kodaira classification. To reduce notational burden and increase readability, we will use the name of Kodaira singularities in the text, but otherwise refer to them in terms of their ADE algebra 
$\fg$, as in Table~\ref{tab:affinized-bases-LST}, unless they are of special type II, III and IV.

We can engineer those singularities using the Tate classification (see e.g. \cite{Katz:2011qp}), which amounts to specifying the vanishing orders of the Tate coefficients $a_i$ of the fiber and $b_i$ of the base. We collect this data in Tate vectors $\vec{n}$ and $\vec{\hat{n}}$ for the fiber and the base ,
\begin{align}
\label{eq:tatedoublevector}
    a_i \rightarrow t^{n_i} \hat{a}_i \, ,\quad \text{ and } \quad b_i \rightarrow t^{\hat{n}_i} \hat{b}_i\, . 
\end{align} 
This allows for a compact notation, and to easily read off the resolution of the fiber and base singularities in terms of toric tops \cite{Bouchard:2003bu} and possible gauge enhancements. 

Let us assume we engineered a fiber a singularity $\fg_F$ and base singularity $\fg_B$ with Tate vectors given in equation ~\eqref{eq:tatedoublevector}. We start by discussing the singularity structure in the base at 
$t=0$. The elliptic fiber $\mathbbm{T}^2_B$ needs to be resolved first by replacing it with a set of $\fg_B\in ADE$ resolution divisors $D_{f_i}$ 
with $i=0, \ldots, r$, with $r=\rk(\fg_b)$ 
such that
\begin{align}
    [\mathbbm{T}^2_B] \sim \sum_{i=0}^r d_i [D_{f_i}]
\end{align}
with $D_{f_i} \cdot D_{f_j} = \widehat{A}_G^{i,j}$ the affine (negative) Cartan
matrix and $d_i$ the Kac labels.\footnote{As the base itself is complex one-dimensional, there is no monodromy that could act on the $\mathbbm{T}^2_B$ fibers, which means they are all split and of ADE type.} The latter are collected in Table
\ref{tab:affinized-bases-LST} for the reader's convenience. At the same time,
we need to replace $t$ by
\begin{align}
    t \rightarrow\hat{t}= \prod_{i=0}^r f_i^{d_i} \,.
\end{align} 
This is required to ensure $\hat{t}$ stays invariant under $\mathbbm{C}^*$ scalings of the respective resolution divisors $D_{f_i}$.
Since the fiber $\{p_F=0\}$ is a section in the new variable $\hat{t}$, it is invariant under those scalings as well.

Note that $\{\hat{t}=0\}$ is a reducible divisor with components $f_i$ of vanishing order $d_i$. This is relevant when considering additional fiber singularities $\fg_F$ in $\{p_F =0\}\cap\{\hat{t}=0\}$, engineered with the Tate vector $\vec{n}$. First, note that the singularity $\fg_F$ splits into $r$ components over $\{\hat{t}=0\}$. Since $\hat{t}$ vanishes to order $d_i$ over $\{f_i=0\}$, the original Tate vector is modified to $\vec{n} \rightarrow d_i \cdot \vec{n}$ over each resolution divisor $D_{f_i}$, which enhances the singularity type. This construction is very useful in the study of the more exotic singularities of Kodaira Type II, III, IV.  

We are furthermore required to choose two non-trivial singularities for fiber and base to obtain a threefold with full $SU(3)$ holonomy. If one of the singularities is chosen to be trivial, the geometry becomes a direct product $X_3=\mathbbm{T}^2 \times (\mathbbm{T}^2\times \mathbbm{C})/\Lambda$
and supersymmetry enhances to $\mathcal{N}=2$ in 5D.
The two different F-theory lifts yield $\mathcal{N}=(2,0)$ or $\mathcal{N}=(1,1)$ LSTs, which we review in more detail in Appendix~\ref{app:Enhanced}. Note that there exist non-simply laced $\mathcal{N}=(1,1)$ LSTs that were argued to be T-dual to $\mathcal{N}=(2,0)$ LSTs with an (affine) outer automorphism twist \cite{DelZotto:2020sop}. The geometry and T-duality structure of such theories has been discussed in \cite{Bhardwaj:2022ekc}.

The above structure readily explains our notation of the two singularities defining a Type II LST: the singularity $\fg_B$ denotes the (affine) Dynkin diagram characterising the shape of the quiver at the endpoint configuration. The fibral singularity $\fg_F$ then encodes the gauge algebra factor over Dynkin multiplicity 1 nodes over the affine quiver base, which is also the flavor symmetry gauged during minimal affinization. This is the geometric avatar of the partial tensor branch description discussed in Section~\ref{sec:ADE-orbifolds}.

\subsection{ADE Singularities}

In the following, we show how to construct LSTs of type $\mathcal{K}^{\text{II}}(\fg_F,\fg_B)$. The strategy is to first tune a singularity $\fg_B$ in the base, resolve it, and then discuss the possible choices for $\fg_F$. We engineer all models discussed in Section~\ref{sec:BottomUp} and lay the grounds for the more exotic Kodaira singularities of Type II, III, IV. 
 
As a warm-up, let us start with a simple $\text{I}_3$ singularity, i.e., $\fg_B=\fsu_3$ in the base, with Tate vector
\begin{align}
    \vec{n}=\{0,1,1,2,3\} \, .
\end{align}
The resolution requires three divisors, which replace the central fiber with
\begin{align}
    \hat{t} \rightarrow f_0 f_1 f_2 , 
\end{align}
and the resolved hypersurface for the base takes the form 
\begin{align}
    p_B=Y^2+f_1^2 f_2 X^3  + X^2 Z^2 c_2+ f_0^2 f_2 X Z^4 c_4 + X Y Z c_1 + f_0^4 f_2^2 Z^6
c_6 + f_0^2 f_2 Y Z^3 a_3 \, . 
\end{align}
where the $c_i$ are generic polynomials in $\hat{t}$. We can easily compute the various reducible components of the fiber $\mathbbm{P}^1$'s at $\{p_B=0\}$ upon replacing $t$ by $\hat{t}=f_0 f_1 f_2$. Next we engineer an ADE singularity $\fg_F$ in the F-theory fiber over $t=0$. We choose to take a Type $\text{IV}^{*}$ Kodaira singularity---that is $\fg_F=\fe_6$---with Tate vector
\begin{align}
    \vec{\hat{n}}=(1,2,2,3,5) \, .
\end{align}
Recall that upon resolving the base, we have to replace $t \rightarrow
\hat{t}=f_0 f_1 f_2$, leading to three copies of $\fe_6$ in the
F-theory fiber in which each singularity intersects the other two. This leads
to non-minimal singularities that can be resolved as chains of minimal
conformal matter $A_{0}^{\fe_6}$~\cite{DelZotto:2014hpa}, and we obtain the
necklace quiver $\widehat{A}_2^{\fe_6}$ discussed around equation
\eqref{def-necklace}: 
\begin{align}
		\mathcal{K}^{\text{II}}(\fe_6,\fsu_3): \quad    \textcolor{blue}{//} 1\,  \overset{\fsu_3}{3}\, 1\,  \overset{\fe_6}{6} \, 1 \,\overset{\fsu_3}{3}\, 1 \, \overset{\fe_6}{6} \,  1 \, \overset{\fsu_3}{3} \, 1 \, \textcolor{blue}{\overset{\fe_6}{6} //} \, ,
\end{align}
In Section \ref{sec:BottomUp}, we have seen that the curve denoted in blue is
the one involved in minimal affinization. In this section, the fiber
singularity of such curves define $\mathfrak{g}_F$. As mentioned above, in the
geometric construction this is the algebra associated with the affine Kac label
of multiplicity one.

Next, we exchange the role of the fiber and base $\mathbbm{T}^2$ prior to
resolution. We then resolve the $\fe_6$ singularity in the base by replacing
\begin{align}
    t \rightarrow \hat{t} =f_0 f_1 f_2 g_1^2 g_2^2 g_3^2 h_1^3\,,
\end{align}
which leads to the base hypersurface equation 
\begin{align}
    p_B&= f_1 f_2^2 g_3 X^3 + f_0^2 f_1 f_2^2 g_1 g_2^2 g_3^2 h_1^2 X^2 Z^2 a_2 + 
 f_0^3 f_2 g_2^2 g_3 h_1 X Z^4 a_4   \nonumber \\ & +f_0 f_1 f_2 g_1 g_2 g_3 h_1 X Y Z a_1  + 
 f_0^5 f_2 g_1 g_2^4 g_3^2 h_1^3 Z^6 a_6+ f_0^2 g_2 Y Z^3 a_3 + f_1 g_1 Y^2 \, .
\end{align}
Having resolved the central Type IV$^*$ fiber in the base, we turn to the
F-theory fiber.  For simplicity, we only discuss the Weierstrass model
corresponding to the Tate model of the fiber. Since we have an $\text{I}_3$
singularity, $f$ and $g$ do not vanish but the discriminant vanishes to order 3
over $\{t=0\}$. Replacing $t$ by $\hat{t}$, we thus obtain the reducible
discriminant locus  
\begin{align}
    \Delta= (\hat{t})^3 \hat{\Delta} = (f_0 f_1 f_2 g_1^2 g_2^2 g_3^2 h_1^3)^3 \hat{\Delta} \, .
\end{align}
The $\text{I}_3$ singularity over $\{t=0\}$ is hence enhanced to
$(\text{I}_{3})^3\times(\text{I}_6)^3\times \text{I}_9$, resulting in the quiver 
\begin{align}
		\mathcal{K}^{\text{II}}(\fsu_3,\fe_6):  \quad \overset{\mathfrak{su}_{3}}{2}\,\overset{\mathfrak{su}_{6}}{2} \, \overset{\displaystyle\textcolor{blue}{\overset{\mathfrak{su}_{3}}{2}}}{\overset{\displaystyle \overset{\mathfrak{su}_{6}}{2}}{\overset{\mathfrak{su}_{9}}{2}}}\,\overset{\mathfrak{su}_{6}}{2} \, \overset{\mathfrak{su}_{3}}{2} \, .
\end{align}
It is straightforward to generalize this procedure to an $\text{I}_k$
singularity in the fiber, which is then enhanced to an $\text{I}_{d_i k}$
singularity where $d_i$ are the Dynkin multiplicities of the base singularity.
For the $\fe_6$ case, those are the $\widehat{E}_6^{\mathfrak{su}_k}$ theories
given in \eqref{eq:E6SUkQuiver}.

The above construction proves T-duality explicitly for the LSTs
$\mathcal{K}^{II}(\fsu_k, \fe_6)$ and $\mathcal{K}^{II}( \fe_6,\fsu_k)$ as they
are engineered from the very same singular threefold. After full resolution of fibers and base, both would be birational phases of the same (extended) K\"ahler moduli space. It is straightforward to engineer all
$\mathcal{K}^{\text{II}}(\fg_F, \fg_B) $ type of LSTs and their T-duals
$\mathcal{K}^{\text{II}}(\fg_F, \fg_B) $ when choosing $\fg_B=\fsu_{K}$ and
$\fg_F\in$ ADE. This reproduces the list given in
Section~\ref{sec:ADE-orbifolds} and Table~\ref{tab:LST-A-ADE}.\footnote{For a
similar recent LST construction via brane webs, see
\cite{Wei:2022hjx}.} 

From this geometric perspective, it becomes clear why there are no theories
$\mathcal{K}^{\text{II}} (\fg_F, \fg_B)$ where $\fg_F,\fg_B$ are both
singularities of type I$_0^*$ or higher: consider an $\text{I}_{n}^*$
singularity, which engineers a $\fg_F=\fso_{2n+8}$ gauge algebra with
Weierstrass coefficients 
\begin{align}
    f= t^2 \hat{f}\, , \qquad g= t^3 \hat{g} \, .
\end{align}
When combining this with another singularity of type $\text{I}_0^*$ or higher,
the reducible locus $\{\hat{t}=0\}$ is given by a set of irreducible  divisors,
at least one of which has $d_i>1$. This leads to a non-minimal singularity of
vanishing orders $(4,6)$ or higher in the fiber. Such a singularity has no
crepant resolution and does not lead to a 6D supersymmetric theory.

\subsection{Type II, III, IV Fibers}

The above procedure is very useful to discuss the exotic Kodaira singularities
of Type II, III, IV. These singularities correspond to strong coupling
versions of the ordinary $\text{I}_1, \text{I}_2$ and $\text{I}_3$
singularities.  For a given fiber singularity of Type II, III, IV, we can
enumerate all compatible base singularities by identifying the highest integer
$d$ that still leads to a crepantly resolvable threefold, see Table~\ref{tab:IIDESing}.  Hence, we can get the following combinations of
fiber and base singularities:
\begin{table}[t!]
    \centering 
    $
 \begin{array}{cccc}
    \vec{n} & d & \text{Fiber} & \text{Algebra} \\ \toprule 
    (0,0,1,1,1) & d & \text{I}_d & \fsu_d \\ \midrule
     (1,1,1,1,1) & 1 & \text{II} & - \\  
     & 2 & \text{IV}^{s/ns} & \fsp_1/\fsu_3 \\ 
     & 3 & \text{I}_0^{*,s/ss/ns} & \fso_8/\fso_7/\fg_2 \\ 
     & 4 & \text{IV}^{*,s/ns} & \fe_6/\ff_4  \\ 
     & 5 & \text{II}^{*} & \fe_8  \\ 
     & 6 & \text{non-minimal} & - \\ \midrule 
     (1,1,1,1,2) & 1 & \text{III} & \fsp_1\\  
& 2 & \text{I}_0^{*,s/ss/ns}  & \fso_8 /\fso_7/\fg_2  \\ 
& 3 & \text{III}^{*} & \fe_7 \\  
& \geq4 & \text{non-minimal} & - \\ \midrule
(1,1,1,2,3) & 1& \text{IV}^{s} & 
\fsu_3 \\ 
 & 2& \text{IV}^{*} & \fe_6 \\
 & \geq3 & \text{non-minimal} & - \\ \bottomrule
    \end{array}$
    \caption{
    Singularity and gauge algebra structure for Type $\text{I}_1$, II, III, IV singularities and possible enhancements by a factor $d$.
    }
    \label{tab:IIDESing}
\end{table}
\begin{align}
\begin{array}{|c|c|} \hline
    \text{Fiber} & \text{Base} \\ \hline
	\text{II} & A_{M-1}, D_N, E_6,E_7 \\ \hline
     \text{III} &A_{M-1}, D_N, E_6 \\ \hline
     \text{IV} &A_{M-1}, D_N \\
    \hline
     \end{array}
\end{align}
We have left out the possible Kodaira Type II, III and IV base singularities, which we will discuss separately in the next section. We will
also omit a separate discussion of the exotic singularities over $A_{M-1}$
bases, as those are indistinguishable from their I$_k$ counterparts for
$k=1,2,3$. Our main interest here lies in the D and E type singularities, and
their gauge enhancements upon resolution.
We can read off the final (enhanced) gauge algebra from  Table~\ref{tab:IIDESing}.

Note that a Kodaira Type II singularity over a base singularity $\text{I}_{N}^*$ with a $D_{N+4}\simeq\mathfrak{so}_{2N+8}$ algebra enhances the fiber singularity to Type IV on multiplicity 2 nodes. The two endpoints of the multiplicity-two chain (i.e., those with multiplicity-one neighbors), have a non-split fiber of singularity Type $\text{IV}^{ns}$ and the gauge algebra is reduced to $\fsu_2$. In the other cases, the fiber singularities are split and hence host an $\fsu_3$ algebra:
\begin{align}\label{eq:IIDk4}
		\mathcal{K}^{\text{II}}(\text{II},\fso_{2N+10}):\quad  2~~\overset{\displaystyle 2}{\overset{\fsu_2}{2}} \, \, \underbrace{\overset{\fsu_3}{2}\, \, \overset{\fsu_3}{2} \ldots \overset{\fsu_3}{2} }_{\times N}\, \,   \overset{\displaystyle \textcolor{blue}{2}}{\overset{\fsu_2}{2}} ~~ 2   \, , \qquad  (\text{CB},\kappa_R)=(3N+7,6N+12) \, .
\end{align}
The matter above originates from intersections of Type II$-$IV$^{ns}$ and  IV$^{s/ns}-$IV  fiber singularities that have been partially computed in \cite{Arras:2016evy,Grassi:2021ptc}. The Type  IV$-$IV  gives a bifundamental hypermultiplet of $\fsu_3^2$, while the II$-$IV$^{ns}$ collision gives a fundamental of $\fsu_2$.

For generic $N$, the above theory has a $U(1)$ flavor symmetry~\cite{Apruzzi:2020eqi,Ahmed:2023lhj}, necessitated by the small ``ramp'' under which the first and last $\mathfrak{su}_3$ fundamental multiplets are charged.
The $D_4$ and $D_5$ case are special since they have only $\fsu_2$ gauge algebra factors over the middle curves. 

Moving on to $\fe_6$ singularities in the base, we get the quiver
\begin{align}
    \mathcal{K}^{\text{II}}(\text{II}, \fe_6):\quad    2 \, \overset{\fsu_2}{2} \, 
	\overset{ \displaystyle \overset{\displaystyle \textcolor{blue}{2}}{\overset{\fsu_2}{2}} } {\underset{[\fsu_2]}{\overset{\fg_2}{2}}} \overset{\fsu_2}{2}\, 2 
       \, \quad \text{ with } (\text{CB}, \kappa_R)=(11,27)
\end{align} 
The $\fg_2$ gauge algebra requires an $\fsp_4$ flavor symmetry which upon
gauging of an $(\fsu_2)^3$ subalgebra leads to a residual $\fsu_2$ flavor
algebra. Since the LST charge $\ell^\text{LST}_I$ of the curve with a $\fg_2$
algebra is $3$, the induced $\fsu_2$ flavor current on the LST curve lies
within the unitarity bound,
\begin{align}
c_{3,\fsp_1}=\frac95 < 2 \, .
\end{align} 
Geometrically, this is the only type of model we can obtain, as the $I_0^*$
singularity in the middle must always be non-split. 

For an $\fe_7$ base we find 
\begin{align}
\label{eq:IIE7}
    \mathcal{K}^{\text{II}}(\text{II}, \fe_7): \quad  
	\textcolor{blue}{2} \, \, \overset{\fsu_2}{2} \, \, \overset{\fg_2}{3} \, \, 1 \, \, 
    \overset{ \displaystyle 
    \overset{ \displaystyle \overset{\fsu_3}{3}}{1}}{ \overset{\ff_4}{5}} \, \, 1 \,\, \overset{\fg_2}{3} \, \, \overset{\fsu_2}{2} \, \, 2  
       \, , \quad \text{ with }(\text{CB}, \kappa_R)=(22,96)
\end{align}
The resolved geometry above has fixed the singularity on the upper $(-3)$-curve
to be a $\text{IV}^{s}$ singularity\footnote{Field theory may have suggested
an $[\ff_4,\fe_8]=\fg_2 $ subalgebra, due to the attached E-string. The
enhanced $\fsu_2$ flavor symmetry of this $\fg_2$ algebra would in principle be
consistent with the LST unitarity bound. However, there is no T-dual model of
type $\mathcal{K}^{\text{II}}(\mathfrak{e}_7, \text{II})$ with consistent flavor rank and
2-group structure data.}, i.e., an $\fsu_3$ algebra as can be seen in
Table~\ref{tab:IIDESing}.  This is also consistent with the considerations made
in \cite{Merkx:2017jey} for SCFTs.

We move on to Type III singular fibers with $I^*_{n+1}$ base singularities. The cases $I^*_0$ and $I^*_1$ are special and we will discuss them momentarily. The generic quiver is
\begin{align}
\label{eq:IIISo2n}
	\mathcal{K}^{\text{II}}(\text{III},\fso_{2N+10}):  \overset{ \fsu_2}{2} \, \overset{ \displaystyle \overset{\fsu_2}{2}}{\overset{\fso_7}{3}} \, \, 1 \, \, \overset{\fso_8}{4}\, \,1 \ldots 1 \, \, \overset{\fso_8}{4} \, \, 1  \,\overset{\displaystyle \textcolor{blue}{\overset{\fsu_2}{2}}}{\overset{\fso_7}{3}} \, \overset{\fsu_2}{2}   \, , \qquad  (\text{CB},\kappa_R)=(6N+16, 16N+32)\,,
\end{align}
where $N$ is the number of $\fso_8$ factors. Again, the geometry requires the first and last $\text{I}_0^{*}$ singularity to be semi-split, i.e., an $\fso_7$. Field theoretically, one can again wonder whether they could be enhanced to $\fso_8$. However, this would violate the worldsheet unitarity bound, since $\fso_8$ would come with an extra $\fsu_2$ flavor symmetry at level two, which induces an $c_l=3$ left-moving current on the LST worldsheet, and is thus inconsistent with the bound derived in Section~\ref{sec:two}.  

Turning to the special case $\text{I}_1^*$ in the base, we do not get an $\fso_8$ algebra but only
\begin{align}
		\mathcal{K}^{\text{II}}(\text{III},\fso_{10}):  \quad \overset{\fsu_2}{2}\,  \overset{ \displaystyle  \overset{\fsu_2}{2}}{\overset{\fso_7}{3}} \, \,   1 \, \,   \overset{ \displaystyle  \textcolor{blue}{\overset{\fsu_2}{2}}}{\overset{\fso_7}{3}} \, \,     \overset{\fsu_2}{2}   \, , \quad \text{with } (\text{CB},\kappa_R)=(16,32)\,.
\end{align} 
Similarly, for a base singularity of type I$_0^*$, we get
\begin{align}\label{K-III-SO8}
		\mathcal{K}^{\text{II}}(\text{III}, \fso_8):\quad 
    \begin{array}{c}
		\textcolor{blue}{\overset{\fsu_2}{2}} \\
	    \overset{\fsu_2}{2} \, \, \overset{\fso_7}{2} \, \, \overset{\fsu_2}{2} \\
		\overset{\fsu_2}{2} 
\end{array}\, \, \,\text{ with } (\text{CB},\kappa_R)=(11,18)\,.
\end{align}
The $\fso_7$ algebra requires an additional hypermultiplet in the vector representation $\mathbf{7}$, and therefore has an $\fsu_2$ flavor symmetry. This flavor symmetry has $k^{\mathfrak{su}_2}=2$, and is consistent with unitarity of the LST worldsheet. Field-theoretically, we could also further enhance the middle algebra to $\fso_8$, but this would-be quiver has an $\mathfrak{su}_2$ flavor symmetry that violate the unitarity bound. We could also try to replace $\fso_7$ algebra by $\mathfrak{g}_2$. This quiver looks promising, since its invariants would be the $N=-1$ case of $\mathcal{K}^{\text{II}}(\text{III},\fso_{2N+10})$ in equation~\eqref{eq:IIISo2n}. There is, however, no candidate T-dual theory; the invariants of the $\mathcal{K}^\text{II}(\mathfrak{so}_{8}, \text{III})$ LST also exhibit a jump as compared to the rest of the infinite series, as we will see below. In either cases, these algebras cannot be constructed with our methods, where consistency of the geometry forces us to have an $\mathfrak{so}_7$ algebra on that curve. A possibility that is allowed for the quiver~\eqref{K-III-SO8} is to  break its flavor symmetry $\fso_7 \rightarrow \fso_6 \simeq \fsu_4$ via a Higgs mechanism. Geometrically, this corresponds to a deformation of the Type III Kodaira singularity to $\text{I}_2$. 

The final base consistent with a Type III fiber singularity is of Type IV$^*$.
After blowing up the base, we first end up with the seemingly consistent
configuration:
\begin{equation}\label{fake-III-E6}
		\begin{matrix}
		\overset{\mathfrak{su}_{2}}{2}\overset{\mathfrak{g}_{2}}{3}1\overset{\displaystyle \overset{\mathfrak{su}_2}{2}}{\overset{\displaystyle \overset{\mathfrak{g}_2}{3}}{\overset{\displaystyle 1}{\overset{\mathfrak{f}_4}{5}}}}1\overset{\mathfrak{g}_{2}}{3}\overset{\mathfrak{su}_{2}}{2}
		\end{matrix}\qquad?
\end{equation}
However, it was shown in \cite{Merkx:2017jey} that three $(-3)$-curves with
$\mathfrak{g}_2$ algebras cannot be linked to a $(-5)$-curve in a trivalent
pattern. We are therefore forced to perform additional blow-ups, which leads to the quiver
\begin{align} \mathcal{K}^{\text{II}}(\text{III}, \fe_6): \, \,
 \overset{\fsu_2}{2} \, \, \overset{\fso_7}{3}\,\, \overset{\fsu_2}{2} \, \, 1 \,\, 
 \overset{\displaystyle \textcolor{blue}{\overset{\fsu_2}{2}}}{\overset{\displaystyle  \overset{\fso_7}{3}}{ \overset{\displaystyle \overset{\fsu_2}{2}}{\overset{\displaystyle  1 }{  \overset{\fe_7}{8}}}}}
 \, \, 1 \, \, \overset{\fsu_2}{2} \, \, \overset{\fso_7}{3}\,\, \overset{\fsu_2}{2} 
  \, \, \, \text{ with } (\text{CB},\kappa_R)=(34,144) \, .
\end{align}

Finally, for Type IV fiber singularities, we can only have $\text{I}_N^*$
base singularities, which results in the quiver
\begin{align}\label{K-IV-SO2N}
 \mathcal{K}^{\text{II}}(\text{IV},\fso_{2N+8})\!:     \overset{\mathfrak{su}_{3}}{3}1\overset{\displaystyle \overset{\mathfrak{su}_3}{3}}{\overset{\displaystyle 1}{\overset{\mathfrak{e}_6}{6}}}1\overset{\mathfrak{su}_3}{3}1\overset{\mathfrak{e}_6}{6}\cdots 
			\overset{\mathfrak{e}_6}{6}1\overset{\mathfrak{su}_3}{3}1
			\overset{\displaystyle\textcolor{blue}{\overset{\mathfrak{su}_{3}}{3}}}{\overset{\displaystyle 1}{\overset{\mathfrak{e}_6}{6}}}1\overset{\mathfrak{su}_{3}}{3} \text{ with } (\text{CB},\kappa_R)=(12N+22,48N+48),
\end{align}
with $N+1$ corresponding to the number of $\fe_6$ gauge algebra factors. 

\subsection{Type II, III, IV Bases}

While the typical ADE bases are straightforward to discuss and to resolve, the
bases of Kodaira Type II, III and IV need special treatment since we need to
blow up double- or triple-point singularities.  The Tate model for Type II
singularities is given in Table~\ref{tab:IIDESing}, but we repeat it here for
convenience: 
\begin{align}
    p_B=Y^2 +  X^3 + c_1 t X Y Z + c_2 t X^2 Z^2 + c_3 t Y Z^3 + c_4 t X Z^4 + 
 c_6 t Z^6 \, .
\end{align}
At $t=0$, the above model has a double-point singularity at $X=Y=0$. In order
to distinguish the quiver from the regular the regular torus, we recall that we
write it as $0\cdot$. We can then perform a blow-up, which can be done locally
by replacing $Y \rightarrow X e_1$. The curve over $\{t=0\}$ is reducible and
splits into
\begin{align}           
    X^2(e_1^2 + X)=0
\end{align}
with a double intersection at $e_1=X=0$.  
In terms of curves, we write the above configuration over $\{\hat{t}=0\}$ as 
\begin{align}
    4||1 \qquad\text{ with }\qquad \vec{\ell}^\text{LST}=(1,2) \,,
\end{align}
where $||$ denotes the double intersection. Setting the stage for more general
cases, we can also perform another blow-up of the double intersection by taking
$X\rightarrow X e_2$ and $e_1 \rightarrow e_1 e_2$. Upon taking the proper
transform, we end up with the equation
\begin{align}
    X^2(e_1^2 e_2 + X) = 0 \,.
\end{align}
All three components meet at a single point, which we write as the quiver
\begin{align}
\label{eq:IIBL2}
    5  \, \overset{\displaystyle 1}{\Delta} \, 2 \, .
\end{align}
These blow-ups are necessary in all cases except for fiber singularities of type
$\text{I}_n$. We expect them to work out similar to the I$_n$ base case. An interesting case appears in $\mathcal{K}^{\text{II}}(
\fsu_n,\text{II})$ theories, i.e., for I$_n$ fibers over Type II bases: it was proposed in \cite{Bhardwaj:2015oru} that at the cusp, the adjoint is split into a symmetric and an antisymmetric representation of the $\fsu_n$ gauge algebra.
This matter breaks the 1-form symmetry group
$\mathbb{Z}_n$ to either $\mathbb{Z}_2$ if $n$ is even or to nothing if $n$ is odd.
This is, however, at odds with the $\mathbb{Z}_n$ defect group in the
T-dual theory $\mathcal{K}^{\text{II}}(\text{II}, \fsu_n)$. It would be
interesting to return to this puzzle in the future.

Next, we want to give a simple example of an I$_0^*$ fiber singularity for quivers with exotic bases. For simplicity, we only discuss the Weierstrass model here, which has vanishing orders $(f,g,\Delta)=(2,3,6)$ in the F-theory fiber over $\{t=0\}$.
Since the cusp in the base is a self-intersection of a curve, the singularity is enhanced to $(4,6,12)$, which requires one blow-up in the base, resulting in the quiver
\begin{align}
		\mathcal{K}^{\text{II}}(\fso_8, \text{II}):\quad \textcolor{blue}{\overset{\fso_8}{4}} ||1 \, \quad \text{ with } \quad (\text{CB},\kappa_R)=(5,8) \, .
\end{align}
Notably, this theory has no flavor symmetry, since the E-string sees two $\fso_8$ factors due to the double intersection. 

For an I$_N^*$ singularity, the F-theory Weierstrass model has vanishing orders $(2,3,N+6)$. Blowing up the self-intersection point and taking the proper transform, the singularity on the exceptional divisor becomes $(0,0,2N)$. The gauge algebra is fixed by the local monodromy, which is only affected by the I$_N^*$ singularity. In other cases, we have the monodromy cover equation 
\begin{align}
    \psi^2 + \frac{g}{f}=0
\end{align}
which does not split if $f$ and $g$ are (locally) quadratic and cubic monomials, respectively. However, due to the double intersections with the exceptional divisor, the order of $f$ and $g$ is (locally) twice as large and hence $g/f$ is a perfect square. The $\text{I}_{2n}$ singularity is therefore split, and the quiver is
\begin{align}
		\mathcal{K}^{\text{II}}(\fso_{2N+8},\text{II}):\,  \textcolor{blue}{\overset{\fso_{2N+8}}{4}} || \overset{\fsu_{2N}}{\underset{[\Lambda^2=1]}{1}}\, \quad \text{ with }\quad (\text{CB},\kappa_R)=(3N+4, 6N+6)\, .
\end{align} 
Note that there is an extra anti-symmetric representation of $\fsu_{2N}$ for $N>2$ leading to an $\fsu_2$ flavor symmetry consistent with the worldsheet unitarity bound.

For a IV$^*$ singularity, we have to perform additional blow-ups: the first blow-up yields a Type IV fiber over the exceptional divisor, resulting in yet another $(4,6,12)$ singularity at the double intersection with the $\fe_6$. This requires two blow-ups to be fully removed, resulting in the quiver
\begin{align}
	\mathcal{K}^{\text{II}}(\fe_6,\text{II}):\,\, \,   \overset{\fe_6}{4}||\overset{\fsu_3}{1} \rightarrow   
	\textcolor{blue}{\overset{\fe_6}{6}}  \, \, \overset{\displaystyle [\mathfrak{su}_2]}{\overset{\displaystyle 2}{1}} \, \, \overset{\fsu_3}{3}  
    \quad\text{ with }\quad   (\text{CB},\kappa_R)=(11, 27)\, .
\end{align}
Note that the empty $(-2)$-curve is endowed with an enhanced $\fsu_2$ flavor symmetry, as that curve is part of a rank two E-string. As with the T-dual theory, this flavor has level 3 and is consistent with unitarity.

For singularities of Type II$^*$, we also find a non-minimal singularity at the self-intersection point. Upon blow-up, this yields a Type $\text{I}_0^*$ singularity where the double intersection removes the monodromy and hence results in an $\fso_8$ gauge algebra, as in the $\mathcal{K}^{\text{II}}(\fso_{2N+8},\text{II})$ case. Performing the next blow-up of the double intersection results in the quiver~\eqref{eq:IIBL2}, with a Type III fiber on the $(-1)$-curve. At the triple intersection point, we obtain another $\text{I}_0^*$ singularity, which is now semi-split, i.e., an $\fso_7$. Performing the residual blow-ups then results in the quiver
\begin{align}
	\mathcal{K}^{\text{II}}(\fe_7,\text{II}): \, \, \, \overset{\fe_7}{4}||\overset{\fso_8}{1}\rightarrow 
	\textcolor{blue}{{\overset{\mathfrak{e}_7}{8}}}1\overset{\mathfrak{su}_2}{2}\overset{\displaystyle\overset{\mathfrak{su}_2}{2}}{\overset{\mathfrak{so}_7}{3}}1\overset{\mathfrak{so}_8}{4}
	\quad\text{ with }\quad (\text{CB},\kappa_R)=(22, 96) \, ,
\end{align}
which is T-dual to the $\mathcal{K}^{\text{II}}(\text{II}, \fe_7)$ theory discussed in the previous section.

Finally, one might consider the case of an II$^*$, that is an $\fe_8$ singularity, which after the first blow-up gives the quiver $\overset{\fe_8}{4}|| \overset{\fe_6}{1}$. However, upon further blowing up this quiver, we end up with a $(-13)$-curve, and hence an inconsistent geometry. This was  to be expected, since we also encountered a non-crepant singularity when interchanging fiber and base in the putative $\mathcal{K}^{\text{II}}(\text{II}, \fe_8)$ theory discussed in the previous section. 
All other types of quivers such as $\mathcal{K}^{\text{II}}(\text{III},\text{II})$ and $\mathcal{K}^{\text{II}}(\text{IV},\text{II})$ have an $\fsu_2$ and $\fsu_3$ gauge algebra with a symmetric representation.

Moving on to the Type III base, we are required to perform a first resolution over $t=0$ and replace it by $\hat{t}=f_0 f_1$. The resulting Tate Model is
\begin{align}
    p_B=Y^2 + f_1 X^3   + c_1 f_0 f_1 X Y Z + c_2 f_0^2 f_1^2 X^2 Z^2 + 
 c_3 f_0^3 f_1^2 Y Z^3 + c_4 f_0 X Z^4 + c_6 f_0^3 f_1 Z^6 \, ,
\end{align}
with the two resolution divisors $f_0,f_1$ having components
\begin{align}
\begin{split}
    p_B& = f_0=0 : \qquad   Y^2 + f_1 X^3\, .\\ 
   p_B & = f_1=0 : \qquad Y^2 + c_4 f_0 X Z^4 \, .
\end{split} 
    \end{align}
Both components have a double intersection at  $Y=0$ which we write as 
\begin{align}
    2 || 2 \, .
\end{align}
In the following we want to resolve the tangential intersection at $f_0=f_1=Y=0$ by taking
\begin{align}
    \left\{  f_0 \rightarrow f_0 e_1, \quad f_1 \rightarrow f_1 e_1 \, ,\quad Y\rightarrow Y e_1   \right\}\, .
\end{align}
Upon resolving $p_B$ and taking the proper transform, we obtain
\begin{align}
    \begin{split}
    p_B&=f_1 X^3 + e_1 Y^2 + c_1 e_1^2  f_0 f_1 X Y Z + 
 c_2 e_1^3  f_0^2 f_1^2 X^2 Z^2 + c_3 e_1^5   f_0^3 f_1^2 Y Z^3 + \\
 &\phantom{=\;}~c_4 f_0 X Z^4 + c_6 e_1^3 f_0^3 f_1 Z^6 \, .
 \end{split}
\end{align} 
The divisors $D_{f_0}$, $D_{f_1}$ and $D_{e_1}$ all intersect in a single point. Hence, we write the resulting base as
\begin{align}
\label{eq:IIIBL1}
    3 \, \overset{\displaystyle 1}{\Delta} \,  3 \, .
\end{align}
We may perform a second resolution at $f_0= f_1=e_1=0$ and take the proper transform, which yields
\begin{align} 
    p_B=&f_1 X^3 + e_1 Y^2 + c_1 e_1^2 e_2^3 f_0 f_1 X Y Z + 
 c_2 e_1^3 e_2^6 f_0^2 f_1^2 X^2 Z^2 \nonumber \\ &+  c_3 e_1^5 e_2^9 f_0^3 f_1^2 Y Z^3 + 
 c_4 f_0 X Z^4 + c_6 e_1^3 e_2^6 f_0^3 f_1 Z^6 \, .
\end{align} 
The resulting quiver is
\begin{align}
    4 \, \overset{\displaystyle 2}{1} \,  4 \, .
\end{align}
It can be checked that the above quivers are LSTs with base singularities of Type III, which have a defect group $\mathbbm{Z}_2$.

We start by considering an I$^*_k$ fiber singularity, which yields the following  quivers
\begin{align}
  \mathcal{K}^{\text{II}}(\fso_{2k+8},\text{III}):  \overset{\fso_{2k+8}}{2} || \overset{\fso_{2k+8}}{2} \rightarrow 
  \textcolor{blue}{\overset{\fso_{2k+8}}{4}} \overset{ \displaystyle \overset{\fsu_{2k}}{2}}{ \overset{\fsp_{2k}}{1}} \overset{\fso_{2k+8}}{4} \text{ with } (\text{CB},\kappa_R)=(6k+10,16k+16)\,,
\end{align}
whose Coulomb branch dimension and 2-group structure data matches those of equation~\eqref{eq:IIISo2n}. Note that the $N=0$ case is again special,
\begin{align}
		\mathcal{K}^{\text{II}}(\fso_8,\text{III}): \textcolor{blue}{\overset{\fso_{8}}{4}}  \, \overset{\displaystyle \overset{[\fsu_2]}{2}}{ 1} \, \overset{\fso_{8}}{4} \text{ with } (\text{CB},\kappa_R)=(11,18) \, .
\end{align}
The data as well as the $\fsu_2$ flavor symmetry on the $(-2)$-curve matches
that of the $\mathcal{K}^{\text{II}}(\text{III},\fso_8)$ theory. Note that the
empty $(-2)$-curve has LST charge 2 and hence the $\fsu_2$ worldsheet current
contribution is given by $c_{2,\fsu_2}=3/2$, which is within the unitarity
bounds.

Moving on to a IV$^*$ singularity, we find a vanishing order $(f,g,\Delta)=(6,8,16)$ for the associated Weierstrass model. After the first blow-up, this gives a Type IV fiber over the exceptional divisor. At the triple intersection point, there is a singularity with vanishing order $(8,10,20)$, see equation \eqref{eq:IIIBL1}, which results after another blow-up and proper transform in a $(4,4,8)$ singularity, i.e another $\fe_6$. Performing all other residual blow-ups, we find the base quiver as
\begin{align}
\mathcal{K}^{\text{II}}(\fe_6, \text{III}): \overset{\fe_6}{2} || \overset{\fe_6}{2} \rightarrow \quad  
\overset{\fe_6
}{6} \, \overset{}{1} \, \overset{\fsu_3}{3} \, \overset{}{1} \,  \overset{
\overset{
\overset{\fsu_3}{\displaystyle 3}
}{\displaystyle 1}
}{\overset{\fe_6}{6}} \, \overset{}{1} \, \overset{\fsu_3}{3} \, \overset{}{1} \, \textcolor{blue}{\overset{\fe_6}{6}}  \, \qquad (\text{CB}, \kappa_R)=(34,144) \, .
\end{align} 
One may proceed in a similar fashion for type III$^*$ and II$^*$ singularities. However, in both cases we find non-crepantly resolvable singularities at codimension two with vanishing orders $(f,g,\Delta)\geq (8,12,24)$ after the first blow-up. This is consistent with the non-crepant singularities we encountered when exchanging the fiber and base. 
 
Finally, there are two exotic cases, given by (III,III) and (IV,III), neither of which require additional blow-ups of the base. The first case is self-T-dual and is given by
\begin{align}
		(\text{III},\text{III}): \overset{\fsu_2}{2}|| \textcolor{blue}{\overset{\fsu_2}{2}}  \, ,\qquad (\text{CB},\kappa_R)=(3,4) \, .
\end{align}
with two bifundamental matters, respecting the $\mathbb{Z}_2$ 1-form symmetry. 
The second case is
\begin{align}
\label{eq:IV-III}
\mathcal{K}^{\text{II}}(\text{IV},\text{III}): \overset{\fsu_3}{2}|| \textcolor{blue}{\overset{\fsu_3}{2}}  \, ,\qquad (\text{CB},\kappa_R)=(5,6) \, .
\end{align}
which also admits $(\mathbf{3},\overline{\mathbf{3}})$ bifundamental hypermultiplets respecting the diagonal $\mathbb{Z}_3$ 1-form symmetry. The fact that we have two bifundamental hypermultiplets suggests that we an $\fsu_2$ flavor symmetry. This is consistent with the expected $\fsu_2$ flavor symmetry in the  T-dual $\mathcal{K}^{\text{II}}(\text{III},\text{IV})$ theory.

Finally, we discuss the structure of a IV base. The Tate vector is given in Table~\ref{tab:IIDESing} and the model requires two more resolution divisors, which replace $\hat{t}=f_0 f_1 f_2$
and result in the Tate-Model
\begin{align}
p_b=  f_ 2 X^3 + f_ 1 Y^2 + c_1 f_ 0 f_ 1 f_ 2 X Y Z + 
c_2  f_ 0^2 f_ 1 f_ 2^2 X^2 Z^2 + c_3 f_ 0 Y Z^3 + c_4f_ 0^2 f_ 2 X Z^4 + 
c_6 f_ 0^3 f_ 2 Z^6 \, .
\end{align}
The three curves $\{f_i=0\}$ all meet in a single point, which we write as the quiver 
\begin{align}
    2 \overset{\displaystyle 2}{\Delta} 2 \,.
\end{align}
In the following we want to blow up this point by letting $e_i \rightarrow e_i s $ with $s$ the resolution divisor. After taking the proper transform, the hypersurface equation becomes
\begin{align}
    p=f_ 2 X^3 + f_ 1 Y^2 + f_ 0 f_ 1 f_ 2 e_1^2 X Y Z + 
 f_0^2 f_1 f_2^2 e_1^4 X^2 Z^2 + f_0 Y Z^3 + f_0^2 f_2 e_1^2 X Z^4 +
  f_0^3 f_2 e_1^3 Z^6 \, .
\end{align}
The three curves are given by 
\begin{align}
\begin{split}
    f_0=0:&\qquad f_2 X^3 + f_1 Y^2\\
    f_1=0:&\qquad f_2 X^3 + f_0 Y Z^3 + f_0^2 f_2 e_1^2 X   Z^4 + f_0^3 f_2 e_1^3 Z^6 \\ 
    e_2=0:&\qquad f_1  + f_0 Z^3\\
    e_1=0:&\qquad f_2 X^3 + f_1 Y^2 + f_0 Y Z^3 \, .
\end{split}
\end{align}
All curves intersect the exceptional divisor, but not each other. The resulting quiver is given by 
\begin{align}
    3 \overset{ \displaystyle 3 }{1} 3 \, .
\end{align}

Having discussed the minimal bases and their blow-ups, we engineer a I$^*_N$ singularity in the F-theory fiber $p_F$. For simplicity, we take the minimal Weierstrass model with coefficients
\begin{align}
    f=\hat{t}^2 \hat{f} \, \quad g=\hat{t}^3 \hat{g}\, \quad \Delta=\hat{t}^{N+6}  \hat{\Delta}
\end{align}
Since $\hat{t}=f_0 f_1 f_2$, and all three $f_i$ collide in a single point, the singularity is enhanced to vanishing order $(6,9,3k+18)$. Blowing up the singular point in the base and taking the proper transform lowers the vanishing order over the exceptional locus to 
$(2,3,3k+6)$, which corresponds to an $\fso_{6k+8}$ gauge algebra. Performing three more resolutions at the three collisions, we obtain the quiver
\begin{align}
\mathcal{K}^{\text{II}}(\fso_{2k+8},\text{IV}): \, \,  
\overset{\mathfrak{so}_{2k+8}}{2}  \overset{\displaystyle \overset{\mathfrak{so}_{2k+8}}{2}}{\Delta}  \overset{\mathfrak{so}_{2k+8}}{2} \rightarrow \overset{\fso_{2k+8}}{4} \, \overset{\fsp_{2k}}{1} \, \overset{\displaystyle \textcolor{blue}{\overset{\mathfrak{so}_{2k+8}}{4}}}{\overset{\displaystyle\overset{\mathfrak{sp}_{2k}}{1}}{\overset{\fso_{6k+8}}{4}}} \, \, \overset{\fsp_{2k}}{1}\,\overset{\fso_{2k+8}}{4} 
	  \quad 
\end{align} 
with the Coulomb branch dimension and structure constant given by
\begin{align}
	\text{dim}(\text{CB}) = 12k+22\,,\qquad
	\kappa_R = 48k+48 \,.
\end{align}
The above data matches that of the T-dual $\mathcal{K}^{\text{II}}(\text{IV},
\fso_{2N+8})$ theory shown in equation \eqref{K-IV-SO2N}. From this perspective, it is also clear that we can have
no exceptional gauge algebra for this base. For example, with an $\mathfrak{e}_6$ algebra we have a
vanishing order enhancement $(3,4,8) \rightarrow (9,12,24)$ in the fiber, which does not
admit a crepant resolution. This is consistent with the fact that we could not
construct a crepant $\mathcal{K}^{\text{II}}(\text{IV},\fe_n)$ model. 

Two more exotic cases come from the theories
$\mathcal{K}^{\text{II}}(\text{III},\text{IV})$ and
$\mathcal{K}^{\text{II}}(\text{IV},\text{IV})$. The quiver of the former is
given by
\begin{align}
\mathcal{K}^{\text{II}}(\text{III},\text{IV}): \,    \overset{\mathfrak{su}_{2}}{2} \overset{\displaystyle \textcolor{blue}{\overset{\mathfrak{su}_{2}}{2}}}{\Delta}  \overset{\mathfrak{su}_{2}}{2} \, \qquad\text{ with }\qquad (\text{CB},\kappa_R)=(5,6) \, .
\end{align} 
Due to the triple intersection, we expect that the matter content of the theory are two half-hypermultiplets in the
tri-fundamental representation $(\mathbf{2,2,2})$. We therefore expect an $\fsu_2$ flavor symmetry, which is compatible with the T-dual theory. 
The above model is T-dual to the $\mathcal{K}^{\text{II}}(\text{IV},\text{III})$
theory shown in equation \eqref{eq:IV-III}.

Finally, The
$\mathcal{K}^{\text{II}}(\text{IV},\text{IV})$ theory has a $(6,6,12)$
singularity at the triple intersections, which requires a single blow-up. The
resulting self-T-dual theory is
\begin{align}
\mathcal{K}^{\text{II}}(\text{IV},\text{IV}):  \overset{\mathfrak{su}_{3}}{3} \overset{\displaystyle \textcolor{blue}{\overset{\mathfrak{su}_{3}}{3}}}{1}  \overset{\mathfrak{su}_{3}}{3} \, \qquad\text{ with }\qquad (\text{CB},\kappa_R)=(9,12) \, .
\end{align}
A remark on the flavor symmetry of the above theory is in order: due to the
undecorated $(-1)$-curve, we expect to find an additional $\fsu_3$ flavor
factor attached to that curve due to the maximal breaking $\mathfrak{e}_8
\rightarrow \fsu_3^4$. As the LST charge of the curve is $\ell^\text{LST}_I=3$,
the contribution of the induced worldsheet current to the central charge is
\begin{align}
    c_{F}=\frac{3 \cdot 8}{3+3}=4 \, .
\end{align}
This does not satisfy the unitarity bound $c_F\leq 2$. A possible explanation
could be that the flavor symmetry is in fact broken to a smaller sub-algebra.
Furthermore, the associated SCFT has an $\mathfrak{su}_3^2$ flavor symmetry. To
perform minimal affinization, we need to gauge a subalgebra of the full flavor
symmetry, but since the two simple factors are the same, it is not clear
\emph{which} subalgebra must be gauged. At the level of the anomaly polynomial,
adding a vector multiplet transforming in the adjoint representation of either
$\mathfrak{su}_3$ factors leads to consistent results. But so does the choice of having the
vector multiplet transform in a combination of the two. Since the worldsheet bounds seem to indicate a smaller flavor symmetry, it is tempting to conjecture that the correct gauging
is a subalgebra of $\mathfrak{su}_3^2$ with an $\mathfrak{su}_2$ commutant, so
that the $\mathcal{K}^\text{II}(\text{IV}, \text{IV})$ LST is a consistent
theory with an $\mathfrak{su}_2$ flavor symmetry.

\subsection{Higgs Branches}

Although most Type II LSTs do not have a flavor algebra, some do admit
non-trivial Higgs branches. In particular, the cases involving exotic Kodaira
Type II, III and IV singularities discussed in the previous sections have
flavor symmetry factors. From a geometric perspective, these singularities arise from deformations of
classical $\text{I}_n$ Kodaira fibers,
\begin{align}
\text{I}_1 \rightarrow \text{II}\,, \qquad 
\text{I}_2 \rightarrow \text{III}\,, \qquad
\text{I}_3 \rightarrow \text{IV}\,, \qquad 
\end{align}
where the complex structure modulus $\tau$ of the F-theory torus is tuned to a
special point. Having discussed the LSTs of type
$\mathcal{K}^{\text{II}}(\fg,\fsu_N)$, as well as their duals and enhancements,
we now turn to their (partial) Higgs branches in cases where a flavor symmetry
exists.

For instance, the theory
$\mathcal{K}^{\text{II}}(\text{II},\mathfrak{so}_{2k+10})$, whose quiver is
given in equation \eqref{eq:IIDk4}, has a $U(1)$ flavor symmetry
\cite{Apruzzi:2020eqi,Ahmed:2023lhj} under which the $(\mathbf{3},\mathbf{3})$
hypermultiplets are non-trivially charged. The Higgsing breaks the $\fsu_3$
factors to $\fsu_2$, resulting in the LST $\mathcal{K}^{\text{II}}(\text{I}_1,
\mathfrak{so}_{2k+10})$. Its T-dual, on the other hand, is more
interesting: here we have a $U(1)$ flavor symmetry under which the
bifundamentals and anti-symmetric representations are charged. ``Higgsing'' the
singularity of the base from $\text{II}\rightarrow \text{I}_1$ gives an expectation value (vev) to these fields, splitting the double intersection:
\begin{align}
    \overset{\fso_{2k+10}}{4} || \overset{\fsu_{2n+2}}{\underset{[\Lambda^2=1]}{1}} \qquad\overset{\text{Higgs}}{\rightsquigarrow}\qquad // \overset{\fso_{2k+10}}{4} \overset{\fsp_{n+1}}{1}// \, .
\end{align}
The $\mathcal{K}^{\text{II}}(\text{II}, E_6)$ theory and its duals work similarly: 
\begin{align}
    2 \, \overset{\fsu_2}{2} \, 
     \overset{ \displaystyle \overset{\displaystyle 2}{\overset{\fsu_2}{2}} } {\underset{[\fsu_2]}{\overset{\fg_2}{2}}} \overset{\fsu_2}{2}\, 2 \qquad \overset{\text{Higgs}}{\rightsquigarrow} \qquad 2 \, \overset{\fsu_2}{2} \, 
     \overset{ \displaystyle \overset{\displaystyle 2}{\overset{\fsu_2}{2}} }   {\overset{\fsu_3}{2}} \, \overset{\fsu_2}{2}\, 2 
\end{align}

The theory has an $\fsp_1\simeq \mathfrak{su}_2$ flavor symmetry under which the $\fg_2$ fundamental
representations are charged. Giving them a vev breaks $\fg_2\rightarrow
\fsu_3$, preserving the Coulomb branch dimension of the LST while reducing the
2-group structure constant $\kappa_R$. The same transition is again more exotic
in the T-dual. There we have an $\fsu_2$ flavor symmetry on the unpaired
$(-2)$-curve which, under the same transition as above, gives rise to the quiver 
\begin{align}
    \overset{\fe_6}{6} \, \overset{\displaystyle \overset{[\fsu_2]}{2} }{1} \, \overset{\fsu_3}{3} \qquad\overset{\text{Higgs}}{\rightsquigarrow}\qquad // \overset{\fe_6}{6} \,  1 \, \overset{\fsu_3}{3} \, 1 // \, .
\end{align} 

We close by recalling an important point concerning the Higgs branches of LSTs.
An LST is also defined by a point in its tensor-branch moduli space. In
particular, a tensor-branch quiver of an LST comes with the choice of a
``contraction map'' \cite{Aspinwall:1997ye} determining a maximal set of curves
that can be collapsed. While this choice is unique for SCFTs, it is not for LSTs due to the
presence of the curve of self-intersection zero, and there are $n_T$ such choices. Each choice leads to a different theory although the
tensor-branch geometry is the same. Furthermore, the invariants we have
discussed in this work are the same, as they are computed on the tensor branch,
but the structure of their Higgs branch can change from one choice to the
other.  Being associated with the volumes of the curves, different choices for
the curve that remains of finite size at the contracted point will have an
impact on the Higgs branch of the theory. This feature was explored in
\cite{Lawrie:2023uiu} for a variety of Heterotic LSTs. From the fusion
perspective, this correspond to the different ways of fusing SCFTs together to
obtain an LST, and minimal affinization defines a particular choice of such a
contraction map.

The structure of Higgs branches can be probed using the technology of
magnetic quivers. While the precise map between 3D magnetic quivers and 6D
generalized quivers is not fully understood, in particular in the presence of
exceptional algebras, they offer a very potent apparatus to extract information
about the Higgs branch of 6D $\mathcal{N}=(1,0)$ theories. In the context of
Heterotic LSTs, they have been used recently in \cite{DelZotto:2023nrb,
Lawrie:2023uiu, Mansi:2023faa, Bourget:2024mgn}, and for Type II LSTs
in the upcoming work \cite{LawrieII}.

\subsection{Outlier Theories}

So far, we have geometrically engineered LSTs  from a given base and fiber algebra by using the double fibration structure described in
Section ~\ref{sec:LST-double-fibration}. There are a few quivers that look consistent, but have fibers that cannot be reproduced
geometrically. In addition, there are a few more quivers that cannot be thought
of as deformations of other LSTs, some of which have already appeared in the
literature~\cite{Bhardwaj:2015xxa, Bhardwaj:2015oru}. These outliers all elude
the geometric construction and do not admit a candidate T-dual theory with the
same two-group invariants. They satisfy all known field-theoretic
constraints, such as anomaly cancellation and the worldsheet bound given in
equation~\eqref{flavor-bound-WZW}. 

For instance, consider the following quiver, with $(\mathfrak{g}_F,
\mathfrak{g}_B) =
(\mathfrak{so}_{2N+8}, \text{III})$:
\begin{align}\label{D-outlier-I}
  \overset{\fso_{2N+8}}{2} || \overset{\fso_{2N+8}}{2} \rightarrow 
	 \overset{\fso_{2N+8}}{4}  \, \, \overset{\displaystyle \textcolor{blue}{\overset{\fso_{2N+8}}{4}}}{\overset{\fsp_{2N}}{1}}
	 \overset{\fso_{4N}}{4} \, \,\overset{\fsp_{2N-8}}{1} \, \,  \overset{\fso_{4N-16}}{4} \ldots \overset{\fsp_{2N-8k}}{1} \, \, \overset{\displaystyle\overset{\fsp_{N-4k-4}}{1}}{\overset{\fso_{4N-16k}}{4}}
 \overset{\fsp_{N-4k-4}}{1} \, ,
\end{align} 

From the minimal affinization point of view, this quiver can be reached from an
$\mathfrak{so}_{2N+8}$ orbi-instanton with an
$\mathfrak{so}_{2N+8}\oplus\mathfrak{e}_8$ symmetry, albeit through a few
modifications. The $\mathfrak{so}_{2N+8}$ flavor can first be gauged, leading
to another SCFT, the so-called $\frac{1}{2}$-fractional D-type orbi-instanton
\cite{Tachikawa:2015wka, Mekareeya:2016yal, Baume:2023onr}. The
$\mathfrak{e}_8$ symmetry is then broken by moving to the Higgs branch. The
effect of the breaking propagates through the spine to the other side of the
quiver, and the resulting remnant flavor symmetry can then be gauged to obtain
the LST. Both the rank of $\mathfrak{so}_{2N+8}$ and the breaking pattern set the length of the quiver. We are not aware of a T-dual theory for this LST, but it exhibits peculiar
features. For instance, from the M-theory perspective, one would expect it to be constructed
via M5-branes probing a $\mathbb{C}^2/D_{N+4}$ singularity, with a single
M9-brane giving rise  the $\mathfrak{e}_8$ symmetry. However, the partial tensor
branch description of this quiver is that of a Type II LST with $\kappa_P=0$, where no M9-brane is expected.

Furthermore, while the quiver above has a Type III base, there are two further
outliers theories with D-type fibers:
\begin{equation}\label{D-outlier-II}
		\overset{\mathfrak{so}_{2N+8}}{0\cdot}\quad\longrightarrow\quad \textcolor{blue}{\overset{\fso_{2N+8}}{4}} || \overset{\fsp_N}{1}\,,\qquad
		\overset{\mathfrak{so}_{4(p+4)}}{2} \overset{\displaystyle{\overset{\mathfrak{so}_{4(p+4)}}{2}}}{\Delta}  \overset{\mathfrak{so}_{4(p+4)}}{2}\quad\longrightarrow\quad
		\overset{\mathfrak{sp}_{p}}{1} \overset{\fso_{4(p+4)}}{4} \overset{\displaystyle\textcolor{blue}{\overset{\mathfrak{sp}_{p}}{1}}}{\overset{\displaystyle\overset{\mathfrak{so}_{4(p+4)}}{4}}{\overset{\fsp_{3p+8}}{1}}} \, \, \overset{\fso_{4(p+4)}}{4}\,\overset{\fsp_{p}}{1}\,.
\end{equation}
We have not found T-duals for these LSTs. It is, however, quite intriguing that these three quivers all have D-type fibers with a base of Kodaira Type II,
III, or IV. While we constructed the T-dual pairs for these singularity types, the quivers shown in equations \eqref{D-outlier-I} and \eqref{D-outlier-II} are qualitatively different: their duality invariants do not match any of the
$\mathcal{K}^\text{II}(D_N,\mathfrak{g}_B)$ LSTs.

Similarly, quivers with a Type II base also fall in this category, e.g. 
\begin{equation}\label{outlier-III}
		[\mathfrak{su}_2]\textcolor{blue}{1\overset{\mathfrak{e}_8}{8}}1\overset{\displaystyle\overset{\mathfrak{g}_2}{3}}{\overset{\mathfrak{su}_2}{2}}2 \qquad\overset{\text{Higgs}}{\rightsquigarrow}\qquad  \textcolor{blue}{\overset{\mathfrak{e}_8}{7}}1\overset{\displaystyle\overset{\mathfrak{g}_2}{3}}{\overset{\mathfrak{su}_2}{2}}2\,.
\end{equation}
These two theories seem to be on the same Higgs branch, as the second quiver is
obtained by shrinking the left-most $(-1)$-curve. One would expect such a
transition to be described by the closure of the minimal nilpotent orbit of the
$\mathfrak{su}_2$ flavor symmetry. Interestingly, only the second quiver can be
reached through minimal affinization. For the first one, sending any curve to
infinite volume will give rise to an SCFT with a different defect group.

Another LST for which we do not have a dual is a deformation of the
$\mathcal{K}^\text{II}(\mathfrak{e}_6, \text{III})$ theory that we have
realized geometrically in this section:
\begin{equation}
			\overset{\fe_6}{6} \, \overset{}{1} \, \overset{\fsu_3}{3} \, \overset{}{1} \,  \overset{\overset{\overset{\fsu_3}{\displaystyle 3}}{\displaystyle 1}}{\overset{\fe_6}{6}} \, \overset{}{1} \, \overset{\fsu_3}{3} \, \overset{}{1} \, \textcolor{blue}{\overset{\fe_6}{6}} \qquad\qquad\overset{?}{\rightsquigarrow}\qquad\quad
			\overset{\mathfrak{f}_{4}}{5} 1 \overset{\displaystyle\overset{\mathfrak{su}_{2}}{2}}{\overset{\mathfrak{g}_{2}}{3}} 1\textcolor{blue}{\overset{\mathfrak{f}_4}{5}}\,.
\end{equation}
While this configuration does not have apparent gauge anomalies there are no simple
deformation of the T-dual parent theory $\mathcal{K}^\text{II}(\mathfrak{e}_6,
\text{III})$ that can be obtained by further blowing up or down curves, and we
have not been able to engineer it geometrically. The quiver furthermore contains
exceptional algebras, and we are not aware of any magnetic quiver realization
that could shed light on the structure of the Higgs branch of these two theories. We therefore do not
know whether such a deformation is even allowed, or whether it is obstructed
either geometrically or in field theory.

We have collected the outliers presented here in Table \ref{tab:LST-outlier}
along with their T-dual invariant quantities for the reader's convenience. We
do not know whether these quivers are consistent. For instance, certain SCFTs
can be shown to be gauge anomalous due the presence of extra matter arising at
trivalent intersections~\cite{Morrison:2016djb, Merkx:2017jey}, and similar arguments might apply here. Another
possibility is that they are simply not captured by the construction above.
Indeed, we have assumed that the two hypersurface of the double fibrations can
be written in Tate form. It is then tempting to conjecture that, if the
outliers exist at all and are consistent quivers, their T-dual belong to the
frozen phase of F-theory or arise from twisted T-dualities, and are therefore
not part of the analysis performed in this work.

\section{Conclusion and Outlook}\label{sec:outlook}

Novel invariants, such as higher group symmetries, have initiated a recent
exploration of Heterotic LSTs via geometry and uncovered a very rich landscape
connected via T-dualities
\cite{DelZotto:2022ohj,DelZotto:2022xrh,DelZotto:2023ahf,Ahmed:2023lhj}. In
this work, we have considered another class of LSTs, so called Type II LSTs,
which are disconnected from their Heterotic cousins. From a higher-dimensional
perspective, the most direct consequence is the absence of flavor 9-branes that
intersect the spacetime boundary. While the absence of such branes severely
restricts possible flavor symmetries, it allows for non-trivial one-form symmetries and 
defect group structure, characterized by the centers of two algebras
$(\fg_F,\fg_B)$ that define the theory. These higher symmetries are exchanged
under T-duality, and thus give rise to novel invariants of Type II LSTs that do
not exist in Heterotic models.

We have furthermore studied the possible flavor symmetries of these theories
from the perspective of the fundamental 2D little string worldsheet theory
through anomaly inflow. The contributions of the induced worldsheet currents to
the central charge are strongly constrained by unitarity, which allowed us to
derive universal bounds on the flavor symmetry algebra. This results in a new
set of consistency conditions for 6D LSTs, analogous to those derived for SUGRA
models \cite{Kim:2019vuc,Lee:2019skh}, to which they apply as well. The
presence of these flavor symmetries, together with the duality constraints of
the higher form symmetries imply that Type II LSTs are much more constrained
than their Heterotic counterparts. 

To further discuss these conditions, we contrast bottom-up and top-down constructions. Field-theoretically, certain
seed SCFTs can be turned into Type II LSTs by gauging (part of) their flavor
symmetry while preserving the defect group, and without changing its matter
spectrum at a generic point of the tensor branch. We have referred to this
operation as \emph{minimal affinization}, and this enabled us to explain some
of the factorization of the 2-group invariants in terms of group-theoretical
quantities for a class LSTs. In a top-down approach, we have used F-theory to
engineer Type II LSTs systematically, where we have provided a simple
geometric framework using a non-compact toric complete intersection Calabi--Yau
threefold with a double elliptic fibration structure. An LST is then fixed by
specifying two types of Kodaira singularities for the two elliptic fibers, from
which the resulting 6D quiver can be easily read off. Our results are consistent with the field-theory considerations described above. By exploiting F/M-theory
duality, the double elliptic fibration structures also makes T-duality of
theories manifest. 

These two approaches therefore provide a framework to compare
geometrically-realizable LSTs with those that appear consistent in field
theory. We have also computed
the T-duality invariants for all the theories we have encountered, such as the
dimension of their Coulomb branch, higher group symmetry structure constants,
and higher symmetries. The are collated in Appendix \ref{app:tables} for
convenience.

Note that our exploration of Type II LSTs only considers the unfrozen phase of
F-theory, as well as untwisted T-dualities. The frozen phase has recently been
investigated for Heterotic LSTs from a top-down perspective in
\cite{Oehlmann:2024cyn,Morrison:2023hqx}. Similarly, twisted LSTs can be
engineered via genus-one fibrations \cite{Bhardwaj:2019fzv} that have recently
been investigated using toric geometry methods
\cite{Bhardwaj:2022ekc,Anderson:2023wkr, AhmedII}, for upcoming work on twisted
Heterotic LSTs. Both discussions can be extended to Type II LSTs, which we will
return to in the future.

Moreover, we have shown that there is a very small and exotic class Type II
LSTs with non-trivial flavor symmetries. More generally Type II LSTs are
endowed with  non-trivial Higgs branches, albeit very restricted due to the
constrains described above. Some of these theories admit a description in
terms of magnetic quivers~\cite{Mansi:2023faa, LawrieII}, and it would be interesting to study the Higgs branch of Type II LSTs in more
detail, both field-theoretically and via geometry.

Finally, the toolbox to compute all types of higher form symmetries directly in
M-theory have been fully developed in
\cite{Albertini:2020mdx,DelZotto:2022joo,Hubner:2022kxr,Cvetic:2021sxm,
Cvetic:2022imb}. It would be an important cross-check to compute all 1-form
symmetries directly in M-theory and match them to their 6D origins. This approach might also allows us to access all $p$-form symmetries
(which might also be present in Heterotic LSTs), and could result in yet
additional duality constraints.

 
\subsection*{Acknowledgments}
The authors thank Hamza Ahmed, Max H\"ubner, Craig Lawrie, Lorenzo Mansi and
Timo Weigand for helpful discussions. We are particularly grateful to Craig Lawrie for comments on an early version of the manuscript.
The work of FR is supported by the NSF
grants PHY-2210333 and PHY-2019786 (The NSF AI Institute for Artificial
Intelligence and Fundamental Interactions). The work of FR and PKO is also
supported by startup funding from Northeastern University.  The work of FB is
supported by the German Research Foundation through a German-Israeli Project
Cooperation (DIP) grant “Holography and the Swampland”, and in part by the
Deutsche Forschungsgemeinschaft under Germany’s Excellence Strategy EXC 2121
Quantum Universe 390833306.  PKO would like to thank the KITP and the program
"What is String Theory?  Weaving Perspectives Together" for support during the
completion of this work. This research was supported in part by grant NSF
PHY-2309135 to the Kavli Institute for Theoretical Physics (KITP).

\appendix
\section{Tables of Type II Little String Theories}\label{app:tables}

In this appendix, we collate an exhaustive list of the Type II LSTs discussed
in the main text, along with their 2-group structure constants $\kappa_P$ and $\kappa_R$, their
Coulomb branch dimension $\text{dim}(\text{CB})$, and higher-form data $\mathcal{D}^{(1)}\times \mathcal{D}^{(2)}$. We moreover group them in T-dual pairs. In Table
\ref{tab:LST-A-ADE}, we give the necklace quivers
$\mathcal{K}^\text{II}(\mathfrak{g}_F, \mathfrak{su}_N)$ and their duals; in
Table \ref{tab:LST-D-type} those involving D-type algebras; in
Table \ref{tab:LST-E-type} those of exceptional types---along with
Kodaira type II, III, and IV. In addition, Table \ref{tab:LST-outlier}
shows all outlier theories we could not construct, and who do not have
a known T-dual. Note that in all cases, the theories always satisfy the
worldsheet bound given in equation \eqref{flavor-bound-WZW}.

\begin{longtable}{ccccccc}
		$\mathcal{K}(\mathfrak{g}_F, \mathfrak{b}_B)=\widehat{\Lambda}^{\mathfrak{g}}$ & Quiver & $\mathcal{D}^{(1)}\times \mathcal{D}^{(2)}$ & $\kappa_P$ & $\kappa_R$ & \text{dim}(\text{CB})\\
            \toprule
			${\widehat{A}_{N}}^{\mathfrak{g}}$ & $\textcolor{blue}{//}\underbrace{\overset{\mathfrak{g}}{2}\overset{\mathfrak{g}}{2}\dots\overset{\mathfrak{g}}{2}}_{N-1}\textcolor{blue}{\overset{\mathfrak{g}}{2}//}$& $ Z(\mathfrak{g})^{(1)}\times\mathbbm{Z}_N^{(2)}$ & $0$ & $N\Gamma_{\mathfrak{g}}$ & $h^\vee_{\mathfrak{g}}N-1$\\
			${\widehat{D}_{N}}^{\,\mathfrak{su}_{2k}}([2^k])$ & $\overset{\mathfrak{su}_{k}}{2} \overset{\displaystyle \overset{\mathfrak{su}_{k}}{2}}{\overset{\mathfrak{su}_{2k}}{2}}\underbrace{\,\overset{\mathfrak{su}_{2k}}{2} \cdots\overset{\mathfrak{su}_{2k}}{2}}_{N-5}\overset{\displaystyle\textcolor{blue}{\overset{\mathfrak{su}_{k}}{2}}}{\overset{\mathfrak{su}_{2k}}{2}} \overset{\mathfrak{su}_{k}}{2}$ & $\mathbbm{Z}_k^{(1)}\times Z(D_N)^{(2)}$& $0$& $(4N-8)k$ &  $(2N-2)k-1$\\
			${\widehat{E}_{6}}^{\,\mathfrak{su}_k}$ & $\phantom{[\mathfrak{su}_k]}\overset{\mathfrak{su}_{k}}{2}\overset{\mathfrak{su}_{2k}}{2} \overset{\textcolor{blue}{\displaystyle\overset{\mathfrak{su}_{k}}{2}}}{\overset{\displaystyle \overset{\mathfrak{su}_{2k}}{2}}{\overset{\mathfrak{su}_{3k}}{2}}}\,\overset{\mathfrak{su}_{2k}}{2} \overset{\mathfrak{su}_{k}}{2}$ & $\mathbbm{Z}_k^{(1)}\times\mathbbm{Z}_3^{(2)}$ & $0$& $24k$ &  $12k-1$\\
			${\widehat{E}_{7}}^{\,\mathfrak{su}_k}$ & $\textcolor{blue}{\overset{\mathfrak{su}_{k}}{2}}\overset{\mathfrak{su}_{2k}}{2}\overset{\mathfrak{su}_{3k}}{2} \overset{\displaystyle \overset{\mathfrak{su}_{2k}}{2}}{\overset{\mathfrak{su}_{4k}}{2}}\,\overset{\mathfrak{su}_{3k}}{2}\overset{\mathfrak{su}_{2k}}{2} \overset{\mathfrak{su}_{k}}{2}$ 
			& $\mathbbm{Z}_k^{(1)}\times\mathbbm{Z}_2^{(2)}$ & $0$& $48k$ &  $18k-1$\\
			${\widehat{E}_{8}}^{\,\mathfrak{su}_k}$ & $\overset{\mathfrak{su}_{2k}}{2}\overset{\mathfrak{su}_{4k}}{2} \overset{\displaystyle \overset{\mathfrak{su}_{3k}}{2}}{\overset{\mathfrak{su}_{6k}}{2}}\,\overset{\mathfrak{su}_{5k}}{2}\overset{\mathfrak{su}_{4k}}{2} \overset{\mathfrak{su}_{3k}}{2}\overset{\mathfrak{su}_{2k}}{2}\textcolor{blue}{\overset{\mathfrak{su}_{k}}{2}}$
			& $ \mathbbm{Z}_k^{(1)}$ & $0$& $120k$ &  $30k-1$\\
            \bottomrule
	\caption{
			Little String Theories associated with ADE orbifolds. The blue
			curve refers to the node obtained through minimal affinization of
			the corresponding SCFT. For the necklace LSTs
			$\mathcal{K}(\mathfrak{g}, \mathfrak{su}_{N}) =
			\widehat{A}_{N-1}^{\mathfrak{g}}$ obtained from conformal matter,
			only the partial tensor branch description is shown for brevity,
			the general quivers can be found in equation \eqref{def-necklace}.
			Note that $\kappa_R$ and the dimension of the Coulomb
			branch can be written in a closed form, see equation
			\eqref{def-necklace}.
	}
    \label{tab:LST-A-ADE}
\end{longtable}

\newpage
\begin{longtable}{ccccccc}
		 $(\mathfrak{g}_F, \mathfrak{g}_B)$ & Quiver & $\mathcal{D}^{(1)}\times \mathcal{D}^{(2)}$ & $\widehat{\kappa}_P$ & $\widehat{\kappa}_R$ & \text{dim}(\text{CB})\\
            \toprule
			$(\text{II}, \mathfrak{so}_{8})$ & $ 2\underset{\displaystyle2}{\overset{\displaystyle\textcolor{blue}{2}}{\overset{\mathfrak{su}_2}{2}}} 2$ &  $Z(D_4)^{(1)}$ & 0 & $8$ & $5$\\
			$(\mathfrak{so}_{8}, \text{II})$ & $\textcolor{blue}{\overset{\fso_8}{4}} ||1$ & $Z(D_4)^{(2)}$ & 0 & $8$ & $5$\\
			\midrule
			$(\text{II}, \mathfrak{so}_{2k})$ & $2 \overset{\displaystyle 2}{\overset{\fsu_2}{2}} \, \, \underbrace{\overset{\fsu_3}{2}\, \, \overset{\fsu_3}{2} \ldots \overset{\fsu_3}{2} }_{\times (k-5)}\, \,   \overset{\displaystyle\textcolor{blue}{2}}{\overset{\fsu_2}{2}} 2$ & $Z(D_k)^{(2)}$ & 0 &  $6(k-3)$ & $3k-8$\\
			$(\mathfrak{so}_{2k},\text{II})$ & $  \textcolor{blue}{\overset{\fso_{2k}}{4}}||\overset{\fsu_{2k-8}}{1}$ & $Z(D_k)^{(1)}$ & 0 & $6(k-3)$ & $3k-8$\\
			\midrule
			$(\text{III}, \mathfrak{so}_{8})$ & $\begin{array}{c}     \textcolor{blue}{\overset{\fsu_2}{2}} \\    \overset{\fsu_2}{2} \, \,     \overset{\fso_7}{2} \, \, \overset{\fsu_2}{2} \\     \overset{\fsu_2}{2}\end{array}$ &  $\mathbb{Z}_2^{(1)}\times Z(D_4)^{(2)}$ & 0 & $18$ & $11$\\
			$(\mathfrak{so}_{8}, \text{III})$ & $2  \, \overset{\displaystyle \textcolor{blue}{\overset{\fso_{8}}{4}}}{ 1} \, \overset{\fso_{8}}{4}$ & $Z(D_4)^{(1)}\times\mathbb{Z}_2^{(2)}$ & 0 & $18$ & $11$\\
			\midrule
			$(\text{III}, \mathfrak{so}_{2k})$ &  $\overset{\fsu_2}{2} \overset{ \displaystyle\overset{\fsu_2}{2}}{\overset{\fso_7}{3}} \, \underbrace{\, 1 \, \, \overset{\fso_8}{4}\, \,1 \, \, \ldots 1 \, \, \overset{\fso_8}{4} \, \, 1  \overset{ \displaystyle\textcolor{blue}{\overset{\fsu_2}{2}}}{\overset{\fso_7}{3}} \,}_{(k-5)\,\times\,\overset{\mathfrak{so}_8}{4}} \, \overset{\fsu_2}{2}$ & $\mathbb{Z}_2^{(1)}\times Z(D_k)^{(2)}$ & $0$ & $16(k-3)$ & $6k-14$\\
			$(\mathfrak{so}_{2k},\text{III})$ & $\overset{\fsu_{2k-8}}{2}\, \overset{\displaystyle\textcolor{blue}{\overset{\fso_{2k}}{4}} }{ \overset{\fsp_{2k-8}}{1}}\, \overset{\fso_{2k}}{4}$ & $Z(D_k)^{(1)}\times\mathbb{Z}_2^{(2)}$ & 0 & $16(k-3)$ & $6k-14$\\
			\midrule
			$(\text{IV}, \mathfrak{so}_{2k})$ & $\overset{\mathfrak{su}_{3}}{3}1\overset{\displaystyle \overset{\mathfrak{su}_3}{3}}{\overset{\displaystyle 1}{\overset{\mathfrak{e}_6}{6}}}\underbrace{1\overset{\mathfrak{su}_3}{3}1\overset{\mathfrak{e}_6}{6}\cdots \overset{\mathfrak{e}_6}{6}1\overset{\mathfrak{su}_3}{3}1 }_{(k-5)\,\times\,\overset{\mathfrak{e}_6}{6}}\overset{\displaystyle\textcolor{blue}{\overset{\mathfrak{su}_{3}}{3}}}{\overset{\displaystyle 1}{\overset{\mathfrak{e}_6}{6}}}1\overset{\mathfrak{su}_{3}}{3}$ & $\mathbb{Z}_3^{(1)}\times Z(D_k)^{(2)}$ & 0 & $48(k-3)$ & $12k-26$\\
			$(\mathfrak{so}_{2k}, \text{IV})$ & $\overset{\fso_{2N+8}}{4} \overset{\fsp_{2N}}{1} \overset{\displaystyle \textcolor{blue}{\overset{\mathfrak{so}_{2N+8}}{4}}}{\overset{\displaystyle\overset{\mathfrak{sp}_{2N}}{1}}{\overset{\fso_{6N+8}}{4}}} \, \, \overset{\fsp_{2N}}{1}\,\overset{\fso_{2N+8}}{4}$ & $Z(D_k)^{(1)} \times \mathbb{Z}_3^{(1)}$ & 0 & $48(k-3)$ & $12k-26$\\

            \bottomrule
	\caption{
			Type II Little String Theories involving D-type algebras. The blue
			curve refers to the node obtain through minimal affinization of the
			corresponding SCFT.
	}
	\label{tab:LST-D-type}
\end{longtable}

\begin{longtable}{ccccccc}
		$(\mathfrak{g}_F, \mathfrak{g}_B)$ & Quiver & $\mathcal{D}^{(1)}\times\mathcal{D}^{(2)}$ & $\widehat{\kappa}_P$ & $\widehat{\kappa}_R$ & \text{dim}(\text{CB})\\
            \toprule
			$(\text{II},\text{II})$ & $\textcolor{blue}{0\cdot}$ & $\varnothing$ & 0 & $0$ & $0$\\
			\midrule
			$(\text{II},\text{III})$ & $\overset{}{2}|| \textcolor{blue}{2}$ & $\mathbb{Z}_2^{(2)}$ & 0 & $2$ & $1$\\ 
			$(\text{III},\text{II})$ & $\textcolor{blue}{\overset{\fsu_2}{0\cdot}}$ & $\mathbb{Z}_2^{(1)}$ & 0 & $2$ & $1$\\

			\midrule
			$(\text{II},\text{IV})$ & $ 2 \overset{ \displaystyle \textcolor{blue}{2}}{\Delta} 2 $ & $\mathbb{Z}_2^{(2)}$ & 0 & $3$ & $2$\\ 
	$(\text{IV},\text{II})$ & $\textcolor{blue}{\overset{\fsu_3}{0\cdot}}$ & $\mathbb{Z}_2^{(1)}$ & 0 & $3$ & $2$\\ 
			\midrule
			$(\mathfrak{e}_6,\text{II})$ & $[\mathfrak{su}_2]\,2\,\overset{\displaystyle \textcolor{blue}{\overset{\fe_6}{6}}}{1}\, \overset{\fsu_3}{3}$ & $\mathbb{Z}_3^{(1)}$ & 0 & $27$ & $11$\\
			$(\text{II}, \mathfrak{e}_6)$ & $\begin{array}{c}\textcolor{blue}{2} \\ \overset{\fsu_2}{2}  \\ 2 \overset{\fsu_2}{2}  \underset{\displaystyle[\mathfrak{su}_2]}{\overset{\fg_2}{2}} \overset{\fsu_2}{2} 2  \end{array}$ & $\mathbb{Z}_3^{(2)}$ & $0$ & $27$ & $11$ \\
			\midrule
			$(\text{II}, \mathfrak{e}_7)$ & $\begin{array}{c}\overset{\fsu_3}{3} \\ 1 \\ \textcolor{blue}{2} \overset{\fsu_2}{2} \, \, \overset{\fg_2}{3} \, \, 1 \, \, \overset{\ff_4}{5} \, \, 1 \,\, \overset{\fg_2}{3} \, \, \overset{\fsu_2}{2}  2  \end{array}$ & $\mathbb{Z}_2^{(1)}$ & 0 & 96 & 22 \\
			$(\mathfrak{e}_7,\text{II})$ & $\textcolor{blue}{\overset{\mathfrak{e}_7}{8}}1\overset{\mathfrak{su}_2}{2}\overset{\displaystyle\overset{\mathfrak{su}_2}{2}}{\overset{\mathfrak{so}_7}{3}}1\overset{\mathfrak{so}_8}{4}$ & $\mathbb{Z}_2^{(2)}$ & 0 & $96$ & $22$\\
			\midrule
			$(\text{III}, \mathfrak{e}_6)$ & $\begin{array}{c}     \textcolor{blue}{\overset{\fsu_2}{2}} \\    \overset{\fso_7}{3}\\    \overset{\fsu_2}{2} \\    1 \\  \overset{\fsu_2}{2}\, \, \overset{\fso_7}{3}  \, \, \overset{\fsu_2}{2}  \, \, 1 \, \,   \overset{\fe_7}{8}  \, \, 1 \, \, \overset{\fsu_2}{2} \, \, \overset{\fso_7}{3}\,\, \overset{\fsu_2}{2} \end{array}$ & $\mathbb{Z}_2^{(1)}\times\mathbb{Z}_3^{(2)}$ & 0 & 144 & 34\\
			$(\mathfrak{e}_6, \text{III})$ & $\overset{\fe_6}{6} \, \overset{}{1} \, \overset{\fsu_3}{3} \, \overset{}{1} \,  \overset{\overset{\overset{\fsu_3}{\displaystyle 3}}{\displaystyle 1}}{\overset{\fe_6}{6}} \, \overset{}{1} \, \overset{\fsu_3}{3} \, \overset{}{1} \, \textcolor{blue}{\overset{\fe_6}{6}}$ & $\mathbb{Z}_3^{(1)}\times\mathbb{Z}_2^{(2)}$ & 0 & $144$ & $34$\\
			\midrule
			$(\text{III},\text{III})$ & $\overset{\fsu_2}{2}|| \textcolor{blue}{\overset{\fsu_2}{2}}$ & $\mathbb{Z}_2^{(1)}\times\mathbb{Z}_3^{(2)}$ & 0 & $4$ & $3$\\
   \midrule
	 	   $(\text{IV},\text{III})$ & $\overset{\fsu_3}{2}|| \textcolor{blue}{\overset{\fsu_3}{2}}$ & $\mathbb{Z}_2^{(2)}\times\mathbb{Z}_3^{(1)}$ & 0 & $6$ & $5$\\
		   $(\text{III},\text{IV})$ & $\overset{\mathfrak{su}_{2}}{2} \overset{\displaystyle \textcolor{blue}{\overset{\mathfrak{su}_{2}}{2}}}{\Delta}  \overset{\mathfrak{su}_{2}}{2}$ & $\mathbb{Z}_2^{(1)}\times\mathbb{Z}_3^{(2)}$ & 0 & $6$ & $5$\\
			\midrule
			$(\text{IV},\text{IV})$ & $\overset{\mathfrak{su}_{3}}{3} \overset{\displaystyle \textcolor{blue}{\overset{\mathfrak{su}_{3}}{3}}}{1}  \overset{\mathfrak{su}_{3}}{3}$ & $\mathbb{Z}_3^{(1)}\times\mathbb{Z}_3^{(2)}$ & 0 & $12$ & $9$\\
			\midrule
            \bottomrule
	\caption{
			Type II Little String Theories involving only exceptional algebras,
			and the special Kodaira fibers of type II, III, and IV. The blue
			curve refers to the node obtained through minimal affinization of
			the corresponding SCFT.
	}
	\label{tab:LST-E-type}
\end{longtable}

\begin{landscape}
\begin{table}[H]
    \centering
    \begin{threeparttable}   
        \begin{tabular}{ccccc}
				Quiver & $\widehat{\kappa}_R$ & \text{dim}(\text{CB})\\
            \toprule
			$\begin{array}{rcl} 
					\textcolor{blue}{\overset{\fso_{2k}}{4}}\phantom{a} & \qquad & \quad\overset{\fsp_{k-4m}}{1}  \\ 
					\overset{\fso_{2k}}{4}  \, \overset{\fsp_{2(k-4)}}{1}&   
					\overset{\fso_{4(k-4)}}{4} \, \,\overset{\fsp_{2(k-8)}}{1} \, \,  \overset{\fso_{4(k-8)}}{4} \cdots \overset{\fsp_{2(k+4-4m)}}{1} \, \, & \overset{\fso_{4(k+4-4m)}}{4}
				 \overset{\fsp_{k-4m}}{1} 
		    \end{array}$
            & $8 (k - 2 m) ( 2 m-1)$ &  $1 + (2 + 4 k) m - 8 m^2$\\
			\midrule

			$\textcolor{blue}{\overset{\fso_{2k}}{4}} || \overset{\fsp_{k-4}}{1}$ 
			&  $4(k-2)$ &  $2k-3$\\
			\midrule
			$\overset{\mathfrak{sp}_{p}}{1} \overset{\fso_{4(p+4)}}{4} \overset{\displaystyle\textcolor{blue}{\overset{\mathfrak{sp}_{p}}{1}}}{\overset{\displaystyle\overset{\mathfrak{so}_{4(p+4)}}{4}}{\overset{\fsp_{3p+8}}{1}}} \, \, \overset{\fso_{4(p+4)}}{4}\,\overset{\fsp_{p}}{1}$ 
			& $24(p+3)$ &  $12p+38$\\
			\midrule
			$\overset{\mathfrak{f}_{4}}{5} 1 \overset{\displaystyle\overset{\mathfrak{su}_{2}}{2}}{\overset{\mathfrak{g}_{2}}{3}} 1\overset{\mathfrak{f}_4}{5}$
			& $48$ &  $16$\\
			\midrule
			$[\mathfrak{su}_2]\textcolor{blue}{1\overset{\mathfrak{e}_8}{8}}1\overset{\displaystyle\overset{\mathfrak{g}_2}{3}}{\overset{\mathfrak{su}_2}{2}}2$ 
			& $49$ & $15$\\
			$\textcolor{blue}{\overset{\mathfrak{e}_8}{7}}1\overset{\displaystyle\overset{\mathfrak{g}_2}{3}}{\overset{\mathfrak{su}_2}{2}}2$
			&$48$ & $14$\\
			\midrule
			$\overset{\mathfrak{su}_{2}}{2}\underset{\displaystyle\overset{\mathfrak{su}_2}{2}}{\overset{\textcolor{blue}{\displaystyle\overset{\mathfrak{su}_{2}}{2}}}{\overset{\mathfrak{g}_2}2}}\overset{\mathfrak{su}_{2}}{2}$
			& $16$ &  $10$\\
            \bottomrule
        \end{tabular}
    \end{threeparttable}
	\caption{
			Type II LSTs with no known dual six-dimensional quivers. All of them have $\kappa_P = 0$, and satisfy the worldsheet bound given in equation \eqref{flavor-bound-WZW}.
	}
    \label{tab:LST-outlier}
\end{table}
\end{landscape}

\section{Anomalies in Six Dimensions}\label{app:6D-anomalies}

The anomaly of a $D$-dimensional theory is encoded in a formal $(D+2)$-form,
the anomaly polynomial $I_{D+2}$. In six dimensions with $\mathcal{N}=(1,0)$
supersymmetry, it takes the form\footnote{We ignore possible Abelian symmetries
in this work.}
\begin{equation}\label{eqn:I8general-app}
  \begin{aligned}
    I_8 &=  \frac{\alpha}{24} c_2(R)^2+ \frac{\beta}{24}  c_2(R) p_1(T) + \frac{\gamma}{24}  p_1(T)^2 + \frac{\delta}{24} p_2(T) \cr &\quad + \sum_a \Tr F_a^2 \left(\mu^a p_1(T) + k^a c_2(R) + \sum_b \rho^{ab} \Tr   F_b^2\right) + \sum_a \nu^a \Tr F_a^4  \,,
  \end{aligned}
\end{equation}
where $c_2(F)=\frac{1}{4}\Tr F^2$ are one-instanton normalized traces, $c_2(R)$
is the second Chern class associated with the $\mathfrak{su}(2)_R$ R-symmetry
bundle, and $p_{1}(T)$, $p_2(T)$ are the first and second Pontryagin classes of
the spacetime tangent bundle, respectively. Given a generalized quiver, the
anomaly polynomial of an SCFT can be directly computed from its tensor branch
description \cite{Ohmori:2014kda}, and one distinguishing two different types
of contributions,
\begin{equation}
		I_8 = I_8^\text{1-loop} + I_8^\text{GS}\,.
\end{equation}
The first is obtained by summing the contributions of the various
supermultiplets in the spectrum. It is well known that in $D$ spacetime
dimensions, fermions appearing in the various supermultiplets, possibly
transforming in a representation $\mathcal{R}$ of the gauge and
flavor symmetries, lead to a contribution~\cite{AlvarezGaume:1983ig}:
\begin{equation}
		I_{D+2}^\text{fermion} = \left.\widehat{A}(T)\, \text{ch}_{\mathcal{R}}(F)\right|_{(d+2)\text{-form}}\,,\qquad \text{ch}_{\mathcal{R}}(F) = \Tr _{\mathcal{R}}\, e^{i F}\,.
\end{equation}
where $\widehat{A}(T)$ is the A-roof genus and $\text{ch}_{\mathcal{R}}(F)$ is
the Chern character of the field strength of the associated symmetry,
potentially also including the R-symmetry depending on the supermultiplet.
Similar anomalies are induced by self-dual two-forms, and one must also
consider contributions from more complicated objects such as E-strings. The
anomaly polynomial contributions for the supermultiplets that are relevant in
this work are
\begin{align}\label{I8-vector-tensor}
		\begin{split}
		I_8^\text{tensor} = &~ \frac{1}{24}c_2(R)^2 + \frac{1}{48}c_2(R)p_1(T) + \frac{1}{5760}\big(23p_1(T)^2 - 116 p_2(T)\big)\,,\\
				I_8^\text{vec}(F) =& -\frac{1}{24}\big(\tr _{\textbf{adj}}F^4 + 6 c_2(R)\tr _{\textbf{adj}}F^2 + \text{dim}(\mathfrak{g}) c_2(R)^2\big)\\
		& - \frac{1}{48}p_1(T)\big(\tr _{\textbf{adj}}F^2 + \text{dim}(\mathfrak{g})c_2(R)\big) - \frac{\text{dim}(\mathfrak{g})}{5760}\big(7p_1(T)^2 - 4p_2(T)\big)\,.
		\end{split}
\end{align}
Note that the traces $\tr_{\mathcal{R}}F^n$ must be converted to
one-instanton-normalized traces $\Tr _{\mathcal{R}}F^n$. We follow the
conventions of~\cite{Baume:2023onr} which outlines general procedure. The
conversion coefficients for the most common representations appearing in the
F-theory construction can be found in Appendix F of \cite{Heckman:2018jxk}. The
other group-theoretic quantities appearing explicitly in the main text are
summarized in Table \ref{tab:algebra-values}.  Another one-loop contribution
that arises is in this work is
\begin{equation}
    I_\text{sing} = \frac{1}{24}\left(\frac{1}{2}p_1(T)c_2(R) + \frac{1}{8}p_1(T)^2- \frac{1}{2}p_2(T)\right)\,,
\end{equation}
which finds its origin in the M-theory construction of conformal matter, where
it corresponds to the modes localized on the orbifold singularity probed by
M5-branes \cite{Ohmori:2014kda}.

\begin{table}
		\centering
		\begin{tabular}{ccccc}
				$\mathfrak{g}$ & $\rk(\mathfrak{g})$ & $\text{dim}(\mathfrak{g})$ & $h^\vee_\mathfrak{g}$ & $\Gamma$\\
				\toprule
				$\mathfrak{su}_{k}$ & $k-1$ & $k^2-1$ & $k$ & $k$\\
				$\mathfrak{so}_{8}$ & $4$ & $28$ & $6$ & $8$\\
				$\mathfrak{so}_{p\neq8}$ & $\lfloor \frac{p}{2}\rfloor$ & $\frac{1}{2}p(p-1)$ & $p-2$ & $2p-8$\\
				$\mathfrak{sp}_{k}$ & $k$ & $k(2k+1)$ & $k+1$ & ---\\
				$\mathfrak{g}_{2}$ & $2$ & $14$ & $4$ & ---\\
				$\mathfrak{f}_{4}$ & $4$ & $52$ & $9$ & ---\\
				$\mathfrak{e}_{6}$ & $6$ & $78$ & $12$ & $24$\\
				$\mathfrak{e}_{7}$ & $7$ & $133$ & $18$ & $48$\\
				$\mathfrak{e}_{8}$ & $8$ & $248$ & $30$ & $120$\\
				\bottomrule
		\end{tabular}
		\caption{
			Relevant quantities of simple Lie algebras. Note that the order
			$\Gamma$ of the discrete subgroups of $SU(2)$ are only defined
			for ADE algebra
			$\mathfrak{su}_n\,,\mathfrak{so}_{2k}\,,\mathfrak{e}_{6,7,8}$. 
		}
		\label{tab:algebra-values}
\end{table}

The one-loop term will generically be gauge anomalous. Due to the presence of
tensors in the spectrum, there is a Green--Schwarz--West--Sagnotti
mechanism~\cite{Green:1984bx, Sagnotti:1992qw} that harnesses the non-trivial
Bianchi identities to cure anomalies via anomaly inflow. This contribution,
often referred to as the Green--Schwarz or GS term, takes the generic form
\begin{equation}\label{gs}
		I_8^\text{GS} = \frac{1}{2} A_{ij} I^iI^j\,,\qquad I^i = A^{ik} c_2(F_k) + B^{ia}c_2(F_a) - a^i p_1(T) + h^i c_2(R)\,.
\end{equation}
At a generic point of the tensor branch, $A^{ij}=\eta^{ij}$ (we denote the
inverse by $A_{ij}$), $B^{ia}$ is the flavor pairing matrix, $a^{i}=2-A^{ii}$,
and $h^i=h^\vee_{\mathfrak{g}^i}$. When the curve is undecorated, we set
$h^i=1$. However, the GS term is usually computed away from the tensor branch
by blowing down $(-1)$-curves. Since such a move does not break any of the
symmetries of the theory, the complete anomaly polynomial remains invariant by
't Hooft anomaly matching, which we can use to track the changes in the various
quantities appearing in equation \eqref{gs}. One finds that after blowing down
a $(-1)$-curve
\begin{equation}
		\begin{aligned}
				\text{Quiver:}\qquad & \cdots\,\overset{\mathfrak{g_1}}{m_1}\,\overset{\mathfrak{g}_2}{1}\,\overset{\mathfrak{g_3}}{m_3}\,\cdots & & \cdots\overset{\mathfrak{g_1}}{(m_1-1)}~\overset{\mathfrak{g_3}}{(m_3-1)}\cdots\\
				\ell_I^\text{LST}:\qquad & \cdots~\ell_1~ \ell_2~ \ell_3~\cdots & \qquad\longrightarrow\qquad &\cdots \qquad\ell_1\,\qquad~ \ell_3\qquad\cdots\\
				y^I: \qquad & \cdots~y^{1}~y^{2}~y^{3}~\cdots & &\cdots (y_1 + y_2)~(y_2 + y_3)\cdots
		\end{aligned}
\end{equation}
Additional details can be found in e.g.
\cite{Ohmori:2014kda,DelZotto:2015isa,DelZotto:2020sop, Baume:2021qho}.

\subsection[Anomalies of Strings in 6D \texorpdfstring{$\mathcal{N}=(1,0)$}{N=(1,0)} Theories]{Anomalies of Strings in 6D \texorpdfstring{$\boldsymbol{\mathcal{N}=(1,0)}$}{N=(1,0)} Theories}\label{app:6D-strings}

In six dimensions, self-dual tensor fields couple naturally to strings, giving
rise to an $\mathcal{N}=(0,4)$ theory on their worldsheet, whose central charges and flavor levels are related to those of the six-dimensional bulk theory.  We review here how one can derive bounds using unitarity and the central charges of the worldsheet theory, following \cite{Shimizu:2016lbw, Kim:2019vuc, Lee:2019skh}.

From the worldsheet point of view, the strings have the flavor symmetry
\begin{equation}
		\mathfrak{so}(4)_{\mathcal{N}}\oplus\mathfrak{su}(2)_R\oplus_I\mathfrak{g}^I \oplus_A\mathfrak{f}^A\,,
\end{equation}
corresponding to the directions normal to the string, as well as the
R-symmetry, flavor and gauge symmetries of the bulk theory. The latter appear as flavor symmetries on the worldsheet. The indices $I$ and $A$ runs over
the 6D gauge and flavor symmetries, respectively. Furthermore, we decompose the
normal directions as $\mathfrak{so}(4)_{\mathcal{N}} =
\mathfrak{su}(2)_{\mathcal{A}}\oplus\mathfrak{su}(2)_{\mathcal{B}}$.\footnote{The
		normal directions are often denoted as
		$\mathfrak{su}(2)_L\oplus\mathfrak{su}(2)_R$, while the six-dimensional
		R-symmetry is denoted $\mathfrak{su}(2)_I$. Our nomenclature differs
		slightly from that of \cite{Shimizu:2016lbw}, since we are
mainly interested in six-dimensional quantities in the main text. We therefore
reserve uppercase letters for the 6D R-symmetry, and use lowercase letters for the left-/right-handed modes on the worldsheet.} Ignoring the flavor-
and gauge-symmetry factors for a moment, we decompose the supercharges into
two-dimensional quantities,
\begin{equation}
\begin{aligned}
		\mathfrak{so}(6)\oplus\mathfrak{su}(2)_R ~&\longrightarrow~ \mathfrak{u}(1)\oplus\mathfrak{su}(2)_{\mathcal{A}}\oplus\mathfrak{su}(2)_{\mathcal{B}}\oplus\mathfrak{su}(2)_R\,,\\
	(\bm{4}, \bm{2}) ~&\longrightarrow~ (\bm{2},\bm{1},\bm{2})_+ \oplus (\bm{1},\bm{2},\bm{2})_-\,,
\end{aligned}
\end{equation}
where the subscript refers to the chirality of the corresponding worldsheet
fermions; the R-symmetry of the $\mathcal{N}=(0,4)$ theory is then identified
with
$\mathfrak{so}(4)_{R,\text{2D}}=\mathfrak{su}(2)_\mathcal{B}\oplus\mathfrak{su}(2)_R$.

As for their six-dimensional counterparts, the levels of the flavor symmetries,
as well as the gravitational anomalies, are encoded in the anomaly polynomial
of the theory. Given a collection of strings with intersection pairing
$\eta^{IJ}$ associated with two-forms $B_I$, it takes the generic form
\cite{Shimizu:2016lbw}
\begin{equation}
\label{eq:I4TS}
\begin{aligned}
		I_4 &= \frac{1}{2}Q_I\eta^{IJ}Q_J \chi(N) + Q_I I^{I}\,,\\
		I^I &= B^{Ia}c_2(F_a) - a^I p_1(T_6) + h^I c_2(R)\,,
\end{aligned}
\end{equation}
where we defined $a^I = (2-\eta^{II})$, and $B^{Ia}$ is the 6D flavor pairing
matrix. 
The coefficients $Q_I$ are the charges of the string and $h^I =
h^\vee_{\mathfrak{g}^I}$ the dual Coxeter number of the associated algebra. Since the Euler density $\chi(N)$ of the $\mathfrak{so}(4)_{\mathcal{N}}$ normal bundle, and the Pontryagin class of the bulk tangent bundle of the bulk $p_1(T_6)$ can be decomposed as
\begin{equation}
	\chi(\mathcal{N}) = c_2(\mathcal{A})-c_2(\mathcal{B})\,,\qquad 
	p_1(T_6) = p_1(T_2) + p_1(\mathcal{N}) = p_1(T_2) - 2c_2(\mathcal{A}) - 2c_2(\mathcal{B})\,,
\end{equation}
we can rewrite~\eqref{eq:I4TS} in terms of worldsheet quantities as
\begin{equation}\label{I4-2D-UV}
	\begin{aligned}
			I_4 =& -\frac{6Q_I a^I}{24}\,p_1(T_2) + Q_I h^I\,c_2(R) + Q_IB^{Ia}\,c_2(F_a) \\
			&+ \frac{1}{2}\big(Q_Ia^I + Q_I\eta^{IJ} Q_J)c_2(\mathcal{A}) + \frac{1}{2}(Q_Ia^I - Q_I\eta^{IJ} Q_J)c_2(\mathcal{B})\,,
	\end{aligned}
\end{equation}
The levels can then be directly inferred from this expression. In
particular, the level of the
$\mathfrak{su}(2)_r=\mathfrak{su}(2)_{\mathcal{B}}$ is easily found to be
\begin{equation}
		\begin{gathered}
				k_r = k_{\mathcal{B}} =  \frac{1}{2}\left(Q_Ia^I - Q_I \eta^{IJ}Q_J\right)\,.
		\end{gathered}
\end{equation}
In the deep IR, the worldsheet theory flows to a two-dimensional CFT. The coefficients of the UV anomaly polynomial can be related to the central charges. These are the quantities associated with the relevant poles of the OPE of the energy-momentum tensor $(\mathcal{T}, \overline{\mathcal{T}})$ and the holomorphic currents $\mathcal{J}^A$ of the non-Abelian symmetries, see e.g.\ Appendix A of \cite{Benini:2013cda} for a concise review:
\begin{equation}
		\mathcal{T}(z)\mathcal{T}(0) \sim \frac{c_l}{2z} + \dots\,,\quad
		\overline{\mathcal{T}}(\overline{z})\overline{\mathcal{T}}(0) \sim \frac{c_r}{2\overline{z}} + \dots\,,\quad
		\mathcal{J}^A(z)\mathcal{J}^B(0) \sim \frac{k^{A}\delta^{AB}}{z^2} + \dots\,,
\end{equation}
where $(z,\overline{z})$ are the coordinates on the worldsheet. The central
charges $c_l\,,c_r$ are positive by unitarity, and their difference is
given by the gravitational anomaly,
\begin{equation}
		c_r - c_l = k_G\,,\qquad I_4^\text{CFT} \supset -\frac{k_G}{24}p_1(T_2)\,.
\end{equation}
Note that along the flow, some of the modes will decouple as they become
massive, and those must be taken into account when computing the anomaly
polynomial of the worldsheet CFT,
\begin{equation}
		I_4 = I_4^\text{CFT} + I_4^\text{mas}\,,
\end{equation}
where $I_4$ is given in equation \eqref{I4-2D-UV}. In the cases relevant in
this work, the massive modes are those associated with the center-of-mass of
the string, and correspond to the contribution of a universal hypermultiplet
\begin{equation}
		I_4^\text{mas} = -\frac{2}{24}p_1(T_2) - c_2(\mathcal{A})\,.
\end{equation}
The gravitational anomaly of the IR SCFT is therefore given by 
\begin{equation}
		k_G = 6\,Q_Ia^I + 2\,.
\end{equation}
Along the flow, the $\mathfrak{su}(2)_l\oplus\mathfrak{su}(2)_r$ R-symmetry
might mix with the other $\mathfrak{su}(2)$ factors. However, in the cases
relevant here, it was shown through direct computations as well as holographic
considerations that the $\mathfrak{su}(2)_r=\mathfrak{su}(2)_{\mathcal{B}}$
R-symmetry component survives unchanged at the conformal fixed point
\cite{Lawrie:2016axq, Couzens:2017way, Lawrie:2018jut}. We can therefore infer
that the right-handed central charges, related to the level of the corresponding
factor, are given by
\begin{equation}
		c_r = 6\, k_{r}^\text{IR} = 6\, k_{\mathcal{B}}^\text{UV} \,.
\end{equation}
Finally, the presence of the holomorphic current implies that the IR
description is at least that of a Wess--Zumino--Witten (WZW) model with a
Kac--Moody algebra with flavor algebra $\mathfrak{g}=\oplus_a \mathfrak{g}^a$,
each factor being at level $k^a$. Here, the index $a$ runs \emph{a
priori} over both bulk flavor and gauge symmetries.  The central charge of such a
theory is fixed by the symmetry data, and must be smaller than the actual value
of $c_l$, giving a constraint on the possible flavor data from the left-handed
central charge \cite{Kim:2019vuc}:
\begin{equation}
		c_\text{WZW} = \sum_{a} \frac{k^a\,\text{dim}(\mathfrak{g}^a)}{k^a + h^\vee_{\mathfrak{g}^a}} \leq c_l\,.
\end{equation}

\section{LSTs with Enhanced Supersymmetry}
\label{app:Enhanced}
When one of the singularities $\fg$ that defines the LST is trivial, the number of unbroken SUSY generators is doubled. There are two kinds  of supersymmetry enhancement in six dimensions, depending on which singularity is trivial. In terms of $\mathcal{N}=(1,0)$ multiplets, we have
\begin{enumerate}
		\item $\mathcal{N}=(1,0) \rightarrow \mathcal{N}=(1,1)$: an adjoint-valued hypermultiplet and a vector multiplet recombine into a $\mathcal{N}=(1,1)$ vector multiplet.
		\item $\mathcal{N}=(1,0) \rightarrow \mathcal{N}=(2,0)$: singlet hypermultiplets and tensor multiplets recombine into a $\mathcal{N}=(2,0)$ tensor multiplet.
\end{enumerate}
The two theories are related by T-duality \cite{Bhardwaj:2015oru, DelZotto:2020sop}, which becomes evident when viewed from the M-theory geometry perspective. There, the total space is given as 
\begin{align}
    X_3 = (\mathbbm{T}^2_A \times \mathbbm{C})/\Lambda) \times \mathbbm{T}^2_B
\end{align}
which is endowed with $\mathcal{N}=2$ supersymmetry in five dimensions since we do not have the full $SU(3)$ holonomy. The theory has two torus fibrations and thus two 6D lifts: when lifting the theory by sending vol$(\mathbbm{T}^2_B)  \rightarrow \infty$, we obtain the $\mathcal{N}=(2,0)$ IIB limit, which is just the constant F-theory torus over the LST base $B_2=(\mathbbm{T}^2_A \times \mathbbm{C})/\Lambda_\text{ADE}$. When lifting to 6D by sending vol$(\mathbbm{T}^2_A)  \rightarrow \infty$, the resulting base is simply $B_2 = \mathbbm{C} \times \mathbbm{T}^2_B$.
From the F-theory perspective, there is a 7-brane stack at the origin of $\mathbbm{C}$ that wraps $\mathbbm{T}^2_B$, which yields an extra hypermultiplet in the adjoint representation. Combined with the 
vector multiplet of the seven-brane, this enhances SUSY to 
 $\mathcal{N}=(1,1)$ in~6D. 

The above structure can be deformed by deforming the base slightly when moving to a Type $\text{I}_1$ or Type II base \cite{Bhardwaj:2015oru}. In this picture, when the IIA base has a $\text{I}_1$ degeneration, the adjoint representation decomposes as
\begin{align}
    \textbf{Adj} \rightarrow \mathbf{S}^2 + \mathbf{A}^2 \, ,
\end{align}
modulo a singlet. Indeed, the degeneration locus of the F-theory elliptic fiber $\text{I}_m$ signals the precence of a defect that breaks half of the supersymmetries. Similarly, when switching to the IIB dual, we have an $A_{N-1}$ LST, but with $\text{I}_1$ fibers. From the IIB perspective, there is a single $D7$ brane that breaks half of the supersymmetry. In field theory, this corresponds again to an obstruction to enhance $\mathcal{N}=(1,0) \rightarrow \mathcal{N}=(2,0)$. Indeed, while there are $m$ copies of the $\text{I}_1$ D7 brane over each tensor multiplet which do not carry a gauge group, there are nevertheless some matter states. Due to the $\text{I}_2$ enhanced singularity at each intersection locus, a massless hypermultiplet charged under a massive $U(1)$ trapped between each tensor arises. These $\mathcal{N}=(1,0)$ hypermultiplets cannot recombine with the $\mathcal{N}=(1,0)$ tensors and therefore obstructs any possibility of enhancement to $\mathcal{N}=(2,0)$. 

\bibliography{references}
\bibliographystyle{utphys}

\end{document}